\documentclass[preprint,12pt,authoryear]{elsarticle}
\usepackage{epsfig}
\usepackage{graphicx}
\usepackage{amssymb}
\usepackage{amsmath}
\usepackage{times}
\usepackage{blindtext}
\usepackage[utf8]{inputenc}
\usepackage{roussev}
\usepackage{color}
\usepackage{mathrsfs}
\usepackage{xcolor} 
\usepackage{soul}
\usepackage{cancel}

\newcommand{\sSun}{ {\scriptscriptstyle{\rm \odot}} }

\newcount\doepsf
\newcommand{\cf}{{cf.}}
\newcommand{\eg}{{e.g.,}}
\newcommand{\alf}{Alfv\'en }

\newcommand{\be}{\begin{equation}}
\newcommand{\ee}{\end{equation}}
\newcommand{\bea}{\begin{eqnarray}}
\newcommand{\eea}{\end{eqnarray}}
\newcommand{\bean}{\begin{eqnarray*}}
\newcommand{\eean}{\end{eqnarray*}}

\newcommand{\mnras}{{Monthly Notes Roy. Astron. Soc.}}
\newcommand{\planss}{{Planet. Space Sci.}}
\newcommand{\apj}{{The Astrophysical Journal}}
\journal{Journ. Comp. Phys.}
\begin{document}
\begin{frontmatter}
\title{High Resolution Finite Volume Method for Kinetic Equations with Poisson Brackets}
\author[label1]{
Igor V. Sokolov\corref{cor1}}
\ead{igorsok@umich.edu}
\author[label1,label2]{Haomin Sun}
\ead{shm@mail.ustc.edu.cn}
\author[label1]{Gabor Toth}
\ead{gtoth@umich.edu} 
\author[label1]{Zhenguang Huang}
\ead{zghuang@umich.edu}
\author[label1]{Valeriy Tenishev}
\ead{vtenishe@umich.edu}
\author[label1]{Lulu Zhao}
\ead{zhlulu@umich.edu}
\author[label3]{Jozsef Kota}
\ead{kota@lpl.arizona.edu}
\author[label4]{Ofer Cohen}
\ead{ofer_cohen@uml.edu}
\author[label1]{and Tamas I. Gombosi}
\ead{tamas@umich.edu}
\cortext[cor1]{Corresponding author}

\affiliation[label1]{organization={Center for Space Environment Modeling, University of Michigan},
addressline={2455 Hayward St}, 
city={Ann Arbor}, 
state={MI},
postcode={48109},
country={USA}}
\affiliation[label2]{organization={CAS Key Laboratory of Geospace Environment, Department of Geophysics and Planetary Science, University of Science and Technology of China}, city={Hefei}, country={China}} 
\affiliation[label3]{organization={University of Arizona},city=Tucson,state={Arizona},country=USA}
\affiliation[label4]{organization={Lowell Center for Space Science and Technology, University of Massachusetts Lowell},city=Lowel,state={Massachusetts},country=USA}
\begin{abstract}
Simulation of plasmas in electromagnetic fields requires numerical solution of a kinetic equation that describes the time evolution of the particle distribution function. In this paper we propose a finite volume scheme based on integral relation for Poisson brackets to solve the Liouville equation, the most fundamental kinetic equation. The proposed scheme conserves the number of particles, maintains the total-variation-diminishing (TVD) property, and provides high-quality numerical results. Other types of kinetic equations may be also formulated in terms of Poisson brackets and solved with the proposed method including the transport equations describing the acceleration and propagation of Solar Energetic Particles (SEPs), which is of practical importance, since the high energy SEPs produce radiation hazards. The proposed scheme is demonstrated to be accurate and efficient, which makes it applicable to global simulation systems analyzing space weather.
\end{abstract}

\begin{keyword}
finite volume method \sep kinetic equation \sep Liouville equation \sep Poisson bracket \sep TVD scheme
\end{keyword}
\end{frontmatter}

\section{Introduction}
\label{Sec:Introduction}
In astrophysics and space science, a hierarchy of models is used to simulate plasma motions. At the base of hierarchy there is fluid dynamics \cite[or hydrodynamics, \cf,][]{Landau1959}, which treats the moving media as fluids. Being applicable to a wide range of physical and technical problems, Computational Fluid Dynamics (CFD) has been developed to be a powerful applied science  employing a variety of numerical methods \cite[see the review by][]{Hirsch97}. Among them, the  {\it finite volume approach} is widely used. This framework treats the governing equations of CFD as a system of {\it conservation laws}, which are actually Partial Differential Equations (PDEs) of a special kind, mathematically expressing the conservation of physical quantities such as mass, momentum and energy. Specifically, for each of these conserved variables the conservation law reads:
\be\label{eq_conservationlaw}
\partial_tU+\nabla\cdot\mathbf{F}=0
\ee
where $U$ is the density of conserved quantity, $\partial_tU$ is its partial time derivative and the vector $\mathbf{F}$ is the flux function, $\nabla$ being the differential operator with regard to spatial coordinates. Once Eq.~(\ref{eq_conservationlaw}) is integrated over a control volume, the integral of the term, $\nabla\cdot\mathbf{F}$, reduces to a surface integral of the flux function over the boundary of the control volume. Therefore, if the computational domain is decomposed into a set of control volumes (cells), the time derivative of the conserved variable within each control volume reduces to the exchange by the {\it numerical fluxes} between each pair of neighboring cells, these numerical fluxes being essentially the integral of the flux function, $\mathbf{F}$, over the interface (the shared boundary) of the two cells. Gauss' theorem is formulated via the scalar product of the flux function and the external area vector of the boundary for $i$th cell, which is at the same time the negative of the external area vector to the same interface for $j$th neighboring cell, so that the numerical flux from $i$th cell to $j$th cell is always equal to the negative of the flux from $j$th cell to $i$th cell. Therefore, the time derivative of the total integral of the conserved quantity over the computational domain reduces to mutually cancelling contributions from each numerical flux to the neighboring cells, resulting in the automatically conserved  total quantities, unless there is a non-vanishing flux of these quantities through the external boundary of the computational domain.

In the solar-terrestrial environment, the  plasma motion may be affected by the solar, interplanetary or planetary magnetic field. To account for both the magnetic field contribution to the force acting on a plasma and evolution of the magnetic field frozen into the moving plasma, the magnetohydrodynamic (MHD) approximation is used \citep[see][]{Shore1992}. The finite volume scheme for MHD treats eight scalar conservation laws in the three-dimensional (3-D) case, which further increases the complexity compared to the case of five conservation laws in CFD. To solve a 3-D system of conservation laws using \textit{high-resolution} schemes developed for a linear 1-D advection equation, 
\begin{equation}\label{eq:advection}
 \partial_tU+c\partial_xU=0,   \qquad c={\rm const},
\end{equation} 
one can  employ  the concept of \textit{characteristics} associated with different types of eigen waves, e.g., MHD waves \citep[see][]{Hirsch97}. Usually the characteristics are the lines in time-space along which the \textit{Riemann invariants} are conserved, which are some functions of the conserved variables. \cite{powell99} demonstrated how to use characteristics to construct high resolution numerical flux for computational MHD, including a magic ``8th'' wave, which flushes away non-zero $\nabla\cdot\mathbf{B}$, if any.    

The top level in the model hierarchy is represented by two main families of numerical models providing a kinetic  description \cite[see][]{Lifshitz1981} for plasmas in the presence of electromagnetic fields. Both categories of numerical methods in fact solve the same mathematical PDE, describing the evolution of the velocity distribution function (VDF). An approach currently becoming popular is to solve VDF numerically from the kinetic PDE by applying directly the {\it finite difference scheme} to discretize the derivatives over coordinates and momenta, in the equation. A rather advanced numerical framework based on this approach is the hybrid-Vlasov simulator, the ``Vlasiator'' described by \cite{Hoilijoki2019}. An alternative approach is to integrate the same set of PDEs along Hamiltonian trajectories of the charged particles in the phase space of coordinates, $q_l$,  and momenta, $p_l$. The actual computational algorithm for the latter approach, for example, within the framework of particle-in-cell (PIC) scheme  is to compute a huge ensemble of particles, jointly moving in the electromagnetic fields, as described by \cite{Birdsall05}. Examples of application of the particle approach to solve the Boltzmann equation for neutral species and dust particles with a variable electric charge in planetary and space environments are discussed by, \eg \cite{tenishev21}.
From the mathematical standpoint, however, such schemes do not solve the motion of individual plasma particles, but they just sample the averaged value of the VDF function about some point of the phase space and then transport this value along the the particle Hamiltonian trajectory. Note, that the VDF advection along the Hamiltonian trajectory is similar to the Riemann invariant transport along the characteristics in CFD/MHD.  

In the present paper, we propose a new finite volume scheme to solve the kinetic equation. 
We employ a finite volume scheme, thus avoiding a potential drawback of the finite difference approach, which might fail to maintain the important particle number conservation (this drawback is seldom discussed, although it may be pertinent, particularly, to the popular \textit{semi-Lagrangian} schemes including the original algorithm by \cite{Cheng1976}). In this way we can benefit from a variety of useful tools and methods developed for CFD/MHD, \eg\ the Total-Variation-Diminishing (TVD) principle. At the same time we benefit from the characteristics property of the Hamiltonian trajectory, onto which we project the VDF gradient.

At the heart of our new approach is the concept of Poisson bracket, enabling the use of the finite volume approach. While in a canonical Liouville equation the Poisson bracket follows from the Hamiltonian theory, for more practical application to the focused-transport equation describing the acceleration and transport of  Solar Energetic Particles (SEP) in the heliosphere, the possibility to re-write the equation via the Poisson brackets is non-evident. However, once introduced, Poisson brackets greatly facilitate the numerical model, allowing us to efficiently produce high-quality simulation results. Note, that both the close connection between the Poisson bracket and conservation laws \cite[see, e.g.,][Ch.17]{Landau1976} and successful application of the Poisson bracket to construct energy-conserving schemes for the shallow water equations by \cite{Salmon2004} are well-known, so the idea to use Poisson brackets for computations is not quite new and was directly exploited before \cite[see, e.g.,][]{Hammett2019a,Hammett2019b}.
\section{The Liouville Equation and Poisson Brackets}
\label{Sec:Liouville}
The general equation describing evolution of a velocity distribution function, $f(t,q_l,p_l)$, for a dynamical system with a Hamiltonian function, $H(q_l,p_l)$, with $q_l,p_l$, $l=1,2,3$, being the generalized coordinates and momenta for $l$th degree of freedom, 
has a form as follows:
\be\label{eq:Liouville}
\partial_tf+\sum_{l=1}^3{\left(\frac{\partial f}{\partial q_l}\frac{\partial H}{\partial p_l}-\frac{\partial f}{\partial p_l}\frac{\partial H}{\partial q_l}\right)}=0.
\ee
In application to the classical particle methods $q_1,q_2,q_3$ may be three Cartesian coordinates of a particle location, $\mathbf{x}$, while $p_1,p_2,p_3$ being three components of its momentum, $\mathbf{p}$, so that Eq.~(\ref{eq:Liouville}) claims that the value of $f$ is constant along the particle Hamiltonian trajectory:
\begin{equation}\label{eq:fadvection}
\left[\frac{\mathrm{d}f\left(t,\mathbf{x},\mathbf{p}\right)}{\mathrm{d}t}\right]_{\frac{\mathrm{d}\mathbf{x}}{\mathrm{d}t}=\frac{\partial H}{\partial \mathbf{p}},\,\frac{\mathrm{d}\mathbf{p}}{\mathrm{d}t}=-\frac{\partial H}{\partial \mathbf{x}}}=0.
\end{equation}
Eq.~(\ref{eq:fadvection}) justifies the particle methods to solve Eq.~(\ref{eq:Liouville}): if at the initial time instant the distribution function is sampled with {\it macroparticles}, $N_p,\mathbf{x}_p,\mathbf{p}_p$, which are the number of real particles, average coordinates and average momentum per macroparticle enumerated with a subscript index, $p$, then on solving the Hamiltonian trajectory equation for each macroparticle we sample the evolving distribution function. Even though in the present paper we do not employ particle methods at all, we benefit from the observation that Eq.~(\ref{eq:Liouville}) in effect reduces to a linear advection equation (\ref{eq:fadvection}) transporting the VDF value along the Hamiltonian trajectory.
\subsection{Poisson Brackets and Conservation of Particles}
\label{Sec:Poisson}
In terms of the Poisson brackets, which we define as:
\be\label{eq:Poisson}
\left\{f;H\right\}_{q_l,p_l}=\frac{\partial f}{\partial q_l}\frac{\partial H}{\partial p_l}-\frac{\partial f}{\partial p_l}\frac{\partial H}{\partial q_l},\ee
the Liouville equation ~(\ref{eq:Liouville}) can be re-written as:
\be\label{eq:viaPoisson}
\partial_tf+\sum_{l=1}^3{\left\{f;H\right\}_{q_l,p_l}}=0.
\ee
Note, that we define the Poisson bracket as each term in Eq.~(\ref{eq:Liouville}), rather than the sum of the terms as defined by \cite{Landau1976}. 
In the more general case, Eq.~(\ref{eq:viaPoisson}) determines the time evolution of VDF via a total of $L$ Poisson brackets, for each of them $q_l$, $p_l$, $l=1,2,\dots,L$ being an arbitrary pair of independent phase variables. With no loss in generality of the methods discussed below, the Hamiltonian function may be different in different Poisson brackets and they may or may not have a meaning of energy expressed in terms of coordinates and momenta.

A major advantage of the Poisson brackets is that they explicitly conserve the total number of particles. The particle number is defined as an integral of the distribution function over the entire phase space: $\int{\mathrm{d}\mathbf{\Gamma} f}$  (the phase-space volume element is $\mathrm{d}\mathbf{\Gamma} = \prod_l{\left(\mathrm{d}q_l\mathrm{d}p_l\right)}$). The particle number is conserved, $\frac{\mathrm{d}}{\mathrm{d}t}\int{\mathrm{d}\mathbf{\Gamma}\;f}=-\sum_l{\int{\mathrm{d}\mathbf{\Gamma} \left\{f;H\right\}_{q_l,p_l}}}\equiv0$, since 
\be\label{eq_4}
\int\limits_{-\infty}^\infty\int\limits_{-\infty}^{\infty}
{
\mathrm{d}q_l\mathrm{d}p_l\left(\frac{\partial f}{\partial q_l}\frac{\partial H}{\partial p_l}-\frac{\partial f}{\partial p_l}\frac{\partial H}{\partial q_l}\right)}=
\int\limits_{-\infty}^\infty\int\limits_{-\infty}^{\infty}{
\mathrm{d}q_l\mathrm{d}p_l\left[
\frac{\partial}{\partial q_l}\left(f\frac{\partial H}{\partial p_l}\right)-\frac{\partial }{\partial p_l}\left(f\frac{\partial H}{\partial q_l}\right)
\right]}\equiv0.
\ee
\subsection{Phase-Space Control Volume Formulation: Second Order Flux}
\label{Sec:ControlVolume}
Eqs.~(\ref{eq:Poisson}-\ref{eq_4}) can be combined to find the rate of change of particle number in a {\it control volume}. For simplicity we consider a rectangular one: $V=\Pi_l(\Delta q_l\Delta p_l)$, centered at the point, $(q^{c}_1,\dots, q^{c}_L,$ $p^{c}_1,\dots,p^{c}_L)$:
\be\label{eq_5}
\int\limits_V{\mathrm{d}\mathbf{\Gamma} \partial_tf}=
-\sum_l{\int{\prod\limits_{m\ne l}\left(\mathrm{d}q_m\,\mathrm{d}p_m\right)}\int\limits_{q_l^{c}-\frac{\Delta q_l}2}^{q_l^{c}+\frac{\Delta q_l}2}\mathrm{d}q_l\int\limits_{q_l^{c}-\frac{\Delta q_l}2}^{p_l^{c}+\frac{\Delta p_l}2}{
\mathrm{d}p_l\left[
\frac{\partial}{\partial q_l}\left(f\frac{\partial H}{\partial p_l}\right)-\frac{\partial }{\partial p_l}\left(f\frac{\partial H}{\partial q_l}\right)
\right]}}.
\ee
In terms of a two-component differential operator, $\nabla_l=\left(\frac{\partial}{\partial q_l},\frac{\partial}{\partial p_l}\right)$, the integrand in Eq.~(\ref{eq_5}) reads:
\be
\frac{\partial}{\partial q_l}\left(f\frac{\partial H}{\partial p_l}\right)-\frac{\partial }{\partial p_l}\left(f\frac{\partial H}{\partial q_l}\right)=\nabla_l\times\left(f\nabla_lH\right).
\ee
Now, using Stokes' theorem and the chain rule, we arrive at the finite volume formulation of Eq.~(\ref{eq:viaPoisson}):
\be\label{eq:finitevolume}
\frac{\mathrm{d}}{\mathrm{d}t}\int\limits_V{\mathrm{d}\mathbf{\Gamma} f}=-\sum_l{\int{\prod\limits_{m\ne l}\left(\mathrm{d}q_m\mathrm{d}p_m\right)}\oint{\mathrm{d}H\,f}}.
\ee

\begin{figure}[t]
\centering
\includegraphics[height={3.5in},width={5in}]{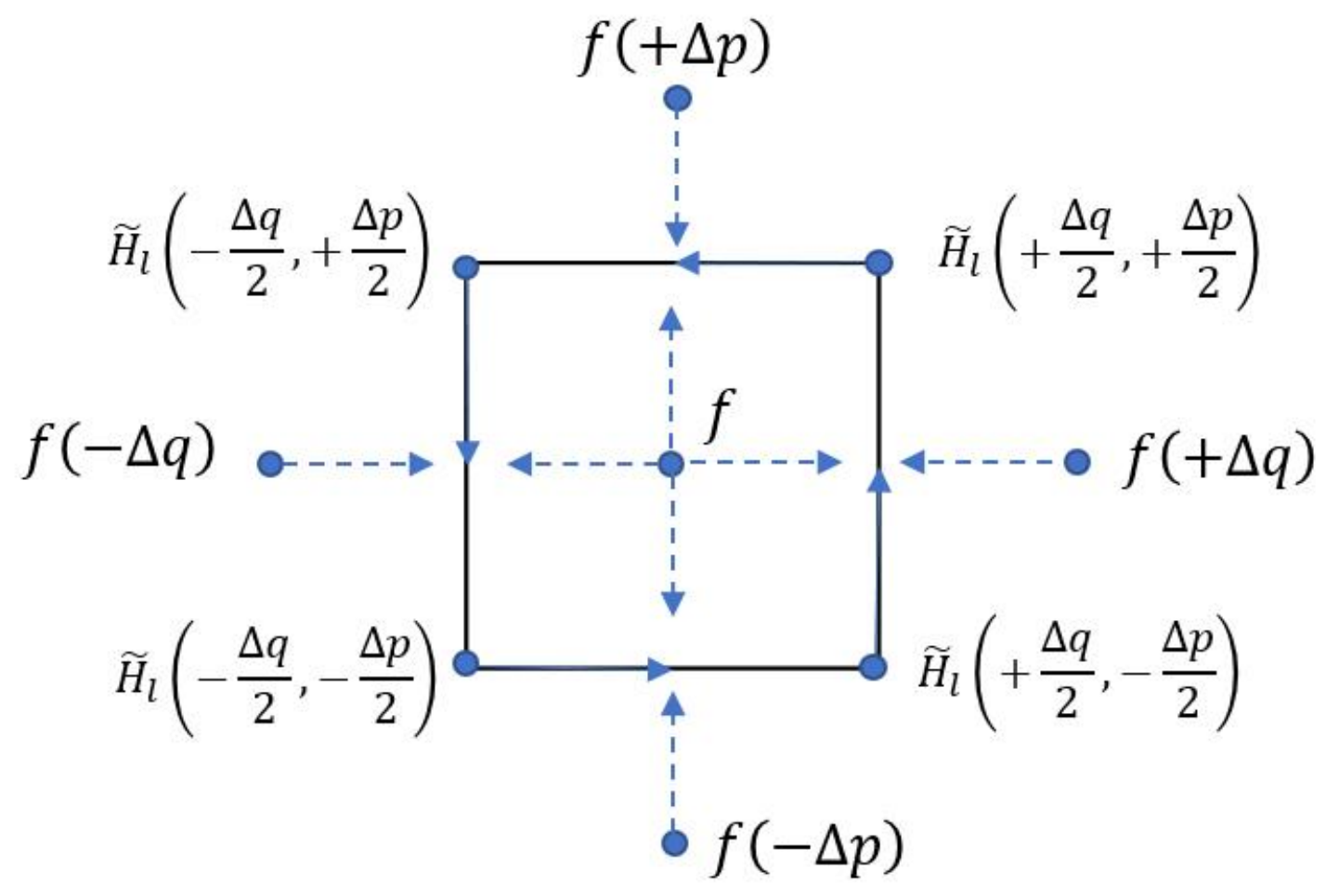}
\caption{Illustration of the control volume method for a rectangular cell. The integration of each Poisson bracket over the control volume reduces to the integral $\int{f\,d\tilde{H}_l}$ on the $q_l,p_l$ plane over the closed contour embracing the volume. To obtain the numerical scheme,     the contour integral is evaluated in terms of differences of the effective Hamiltonian function at the vertices as well as the face-centered values of the distribution function, interpolated from the neighboring cells as shown by the dashed arrows. For the second order scheme, the
arithmetic average of two VDF values is used, while for the first order monotone flux, the upwinded value is chosen as described in Section~\ref{Sec:Upwind}.  
}
\label{fig:scheme}
\end{figure}
The integration contour on the $(q_l,p_l)$ plane consists of four segments (see Fig.~\ref{fig:scheme}), sequentially connecting points $\left(q_l^{c} - {\Delta q_l}/2, p_l^{c} - {\Delta p_l}/2\right)$, $\left(q_l^{c} + {\Delta q_l}/2,p_l^{c} - {\Delta p_l}/2\right)$, $\left(q_l^{c} + {\Delta q_l}/2,p_l^{c} + {\Delta p_l}/2\right)$ and $\left(q_l^{c} - {\Delta q_l}/2,p_l^{c} + {\Delta p_l}/2\right)$. 
Herewith, we do not list those coordinates, which are equal to their cell-centered values.  Thus, similar to a conservative scheme for Eq.~(\ref{eq:Liouville}), the time derivative of the particle number in the control volume may be expressed in terms of numerical fluxes through its faces:
\be
\label{eq:fluxes}
\frac{\mathrm{d}}
{\mathrm{d}t}\int\limits_V{\mathrm{d}\mathbf{\Gamma} f}=-\sum_l{
\left(F_{q^{c}_l+\frac{\Delta q_l}2}-F_{q^{c}_l-\frac{\Delta q_l}2}+F_{p^{c}_l+\frac{\Delta p_l}2}-F_{p^{c}_l-\frac{\Delta p_l}2}\right)},\ee
The integrals of the distribution function are expressed in terms of the face-centered values:
\bea\label{eq:fluxq} 
&&F_{q^{c}_l\pm\frac{\Delta q_l}2}=f\left(q^{c}_l\pm\frac{\Delta q_l}2\right)
\int\limits_{p^{c}_l-\frac{\Delta p_l}2}^{p^{c}_l+\frac{\Delta p_l}2}{
\mathrm{d}p_l \frac{\partial \widetilde{H}_l\left(q^{c}_l\pm\frac{\Delta q_l}2,p_l\right)}{\partial p_l}}=\nonumber\\
&=&
f\left(
q^{c}_l\pm\frac{\Delta q_l}2\right)
\left[
\widetilde{H}_l\left(
q^{c}_l\pm\frac{\Delta q_l}2,p^{c}_l+\frac{\Delta p_l}2\right)-\widetilde{H}_l\left(q^{c}_l\pm\frac{\Delta q_l}2,p^{c}_l-\frac{\Delta p_l}2\right)\right],
\eea
\bea\label{eq:fluxp}
&&F_{p^{c}_l\pm\frac{\Delta p_l}2}=-f\left(p^{c}_l\pm\frac{\Delta p_l}2\right)
\int\limits_{q^{c}_l-\frac{\Delta q_l}2}^{q^{c}_l+\frac{\Delta q_l}2}{\mathrm{d}q_l \frac{\partial \widetilde{H}_l\left(q_l,p^{c}_l\pm\frac{\Delta p_l}2\right)}{\partial q_l}}=\nonumber\\
&=&-f\left(p^{c}_l\pm\frac{\Delta p_l}2\right)\left[
\widetilde{H}_l\left(
q^{c}_l+\frac{\Delta q_l}2,p^{c}_l\pm\frac{\Delta p_l}2\right)-\widetilde{H}_l\left(q^{c}_l-\frac{\Delta q_l}2,p^{c}_l\pm\frac{\Delta p_l}2\right)\right],
\eea
where the {\it effective} Hamiltonian functions are introduced, each depending only on two variables:
\be\label{eq:effective}
\widetilde{H}_l(q_l,p_l)=\int{\prod\limits_{m\ne l}\left(\mathrm{d}q_m\mathrm{d}p_m\right)H(q_1,\dots,q_L,p_1,\dots,p_L)}.
\ee
On a simple grid consisting of identical rectangular boxes, the effective Hamiltonian is approximately equal to the full Hamiltonian function in which all arguments 
except for $q_l,p_l$ are set to cell-centered values: $q_m=q_m^{c},p_l=p^c_l$, and multiply this by $\prod_{m\ne l}\left(\Delta q_m\Delta p_m\right)=V/\left(\Delta q_l\Delta p_l\right)$. However, on more complicated grids, the integral formulation in Eq.~(\ref{eq:effective}) may be needed.

The numerical flux of particles along the coordinate $q_l$ is proportional to the corresponding component of the 3-D particle velocity vector $\frac{\partial \tilde{H}_l}{\partial p_l}$, while along the momentum axis $p_l$, the flux is proportional to the $-\frac{\partial \tilde{H}_l}{\partial q_l}$ component of the 3-D force vector.
The 3-D vectors of velocity and force may be combined into a 6-D phase space velocity vector $\mathbf{u}_6=(\frac{\partial H}{\partial \mathbf{p}}, -\frac{\partial H}{\partial \mathbf{q}})$, 
the product $\mathbf{u}_6 f$ being the particle flux density in the phase space. The conservation of particle number can be written as 
\be\label{eq:cons6}
\partial_tf +\nabla_6\cdot(f \mathbf{u}_6) =  \partial_tf + \left(\mathbf{u}_6 \cdot \nabla_6\right) f = \left(\frac{\mathrm{d}f}{\mathrm{d}t}\right)_H = 0
\ee
which is the same as (\ref{eq:fadvection}) with a different notation, $(\mathrm{d}/\mathrm{d}t)_H=\partial_t + \mathbf{u}_6 \cdot \nabla_6$ is the time derivative along the Hamiltonian trajectory. Here we used the divergence free property of the 6-D
phase space velocity
\be\label{eq:divfreeu6}
\nabla_6\cdot\mathbf{u_6} = \sum_{l=1}^3{ \frac{\partial^2 H}{\partial q_l \partial p_l}} - \sum_{l=1}^3{\frac{\partial^2 H}{\partial p_l \partial q_l}} = 0. 
\ee 
Such ``incompressibility'' of the 6-D velocity field in the phase space results in the Liouville theorem: the ``water bag'', which is the phase space element $\mathrm{d}\mathbf{\Gamma}$ occupied at some time instant by a chosen group of particles, moves with the particles and changes its form, but not its volume. The {\it water bag numerical methods} \cite[see, e.g.,][]{Andreev1979} employ this property. 

If the formulation (\ref{eq:fluxes}) is applied to a control volume in the phase space ({\it cell}) the conservative numerical scheme may be derived from the flux given by Eqs.~(\ref{eq:fluxq}-\ref{eq:fluxp}): the distribution function value needed to calculate numerical flux at each face is an arithmetic average of its values in cells neighboring across this face, the integrals of the Hamiltonian function just reducing to edge value differences of the Hamiltonian function. 
Within this framework, a \textit{semi-discrete} (i.e.,  combining an infinitesimal time interval with finite control volume) second order {\it numerical scheme}, $\left[\partial_tf\right]^{(2)}$, for solving the distribution function 
reads as follows:
\bea\label{eq:4deltaH}
\left[\partial_tf\right]^{(2)}=-\frac{1}{V}
\sum_{l=1}^L \left\{
\left[\tilde{H}_l\left(+\frac{\Delta q_l}2,+\frac{\Delta p_l}2\right)-\tilde{H}_l\left(+\frac{\Delta q_l}2,-\frac{\Delta p_l}2\right)\right]\frac{f(+\Delta q_l)+f}2+\right. \nonumber\\
+\left[\tilde{H}_l\left(-\frac{\Delta q_l}2,+\frac{\Delta p_l}2\right)-\tilde{H}_l\left(+\frac{\Delta q_l}2,+\frac{\Delta p_l}2\right)\right]\frac{f(+\Delta p_l)+f}2+\nonumber\\
+\left[\tilde{H}_l\left(-\frac{\Delta q_l}2,-\frac{\Delta p_l}2\right)-\tilde{H}_l\left(-\frac{\Delta q_l}2,+\frac{\Delta p_l}2\right)\right]\frac{f(-\Delta q_l)+f}2+\nonumber\\
\left.
+\left[\tilde{H}_l\left(+\frac{\Delta q_l}2,-\frac{\Delta p_l}2\right)-\tilde{H}_l\left(-\frac{\Delta q_l}2,-\frac{\Delta p_l}2\right)\right]\frac{f(-\Delta p_l)+f}2\right\},
\eea
where arguments for the distribution function value $f$ related to a considered control volume are not listed, while for other functions only differences in arguments with respect to the center of the considered cell are listed. Fig.~\ref{fig:scheme} illustrates with dashed arrows which cell-centered values of the distribution function are employed in the numerical fluxes. 

In Eq.~(\ref{eq:4deltaH}), for each Poisson bracket there is a sum of four terms, $\sum_j{\delta \tilde{H}_{l,j}\frac{f_j^{\rm ext}+f}2}$, where $\delta \tilde{H}_{l,j}$ is a properly signed difference of the reduced Hamiltonian functions $\tilde{H}_l$ at the two ends of the cell face $j$, while $f_j^{\rm ext}$ is the distribution function value in the neighboring cell
on the external side of cell face $j=1,\ldots,4$. 
It is important that for each Poisson bracket (i. e. for each $l$) the total of four Hamiltonian-dependent multipliers in this sum vanishes: $\sum_j\delta \tilde{H}_{l,j}=0$,
since they constitute the vanishing integral $\oint{dH}=0$ over the closed contour. For this reason,  the discretization in Eq.~(\ref{eq:4deltaH}) keeps a uniform solution $f={\rm const}$ to be a steady-state: $\partial_t f=-\frac{f}V\sum_l\sum_j\delta \tilde{H}_{l,j}=0$.

To simplify Eq. ~(\ref{eq:4deltaH}), we drop the straightforward summation over the Poisson Brackets (over $l$), and write the 4 terms as a summation over the four faces indexed by $j$:
\be\label{eq:deltaHj}
\left[\partial_tf\right]^{(2)}=-\frac{1}{V}
\sum_j{\delta \tilde{H}_j\frac{f+f^{\rm ext}_j}2},
\ee
where $f^{\rm ext}_j$ is the cell-centered value of distribution function in the cell across the $j$th face and $\delta \tilde{H}_j$ is the difference of the effective Hamiltonians along the $j$th face. For a canonical distribution function, per each face  of the control volume there is one and only one $\delta\tilde{H}_l$. Therefore, in this case Eqs.~(\ref{eq:deltaHj}) and (\ref{eq:4deltaH}) differ only by order of summation. However, in more complicated cases (see Section~\ref{Sec:NonCanonical}), more than one $\delta\tilde{H}_l$ may contribute to $\delta\tilde{H}_j$ for a given face.

As mentioned above,
the sum of $\delta\tilde{H}_j$ over all faces equals zero: $\sum_j\delta\tilde{H}_j=0$, which is an integral analog of Eq.~(\ref{eq:divfreeu6}). 
Therefore, we can introduce two groups, $\delta \tilde{H}^+_j$ and $\delta\tilde{H}^-_j$, of positive and negative $\delta \tilde{H}_j$ and partial sums, $\sum_{j,+}$ and $\sum_{j,-}$, over faces with positive and negative $\delta\tilde{H}$, respectively, so that the sum of positive contributions balance the sum of negative contributions: 
\begin{figure}[t]
\centering
\includegraphics[height={4.0in},width={4.0in}]{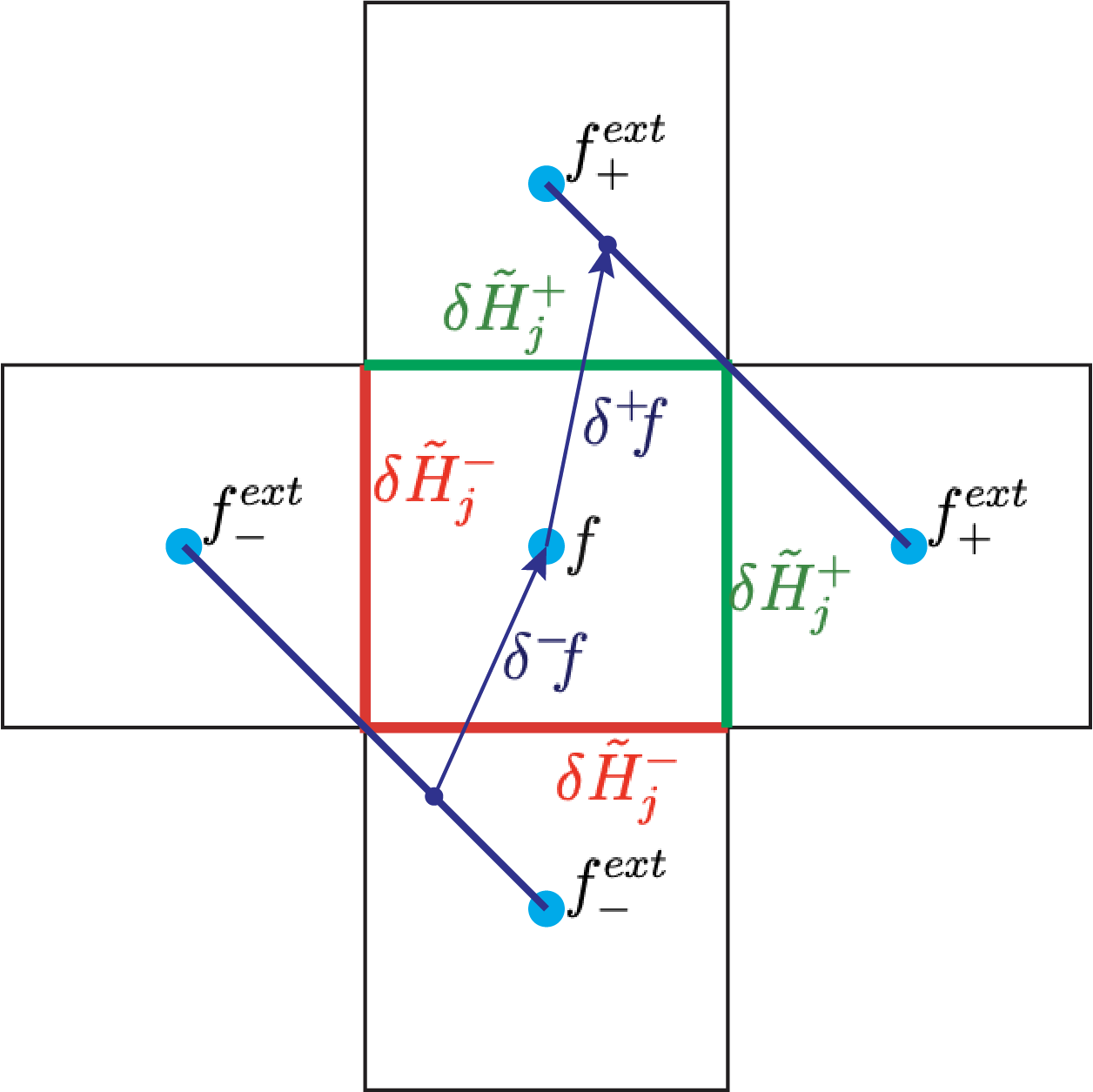}
\caption{This figure illustrates the definitions of $\delta^+f$ and $\delta^-f$. The green and red colors of the faces of the central cell indicate positive and negative signs of $\delta \tilde{H}_j$, respectively. The endpoint of arrow $\delta^+f$ and the start point of arrow $\delta^-f$ lay on the straight lines connecting the neighboring cell centers, since the distribution function at these points equals the weighted averages of $f_+^{\rm ext}$ and $f_-^{\rm ext}$, respectively. The $\delta^+f$ and $\delta^-f$ denote the downwind and upwind estimates for the  distribution function gradient along the Hamiltonian trajectory. The figure shows a typical case with 2 adjacent inflow ($\tilde{H}^-_{1,2}<0$) and 2 adjacent outflow fluxes ($\tilde{H}^+_{1,2}>0$), but in general there are many other possibilities.
}
\label{fig:Cartoon2}
\end{figure}
\be\label{eq:Hplusminus}
\sum_{j,+}{\delta \tilde{H}^+_j}=-\sum_{j,-}{\delta \tilde{H}^-_j}.
\ee
It is also convenient to introduce downwind and upwind estimates for the distribution function gradient along the trajectory of the Hamiltonian system (see Fig.~\ref{fig:Cartoon2}):
\be\label{eq:deltaf}
\delta^+ f=
\frac{\sum_{j,+}{\delta\tilde{H}^+_jf^{\rm ext}_j}}{\sum_{j,+}{\delta\tilde{H}^+_j}}-f,
\qquad 
\delta^- f= f-\frac{\sum_{j,-}{\delta\tilde{H}^-_jf^{\rm ext}_j}}{\sum_{j,-}{\delta\tilde{H}^-_j}}\equiv \frac{\sum_{j,-}{\delta\tilde{H}^-_j\left(f-f^{\rm ext}_j\right)}}{\sum_{j,-}{\delta\tilde{H}^-_j}}.
\ee
The second order scheme Eq.~(\ref{eq:deltaHj}) with the use of Eqs.~(\ref{eq:Hplusminus}) and (\ref{eq:deltaf}) may be formulated as follows:
\bea\label{eq:secondorder}
\left[\partial_tf\right]^{(2)}=-
\frac{\left(\sum_{j,+}{\delta \tilde{H}^+_j}\right)}V\frac{\delta^- f+\delta^+ f}2,
\eea
which is similar to the numerical flux $\left[\partial_t f\right]^{(2)}=-\frac{c}{\Delta x}\frac{\delta f^-+\delta f^+}2$ of the 1-D 
advection equation (\ref{eq:advection}).
Here, we employ high-resolution methods developed for the latter equation, thus benefiting from the characteristics property of the Hamiltonian trajectory.
\subsection{Phase-Space Control Volume Formulation: Upwind Monotone Flux}
\label{Sec:Upwind}
To convert a second order scheme shown by Eq. (\ref{eq:secondorder}) to a monotone first order flux, one needs to add to the right hand side (RHS) a minimal numerical diffusion, $D$, expressed as follows:
\be\label{eq:diffusion}
D=
\frac1{V}\sum_j{
\left\vert \delta\tilde{H}_j\right\vert\frac{f^{\rm ext}_j -f}2}=\frac{\left(\sum_{j,+}{\delta\tilde{H}^+_j}\right)}V\frac{\delta^+ f-\delta^- f}2.
\ee
The resulting first order numerical scheme, $\left[f_t\right]^{(1)}=\left[f_t\right]^{(2)}+D$, becomes:
\bea\label{eq:firstorder}
\left[\partial_tf\right]^{(1)}=-\frac{\left(\sum_{j,+}{\delta\tilde{ H}^+_j}\right)}V\delta^- f=-\frac1V\left[\left(\sum_{j,+}{\delta\tilde{ H}_j^+}\right)f+\sum_{j,-}{\delta\tilde{H}^-_jf^{\rm ext}_j}\right].
\eea
For faces with positive $\delta\tilde{H}_j$, the distribution function value, $f$, from the given control volume is used, otherwise the distribution function value, $f^{\rm ext}_j$, from the neighboring cell is involved. This choice provides the {\it upwinded flux}  \cite[see][]{Hirsch97}, since the sign of $\delta\tilde{H}_j$ determines the local direction of velocity or force, i.e. the direction of particle motion in the phase space (``wind'').

The first order explicit numerical flux, which advances the numerical solution for the distribution function from the time level $t$ through the time step, $\Delta t$, to the time level, $t+\Delta t$, is: 
\bea\label{eq:CFL}
f(+\Delta t)=f-C\,\delta^-f=(1-C)f-
\frac{\Delta t}V
\sum_{j,-}{\delta\tilde{H}^-_jf^{\rm ext}_j},\quad  C=\frac{\Delta t}V\sum_{j,+}{\delta \tilde{H}^+_j}.
\eea
Herewith, we denote the quantities related to the time level $t+\Delta t$ by a presence of  $+\Delta t$ in the argument list. The flux in Equation~(\ref{eq:CFL}) is monotone, since all coefficients of the distribution function values are non-negative ($-\tilde{H}^-_j>0$ and $1-C\ge0$) as long as the CFL condition is satisfied:
\be\label{eq:CFL<1}
C\le 1.
\ee
\subsection{Phase-Space Control Volume Formulation: Total Variation and TVD Property}
\label{Sec:TVD}
We want to  extend the first order monotone flux to second order of accuracy and at the same time avoid spurious oscillations in the numerical solution. In particular,
we are concerned with the monotonicity of the distribution function along the Hamiltonian trajectory. In 1-D case a \textit{total variation} (TV) can be introduced, and the TV diminishing (TVD) numerical schemes constructed to prevent  spurious oscillations. Particularly, for 1-D  advection equation (\ref{eq:advection}) the TV is introduced as
\be
\label{eq:TVadvection}
T_1 = \int \mathrm{d}x |\partial_xU|.
\ee
On splitting 1-D (computational) domain into monotonicity regions, in which $\mathrm{sign}\left(\partial_xU\right)=\pm1$, and integrate $\pm\int{\mathrm{d}x \partial_xU}$ over each region, one can express $T_1$ as twice the total of $U$ values at local maximums minus twice the total of $U$ values at local minimums, with the contributions from the values of $U$ at the endpoints of the domain included with positive or negative sign depending on the sign of $\partial_xU$ near the endpoint. The contributions to the TV from the local extrema do not vary in time, since at these points $\partial_tU=-c\partial_xU=0$, thus ensuring TV conservation. If the numerical scheme also conserves the TV or if it has the TVD property, it does not allow breaking monotonicity by generating new local extrema and/or increasing or decreasing the existing extrema thus ensuring that the numerical solution will not generate spurious oscillations or overshoots.

In 2-D the TV can be chosen in a form of 2-D integral of $|\nabla U|$: $T_2=\int{\mathrm{d}V\left(\left|U_x\right|+\left|U_y\right|\right)}$ \cite[as ][did]{Goodman:1985} or in some similar form tying all components of $\nabla U$ \cite[see, e.g.,][]{Krivodonova:2021}, for the conservation law, $\partial_tU+\left(g(U)\right)_x+\left(h(U)\right)_y=0$, in which the fluxes, $g(U),\,h(U)$, are functions of the conserved variable, $U(t,x,y)$. However, the only case when our Eq. (\ref{eq:cons6}) belongs to this class of conservation laws, is the equation:
\be\label{eq:diradvection}
\partial_tU+c(\mathbf{n}_2\cdot\nabla_2)U=0,
\ee
{which describes advection with the constant speed along the direction of $\mathbf{n}_2$. Eq.~(\ref{eq:diradvection}) is nothing but Eq.~(\ref{eq:advection}) as applied to the function, $U(t,x,y)$, of two spatial coordinates and transformed to the rotated coordinate frame, in which the direction of advection is not necessarily aligned with $x$-axis. A natural generalization of TV (\ref{eq:TVadvection}) for Eq.~(\ref{eq:diradvection}) is a 2-D integral of uni-directional gradient, $T^\prime_2=\int{
\mathrm{d}\mathbf{x}_2|\left(\mathbf{n}_2\cdot\nabla\right)U|}$, which integrates (or sums up in discrete case) a multitude of 1-D variations, $T_1=\int{\mathrm{d}s\left|\partial_sU\right|}$, along straight lines, $\mathrm{d}\mathbf{x}/\mathrm{d}s=\mathbf{n}$. Each of these line integrals is separately and independently conserved in the course of advection along this line and can be applied to constructing the TVD scheme, as long as we do not include not relevant transverse gradient components in the TV} \cite[in contrast with the TV definitions by][]{Goodman:1985,Krivodonova:2021}.

In the general case, our Eq.~(\ref{eq:cons6}) combines only derivatives along a single direction of the vector, $\bf{u_6}$, which is the local direction of the Hamiltonian trajectory. Therefore,  
the TV defined in this paper integrates uni-directional gradients along the Hamiltonian trajectories:
\be\label{eq:TVPoisson}
T=\int\limits_V \mathrm{d}\mathbf{\Gamma}\left\vert\sum_l \left\{f;H\right\}_{q_l,p_l}\right\vert 
= \int\limits_V \mathrm{d}\mathbf{\Gamma} \left|\mathbf{u}_6 \cdot \nabla_6 f\right|.
\ee
Now, we can split the computational domain $V$ into two regions of monotonicity $V^+$ and $V^-$, in which the distribution function increases,  $\mathbf{u}_6\!\cdot\!\nabla_6 f>0$, or decreases, $\mathbf{u}_6\!\cdot\!\nabla_6 f<0$, along the Hamiltonian trajectories, respectively. 
We can evaluate the integration separately in $V^+$ and $V^-$:
\be\label{eq:TVpmPoisson}
T=T^+ + T^-,\qquad T^\pm
= \pm\int\limits_{V^\pm} \mathrm{d}\mathbf{\Gamma}\sum_l{\left\{f;H\right\}_{q_l,p_l}}
= \pm\int\limits_{V^\pm} \mathrm{d}\mathbf{\Gamma}\,\mathbf{u}_6 \cdot \nabla_6 f.
\ee
Using the Gauss theorem reduces (\ref{eq:TVpmPoisson}) to an integral over the boundary $\partial V^\pm$ of the monotonicity region:
\be\label{eq:TVboundaries}
T^\pm=\pm\int\limits_{\partial V^\pm}{ \sum_l\left(\mathrm{d}S_{q_l}\frac{\partial H}{\partial p_l}-\mathrm{d}S_{p_l}\frac{\partial H}{\partial q_l}\right)f}
=\pm\int\limits_{\partial V^\pm} 
\mathrm{d}\mathbf{S}_6 \cdot \mathbf{u}_6 f,
\ee
where $\mathrm{d}S_{q_l}$ and $\mathrm{d}S_{p_l}$ are the individual components of the 6-D outward oriented face area vector $\mathrm{d}\mathbf{S}_6$ of the boundary of the monotonicity region. At the boundary, the gradient of the distribution function along the Hamiltonian trajectory 
$\mathbf{u}_6\cdot\nabla_6 f$ vanishes, because it changes sign (and so does $\partial_tf=-\mathbf{u}_6\cdot\nabla_6 f$). Therefore, similarly to the TV defined by (\ref{eq:TVadvection}), only the local extrema of the distribution function along the Hamiltonian trajectories contribute to the TV defined by (\ref{eq:TVPoisson}). 

The surface integrals in (\ref{eq:TVboundaries}) can be evaluated using {\it flux tubes}.  Any Hamiltonian trajectory may be embraced with an infinitesimally small contour. A multitude of Hamiltonian trajectories passing through each point of the contour shapes the surface enclosing the flux tube. The integration of Eq.~(\ref{eq:divfreeu6}) over the part of such tube bounded by two small cross-sections, $\mathrm{d}\mathbf{S}_6^{(1)}$ and $\mathrm{d}\mathbf{S}_6^{(2)}$ gives: $\left(\mathrm{d}\mathbf{S}_6\cdot\mathbf{u}_6\right)^{(1)}=\left(\mathrm{d}\mathbf{S}_6\cdot\mathbf{u}_6\right)^{(2)}$, i.e. the scalar product, $\left(\mathrm{d}\mathbf{S}_6\cdot\mathbf{u}_6\right)$, is a flux tube invariant:
\be\label{eq:ftinvariant}
\left(\mathbf{u}_6\cdot\nabla_6\right)\left(\mathrm{d}\mathbf{S}_6\cdot\mathbf{u}_6\right)=0.
\ee
Using this flux tube invariance, the surface integral in Eq.~(\ref{eq:TVboundaries}) may be expressed as an integral over the outflow  subsurface $\partial V^\pm_\textrm{out}$ through which the Hamiltonian trajectories leave the monotonicity region, i.e. $\mathrm{d}\mathbf{S}_6\cdot\mathbf{u}_6>0$. At the subsurfaces $\partial V_\textrm{out}^\pm$, the distribution function $f_\mathrm{out}$ reaches a local minimum  (for $\partial V^-_\mathrm{out}$) or local maximum (for $\partial V^+_\mathrm{out}$) along the Hamiltonian trajectory. Through each subsurface area element $\left(\mathrm{d}\mathbf{S}_6\right)_\mathrm{out}$ one can find the other end of the corresponding flux tube segment where it
enters the monotonicity region across the inflow subsurface $\partial V^\pm_\mathrm{in}$ at which $
\left(\mathrm{d}\mathbf{S}_6\cdot\mathbf{u}_6\right)_\mathrm{in}<0$. For the area vector of the flux tube cross-section by the subsurface $\partial V^\pm_\textrm{in}$ Eq.~(\ref{eq:ftinvariant}) gives $\left(
\mathrm{d}\mathbf{S}_6\cdot\mathbf{u}_6\right)_\mathrm{in}=-\left(\mathrm{d}\mathbf{S}_6\cdot\mathbf{u}_6\right)_\mathrm{out}$, while the distribution function  $f_\mathrm{in}$ reaches a local minimum (throughout $\partial V^+_\mathrm{in}$) or local maximum (throughout $\partial V^+_\mathrm{in}$) along the considered flux tube. With these considerations, 
\be\label{eq:Tpminout}
T^\pm = \pm\int\limits_{\partial V^\pm_\mathrm{out}}
{\left(\mathrm{d}\mathbf{S}_6\cdot\mathbf{u}_6\right)_\mathrm{out}\left(f_\mathrm{out} - f_\mathrm{in}\right)}=\int\limits_{\partial V^\pm_\mathrm{out}}
{\left(\mathrm{d}\mathbf{S}_6\cdot\mathbf{u}_6\right)_\mathrm{out}\left\vert f_\mathrm{out} - f_\mathrm{in}\right\vert}.
\ee
We see that $T^\pm$ is, indeed, the total variation, since it integrates the difference between the maximal and minimal values of the distribution function along monotonous parts of Hamiltonian trajectories. 

We now prove that the TV
does not increase in time. 
Using \textit{the Reynolds transport theorem} as well as Eqs.~(\ref{eq:cons6},\ref{eq:divfreeu6}) and the fact that $\mathbf{u}_6$ does not explicitly depend on time we can write the time derivative of $T^\pm$ defined in (\ref{eq:TVpmPoisson}) as:
\bea\label{eq:Reynolds}
\frac{\mathrm{d}T^\pm}{\mathrm{d}t} 
&=& \pm \frac{\mathrm{d}}{\mathrm{d}t} \int\limits_{V^\pm} \mathrm{d}\mathbf{\Gamma}\,\nabla_6\cdot\left(\mathbf{u}_6f\right)=\pm\int\limits_{V^\pm} \mathrm{d}\mathbf{\Gamma}\,\nabla_6\cdot\left(\mathbf{u}_6\partial_tf\right)\pm \int\limits_{\partial V^\pm} \mathrm{d}\mathbf{S}_6\cdot\mathbf{u}_B\nabla_6\cdot\left(\mathbf{u}_6 f\right)=\nonumber\\
&=& \pm \int\limits_{\partial V^\pm} \mathrm{d}\mathbf{S}_6\cdot\left(\mathbf{u}_6 \partial_tf - \mathbf{u}_{\mathrm{B}} \partial_tf\right)
\eea
The first term in parentheses, $\mathbf{u}_6 \partial_tf$, results from Gauss' theorem applied to the volume integral, while the second $-\mathbf{u}_\mathrm{B} \partial_tf$ is the boundary 6-velocity $\mathbf{u}_\mathrm{B}$ times the integrated quantity $\nabla_6\cdot\left(\mathbf{u}_6f\right)=-\partial_tf$. At 
the interfaces separating $V^\pm$ regions, which are advected with the velocity $\mathbf{u}_\mathrm{B}=\mathbf{u}_6$ along the Hamiltonian trajectories together with the local extrema of the distribution function,
both terms are zero too, because $\partial_tf$ is zero at such 
interfaces. At 
the parts of $\partial V^\pm$ constituting the external boundary, $\partial V$, 
that do not move, one has $\mathbf{u}_B=0$, so that:
\be\label{eq:dTdtfixed}
\frac{\mathrm{d}T}{\mathrm{d}t} 
= \int\limits_{\partial V} {\mathrm{d}\mathbf{S}_6\cdot\mathbf{u}_6 \mathrm{sign}\left(\mathbf{u}_6\cdot\nabla_6f\right)\partial_tf}.
\ee

The contribution from the external boundary to the TV as well as its role in the TVD property are usually not discussed. However, the generalization of the TV for 6-D phase space results in a variety of different kinds of the external boundary which deserve attention. Particularly, at the outflow part of the external boundary, $\partial V_\mathrm{out}$ at which $\mathrm{d}\mathbf{S}_6\cdot\mathbf{u}_6>0$ the Hamiltonian trajectories leave the domain, bringing the values of distribution function from inside the domain to the boundary, so that $(\partial_tf)_\mathrm{out}=-\mathbf{u}_6\cdot\nabla_6f$. However, at the inflow part of external boundary, $\partial V_\mathrm{in}$, at which $\mathrm{d}\mathbf{S}_6\cdot\mathbf{u}_6<0$, the Hamiltonian trajectories enter the domain bringing the values of distribution function controlled by the \textit{boundary condition}. Particularly, 
the fixed (time-independent) boundary condition requires that $(\partial_tf)_\mathrm{in}\equiv0$, hence:
$$
\frac{\mathrm{d}T}{\mathrm{d}t}=-\int\limits_{\partial V_\mathrm{out}} \mathrm{d}\mathbf{S}_6\cdot\mathbf{u}_6 \left\vert\partial_tf\right\vert\le0,
$$
With this, we have proved that $\mathrm{d}T/\mathrm{d}t
\le 0$ 
and $T$ can be indeed regarded as a Total Variation that does not increase in time.

The property of TV not to increase with time 
prevents breaking monotonicity. 
Indeed, 
the formation of a small domain $\delta V$ with the opposite sign of $\mathbf{u}_6\cdot\nabla_6 f$
than the surrounding monotonicity region would increase $T$ by a positive increment 
$\delta T=2\int_{\delta V}{\mathrm{d}\mathbf{\Gamma} |\mathbf{u}_6\cdot\nabla_6 f|}$. 
The factor of 2 accounts for the contribution from the newly formed boundary of the outside monotonicity region.

The generalization of Eq.~(\ref{eq:TVPoisson}) to a discrete set of cells is given by the following
sum:
\be\label{eq:TV}
T=\sum_{\rm cells}{
\left(\sum_{j,+}{\delta\tilde{H}^+_j}\right)\vert \delta^- f\vert}= \sum_{\rm cells}{\left[\left(\sum_{j,+}{\delta\tilde{H}^+_j}\right)f+\sum_{j,-}{\delta\tilde{H}^-_jf^{\rm ext}_j}\right]\mathrm{sign}\left(\delta^-f\right)}.
\ee
where we used (\ref{eq:Hplusminus}) and (\ref{eq:deltaf}) to obtain the second equality. Inside a region of monotonicity, the sign of $\delta^-f$ in a given cell is the same as that of $\left(\delta^-f\right)^\mathrm{ext}$ in its plus-neighbors. Under these assumptions, the value of distribution function $f$ in the given cell does not contribute to $T$. Indeed, each cell interface with positive $\delta \tilde{H}^+_j$ provides a contribution, $\delta \tilde{H}^+_j\,f\mathrm{sign}(\delta^-f)$ to $T$ due to the first term on the RHS of (\ref{eq:TV}). However, the same cell is, at the same time, a minus-neighbor of its plus-neighbor, hence, the distribution function value $f$ should be also accounted for as $f^{\rm ext}$ in the plus-neighbor cell, providing a contribution $\delta\tilde{H}^-_jf\mathrm{sign}\left(\delta^-f\right)^\mathrm{ext}$ coming from the second term on the RHS. However, $\delta\tilde{H}^-_j$ in the neighboring cell is the negative of $\delta \tilde{H}^+_j$, resulting in cancellation of these contributions. 
Hence, $T^+$ over the cluster of cells in which $\mathrm{sign}\left(\delta^-f\right)=+1$ (herewith, $\mathrm{cells}^+$), and analogously, $T^-$ over the cluster of cells with $\mathrm{sign}\left(\delta^-f\right)=-1$ ($\mathrm{cells}^-$), reduce to the sum over the boundary cells:
\be\label{eq:TVPlusminus}
T^\pm=\pm\sum_{
\mathrm{cells}^\pm}
{\left(\sum_{j,+}{\delta\tilde{H}^+_j}\right)\delta^- f}=\pm\sum_\mathrm{boundary\, cells^\pm}{\left[\left({\sum_{j,+}}^\prime{\delta\tilde{H}^+_j}\right)f+{\sum_{j,-}}^\prime{\delta\tilde{H}^-_jf^{\rm ext}_j}\right]}.
\ee
In the last expression, the outer sum is taken over the boundary cells, which belong to the cluster, but at least one of their neighbors does not, and the inner sums ${\sum}^\prime_{j,+}$ and ${\sum}^\prime_{j,-}$ count only the boundary faces of the cluster and the external neighbors outside of it. We can notice the similarity between the definition of $\delta^-f$ for a single control volume in Eq.~(\ref{eq:deltaf}) and the TV over the whole monotonicity region. 

Another important property of $T$ is that it is equal to zero for a uniform solution, $f\equiv f_0=\mathrm{const}$. If there is an isolated cell with a local maximum or minimum surrounded by a uniform state ($f\ne f_0$), then $T$ increases and becomes positive: $T=2\left(\sum_{j,+}\delta^+H\right)\left\vert f-f_0\right\vert$.  Therefore, to avoid generation of spurious oscillations in the numerical solution with local extrema, the numerical scheme should possess the TVD property.

To derive a criterion for a numerical scheme to have the TVD property, in the time derivative,
\be
\frac{\mathrm{d}T}{\mathrm{d}t}= \sum_{\rm cells}{
\mathrm{sign}\left(\delta^- f\right)\left[
\left(\sum_{j,+}{\delta\tilde{H}^+_j}\right)\partial_tf
+\sum_{j,-}{
\delta\tilde{H}^-_j\left(\partial_tf\right)^{\rm ext}_j}\right]},
\ee
we change the order of summation to group all multipliers of $f_t$ in the given volume:
\be\label{eq:dTVdt}
\frac{\mathrm{d}T}{\mathrm{d}t}= \sum_{\rm cells}{\left\{\partial_tf\sum_{j,+}{\delta\tilde{H}^+_j}\left[\mathrm{sign}\left(\delta^-f\right)-\mathrm{sign} \left(\delta^-f\right)^{\rm ext}_j\right]\right\}}.
\ee
Only those control volumes contribute to the time derivative of the TV, in which $\left(\delta^-f\right)^{\rm ext}_j$ in any ``plus-neighbor'' has the sign opposite to that of $\delta^-f$ in the control volume. The numerical scheme possesses the TVD property, if in such control volumes the sign of $\partial_t f$ is opposite to that of $\delta^-f$:
\bea\label{eq:TVDCrit}
\mathrm{sign}\left(\partial_tf\right)=-\mathrm{sign}\left(\delta^-f\right).
\eea
Eq.~(\ref{eq:TVDCrit}) also requires that $\partial_tf=0$ at $\delta^-f=0$. 
Particularly, the first order numerical flux given by Eq.~(\ref{eq:firstorder}), possesses the TVD property, since 
\be
\mathrm{sign}\left(\left[ \partial_tf\right]^{(1)}\right)=\mathrm{sign}\left[-\left(\frac{\sum_{j,+}{\delta\tilde{ H}^+_j}}{V}\right)\delta^-f\right]= -\mathrm{sign}\left(\delta^-f\right).
\ee

The properties of $T$ are similar to those of the total variation for the 1-D advection equation \citep[see][]{Hirsch97}. This similarity allows us to apply TVD schemes to the Liouville equation.
\subsection{Phase-Space Control Volume Formulation: Second Order TVD Scheme}
\label{Sec:TVDScheme}
To construct the second order TVD numerical flux, one needs to modify the first order monotone flux (\ref{eq:firstorder}) by: (1) adding the difference between the second order and first order numerical fluxes (i.e., \textit{anti-diffusion}, which is the negative of Eq.~\ref{eq:diffusion}), to achieve the high accuracy; and (2) limiting the added anti-diffusion by applying proper \textit{limiter} function $\Psi_j$, to maintain the TVD property:
\bea\label{eq_22}
&&\left[\partial_tf\right]^{ \mathrm{(TVD)} }=\left[\partial_tf\right]^{(1)}-\frac1{V}
\left[\sum_{j,+}{\delta\tilde{H}^+_j\Psi\left(\frac{f_j-f}2\right)}+\sum_{j,-}{\delta\tilde{H}^-_j\Psi\left(\frac{f-f_j}2\right)}\right]=\nonumber\\
&&=\frac1{V}
\left\{-\sum_{j,+}{\delta\tilde{H}^+_j\Psi\left(\frac{f_j-f}2\right)}+\sum_{j,-}{\delta\tilde{ H}^-_j\left[f-f_j-\Psi^{\rm ext}\left(\frac{f-f_j}2\right)\right]}\right\}.
\eea
The RHS of Eq.~(\ref{eq_22}) may be thought of as the divergence of fluxes, $\delta\tilde{H}^+_jf^\mathrm{(f)}_j$, where the distribution function at the plus-face, $f^\mathrm{(f)}_j=f+\Psi\left(\frac{f_j-f}2\right)$, is the sum of the cell-centered value $f$ (which contributes to the first-order monotone flux) and the \textit{limited} difference, $f^\mathrm{(f)}-f$. In smooth regions the limiting function is close to the argument value:
\be\label{eq:secondorderrange}
\frac{\Psi\left[\left(f_j-f\right)/2\right]}{\left(f_j-f\right)/2}\approx1 .
\ee
In this case we can evaluate:
\be
\sum_{j,+}{\delta\tilde{H}^+_j\Psi\left(\frac{f_j-f}2\right)}
\approx \sum_{j,+}{\delta\tilde{H}^+_j}\frac{\delta^+f}2,\qquad \sum_{j,-}{\delta\tilde{H}^-_j\Psi^{\rm ext}\left(\frac{f-f_j}2\right)}
\approx \sum_{j,-}{\delta\tilde{H}^-_j}\frac{\delta^-f}2,
\ee
so that 
Eq.~(\ref{eq_22}) tends to the second order flux as in Eq.~(\ref{eq:secondorder}), ensuring high-quality numerical results. However, near extrema the distribution function gradients in the neighboring cells may differ either in magnitude or in sign. In these cells the limiters reduce the anti-diffusion or even set it to zero, so that Eq.~(\ref{eq_22}) approaches the monotone numerical flux (Eq.~\ref{eq:firstorder}). The TVD criterion given by Eq.~(\ref{eq:TVDCrit}) is satisfied, thus avoiding spurious oscillations in numerical solutions of Eq.~(\ref{eq_22})  if the limiting functions $\Psi\left(\frac{f_j-f}2\right)$ satisfy the following two
conditions. First, we require, that:  
\be\label{eq:sumtvdplus}
\delta^-f(\delta^+f)^\mathrm{lim}\ge0,\quad\mathrm{where}\quad\frac{(\delta^+f)^\mathrm{lim}}2=\frac{\sum_{j,+}{\delta\tilde{H}^+_j\Psi\left(\frac{f_j-f}2\right)}}{\sum_{j,+}{\delta\tilde{H}^+_j}}.
\ee
Otherwise the first term in Eq.~\ref{eq_22} has the same sign as $\delta^-f$, potentially breaking the TVD property). Then, under the requirement that inequality,
\be\label{eq:downwindlimiter}
0\le\frac{\Psi\left[\left(f-f_j\right)/2\right]}{\left(f-f_j\right)/2}\le\beta_j\le2.
\ee
holds in any control volume, the second term in Eq.~\ref{eq_22}
\be\label{eq:sumtvdminus}
\left[\partial_tf\right]^{ \mathrm{(TVD)} }=
\dots+
\frac1{V}
\sum_{j,-}{\left(-
\delta\tilde{ H}^-_j\right)
\left\{\left(+f_j-f\right)\left[1-\frac{\Psi^{\rm ext}\left(\frac{f-f_j}2\right)}{f-f_j}\right]\right\}},
\ee
 within infinitesimal time interval,  $\Delta t\rightarrow0$,  replaces in the minor fraction of the control volume,
\be\label{eq:VolumeFraction}
\zeta_j=\frac{\Delta t\left(-\delta\tilde{ H}^-_j\right)}V\left[1-\frac{\Psi^{\rm ext}\left(\frac{f-f_j}2\right)}{f-f_j}\right],
\ee
the initial distribution function  ($-f$) with its value in the upwind  neighbor cell ($+f_j$). In the first order scheme (all $\Psi_i=0$), the leftover fraction of the control  volume, still occupied with the initial state is $\zeta=1-\sum_{j,-}\zeta_j=1-C>0$ which is positive at $C<1$, while for infinitesimal time step $C\rightarrow0$ and $\zeta=1-O[C]>0$. The state in the control volume resulting from this term is thus a positively weighted average  of the states in the given cell  and its upwind neighbors,  which property, as far as it concerns maintaining distribution function positiveness and avoiding  spurious oscillations is fully analogous to Eq.~(\ref{eq:CFL}) and not any worse than monotonicity. 

The requirements Eqs.~(\ref{eq:sumtvdplus},\ref{eq:downwindlimiter}) are similar to those derived for the TVD scheme for 1-D advection equation \citep[see ][]{Hirsch97}. The distinction of the  multi-dimensional case is that the contributions to sums in Eqs.(\ref{eq_22},\ref{eq:sumtvdplus}) may have different signs because of gradients in the directions transverse to that of the Hamiltonian trajectory and ruling the contributions of ``improper'' sign out of numerical scheme would reduce the order of approximation and result in excessive numerical particle diffusion in these directions.

The choice, $\beta_j=2$  
leaves the maximal wide range for $\Psi(r)/r$ ratio around the second order approximation point given by Eq.(\ref{eq:secondorderrange}). 
A limiter satisfying Eq.(\ref{eq:downwindlimiter})  may be chosen to add an extra anti-dissipation wherever possible 
to mitigate the effect of numerical dissipation inherently present in a TVD scheme, for example, by using a \textit{$\beta$-limiter} \cite[see, e.g.][]{Hirsch97} in the  following  way:
\bea\label{eq:betalimiter}
\Psi_\beta\left[\frac{f_j-f}2\right]&=&\frac{s_j}2\min\left\{\max\left[\left|f_j-f\right|,s_j\left(f-f^\mathrm{opp}_j\right)\right],\beta_j\left|f_j-f\right|\right\},\nonumber\\
\beta_j\equiv2,
\eea
where $s_j=\mathrm{sign}\left(f_j-f\right)$. For Cartesian (or, at least, for logically Cartesian) grid the difference, $f-f^\mathrm{opp}_j$ across the opposite face of the control volume is used to evaluate anti-diffusion. In the region of smoothness  and for $\left(f_j-f\right)\approx\left(f-f^\mathrm{opp}_j\right)$ the limiter Eq.(\ref{eq:betalimiter}) tends to the second order accurate solution according to Eq.(\ref{eq:secondorderrange}). Another good limiter follows the idea by \cite{Koren:1993}: 
\be\label{eq:Koren}
\Psi_\beta\left[\frac{f_j-f}2\right]=\frac{s_j}2\max\left\{\min\left[\frac{s_j}{3}\left[2(f_j-f)+(f-f^\mathrm{opp}_j)\right],\beta_j\left|f_j-f\right|\right],0\right\}.
\ee
which ensures the approximation of the face flux in 1D advection equation with the third order of accuracy for monotonic and sufficiently smooth regions.

With chosen $\beta$-limiter, the preliminary estimate for the downwind gradient in the distribution function $\left(\delta^+f\right)^\mathrm{lim}$ as present in Eq.(\ref{eq:sumtvdplus})  reads: 
\be\label{eq:lim1}
\frac{(\delta^+f)^\mathrm{lim,1}}2=\frac{\sum_{j,+}{\delta\tilde{H}^+_j\Psi_\beta\left(\frac{f_j-f}2\right)}}{\sum_{j,+}{\delta\tilde{H}^+_j}},
\ee
At smooth solutions, $(\delta^+f)^\mathrm{lim,1}\approx\delta^+f\approx\delta^-f$, so that the LHS in Eq.~(\ref{eq:lim1}) has the same sign as $\delta^-f$ (and smaller in magnitude by a factor of $2$), thus satisfying the TVD criterion Eq.(\ref{eq:sumtvdplus}). However, near extremal points even the estimates of forward and backward derivatives,
$(\delta^+f)^\mathrm{lim,1}$ and $\delta^-f$, may differ in sign. 
If $(\delta^+f)^\mathrm{lim,1}$ goes beyond the range $0\le\frac{(\delta^+f)^\mathrm{lim,1}/2}{\delta^-f}\le1$, we apply extra limiting factors $0\le\gamma_j\le1$ to the limiting functions to keep TVD property. The ultimate limiters have the form $\gamma_j\Psi_\beta\left(\frac{f_j-f}2\right)$, where $\gamma_j$ factors are solved from the following equations:
\be\label{eq:gammageneral} 
\frac{\sum_{j,+}{\delta\tilde{H}^+_j\gamma_j\Psi_\beta\left(\frac{f_j-f}2\right)}}{\sum_{j,+}{\delta\tilde{H}^+_j}}=\frac{(\delta^+f)^\mathrm{lim}}2,\quad\text{where}\quad\frac{(\delta^+f)^\mathrm{lim}}2=\mathrm{minmod}\left(\frac{(\delta^+f)^\mathrm{lim,1}}2,\delta^-f\right).
\ee
The first equation in Eqs.~(\ref{eq:gammageneral}) is used to derive the reduction coefficients,  $\gamma_j$, while the second one defines the desired result of such reduction. If the sum in numerator in Eq.(\ref{eq:lim1}) includes multiple contributions with different signs, upon evaluating a partial sum of dominant contributions, having the same sign as the total sum:
\be\label{eq:fplusmajor}
\frac{\widehat{\left(\delta^+f\right)^\mathrm{lim,1}}}2
=\mathrm{sign}\left[\left(\delta^+f\right)^\mathrm{lim,1}\right]
\frac{\sum_{j,+}{\delta\tilde{ H}^+_j
\max\left\{
\mathrm{sign}\left[\left(\delta^+f\right)^\mathrm{lim,1}\right]\Psi_\beta\left(\frac{f_j-f}2\right),0
\right\}
}}{\sum_{j,+}{\delta\tilde{ H}^+_j}},
\ee
the $\gamma_j$-factors all equal to,
\be
\gamma=1+\frac{
\left(\delta^+f\right)^\mathrm{lim}-\left(\delta^+f\right)^\mathrm{lim,1}}{\widehat{\left(\delta^+f\right)^\mathrm{lim,1}}}\le1
\ee
may be applied only to dominant terms contributing to Eq.(\ref{eq:fplusmajor}),  keeping unchanged the terms of opposite sign. In this case  $\gamma_j\ne0$  and even in the control volume with vanishing $\delta^-f$  the limited second order flux  corrections do not vanish, only their total over all $\delta\tilde{H}^+$ faces does. Otherwise, if all terms in the sum in to Eq.(\ref{eq:fplusmajor}) have the same sign,  more evident expression for $\gamma$-factor reads:
\be
0\le\gamma=\frac{
\left(\delta^+f\right)^\mathrm{lim}}{\left(\delta^+f\right)^\mathrm{lim,1}}\le1.
\ee
The presented limiters seem to be the least restrictive semi-discrete scheme. Now, we discuss the TVD property for a  second order in time explicit scheme,  which may be achieved if Eqs.~(\ref{eq:firstorder}) and (\ref{eq:CFL}) are applied to approximately update the solution to the intermediate time level $t+\Delta t/2$:
\be\label{eq:fdeltat2}
f\left(+\Delta t/2\right)\approx f-C
\frac{\delta^-f}2\approx f-C
\frac{\left(\delta^+f\right)^\mathrm{lim}}2.
\ee
By using Eq.~(\ref{eq:fdeltat2}) and changing in Eq.~(\ref{eq_22}) the face-plus value of the distribution function from $f^\mathrm{(f)}_j=f+\gamma_j\Psi_\beta\left(\frac{f_j-f}2\right)$
to 
$f^\mathrm{(f)}_j=f\left(+\Delta t/2\right)+\gamma_j\Psi_\beta\left(\frac{f_j-f}2\right)$
we obtain:
\bea\label{eq:TVDScheme}
&&f(+\Delta t)-f=-C\,\delta^-f-\frac{\Delta t}V\sum_{j,-}{\delta\tilde{ H}^-_j\left\{\gamma^{\rm ext}_j\Psi^{\rm ext}_\beta\left(\frac{f-f_j}2\right)-\left[C\frac{\left(\delta^+f\right)^\mathrm{lim}}2\right]^{\mathrm ext}_j\right\}}-\nonumber\\
&&-\frac{\Delta t}V(1-C)\frac{\left(\delta^+f\right)^\mathrm{lim}}2\sum_{j,+}{\delta\tilde{H}^+_j}=\frac{\Delta  t}{V}\times\\
&\times&
\sum_{j,-}{\left(-
\delta\tilde{ H}^-_j\right)
\left\{\left(f_j-f\right)\left[1-\frac{\gamma^{\rm ext}_j\Psi^{\rm ext}_\beta\left(\frac{f-f_j}2\right)}{f-f_j}\right]-\left[C\frac{\left(\delta^+f\right)^\mathrm{lim}}2\right]^{\mathrm ext}_j-(1-C)\frac{\left(\delta^+f\right)^\mathrm{lim}}2\right\}}\nonumber.
\eea
As long as the CFL condition given by Eq.~(\ref{eq:CFL<1}) is satisfied in all control volumes and the limiters are correctly applied, the factor, $\propto(1-C)\left(\delta^+f\right)^\mathrm{lim}$, in the third term in braces in the RHS, has the same sign as $\delta^-f$,  thanks to the use  of $\gamma$-limiter in Eq.(\ref{eq:gammageneral}). The first term in braces as discussed above (see Eqs.~(\ref{eq:sumtvdminus},\ref{eq:VolumeFraction})) produces a state which is a positively weighted average of the states in the given cell and its upwind neighbors.  Finally  the second  term in braces, $\left[\left(\delta^+f\right)^\mathrm{lim}\right]^\mathrm{ext}_j$  has  the  same sign  as $\left(\delta^-f\right)^\mathrm{ext}_j$ in the neighboring  cell. If this sign is opposite to that of $\delta^-f$ in the given cell, i.e. the local extremum (=sign change in the gradient) of the distribution function is close to the face, this case needs a special treatment. Indeed, the term $-\left[C\frac{\left(\delta^+f\right)^\mathrm{lim}}2\right]^\mathrm{ext}_j\propto-C^\mathrm{ext}_j\left(\delta^-f\right)^\mathrm{ext}_j$ includes the TVD-breaking contribution, $-C^\mathrm{ext}_jf_j$. To balance this contribution, the $\beta$-limiter Eq.~(\ref{eq:betalimiter})) with more restrictive  choice of $\beta_j=2-2C^\mathrm{ext}_j<2$ should be applied, instead of $\beta_j\equiv2$ allowed in the semi-discrete scheme (see Eq.~(\ref{eq:betalimiter})). In this way, the disturbing term is balanced with the first term in braces, $\left(f_j-f\right)\left(1-\frac{\gamma_j\beta_j}2\right)$, which includes $f_j$ term of a favorable sign and sufficient magnitude. The distribution function in the given cell is partially replaced with the states from upwind neighbors of $j$th cell, also contributing to $-C^\mathrm{ext}_j\left(\delta^-f\right)^\mathrm{ext}_j$, with all positive coefficients, instead of $f_j$ state, and Eq.~(\ref{eq:VolumeFraction}) together with the restriction on $\beta_j$ ensures that the volume fraction, $\xi_j$, is sufficient for this replacement. At the faces which separate cells with the same sign of $\delta^-f$, i.e. within a monotonicity region,  the sign of a second term in braces meets the TVD criterion, and the restriction on $\beta_j$ is only needed to get a gradual transition between the restricted and unrestricted $\beta_j$, which is achieved with the following equation,
\be\label{eq:betalimiter2}
\frac{\beta_j}2=1-C^\mathrm{ext}_j\left\{1-
\mathrm{minmod}
\left[
\frac{C\delta^-f}
{C^\mathrm{ext}_j\left(\delta^-f\right)^\mathrm{ext}_j},1\right]\right\},
\ee
chosen in such way that the weight of the initial state contribution to the final state is positive at $C<1$, $C^\mathrm{ext}_j<1$. The need to use more restrictive limiter Eq. ~(\ref{eq:betalimiter2}) comparing with $\beta_j\rightarrow2$ at $C\rightarrow0$ is due to the use of a single stage scheme up to $C\le1$. 

The scheme (\ref{eq:TVDScheme}) may be implemented as follows. With calculated $\delta H$ for each face, $V$ for each control volume and the known time step, $\Delta t$, the local values of $C$ may be calculated as well as $\delta^- f$ for each cell, in terms of the val ues of distribution function in the given cell and in its ``minus'' neighbors. This is sufficient to calculate the first order contribution to expression (\ref{eq:TVDScheme}) (the first term in the RHS). Then, employing the values of $\delta^-f$ in ``plus'' neighboring cells, the $\beta$-limited anti-diffusion flux through the ``plus'' faces are calculated using (\ref{eq:betalimiter},\ref{eq:betalimiter2}). Where needed, $\gamma$-limiters are calculated. The anti-diffusion from the neighboring cell to the given one are calculated in the ``minus'' neighbors and do not need to be recalculated for the given cell.
\subsection{Numerical Test}
\label{Sec:Result}
\begin{figure}[t]
\includegraphics[width=\textwidth]{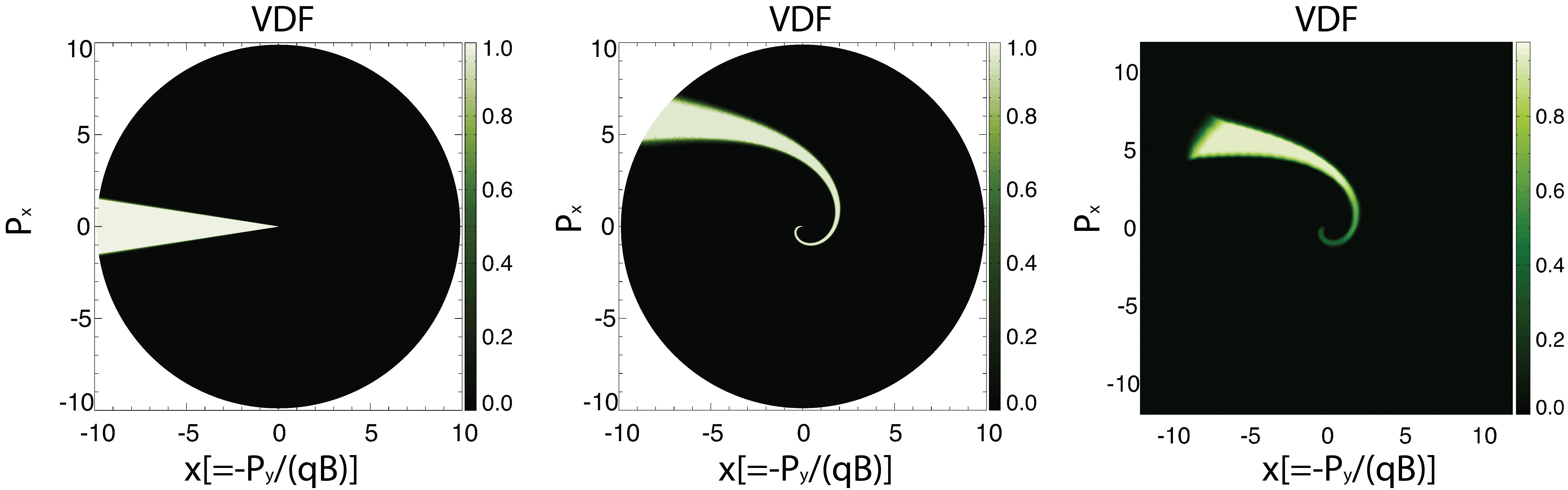}
\caption{Particle gyration in the magnetic field aligned with $z$-axis (out of plane). \textbf{Left panel:} Initial distribution in the phase space, $(x,p_x)$, with $x-$ coordinate  is equal to $-p_y$. \textbf{Middle panel:} Distribution at time instant, $t=2\pi$. Particles of lower energy, $p_x^2+p_y^2\ll1$, completed a full clockwise rotation, while the higher energy particles, $p_x^2+p_y^2\gg 1$, passed through only a small fraction of rotation. A uniform $300\times360$ grid in $\log p$ and polar angle is used; \textbf{Right panel:} Same problem simulated on a uniform rectangular $300\times300$ grid.}
\label{fig:Tenishev}
\end{figure}
To illustrate the advantages of our TVD scheme, we perform a test with the Hamiltonian function 
\be\label{eq:HamiltonianViaA}
H=c\sqrt{m^2c^2+\left({\cal P}_x-qA_x\right)^2+\left({\cal P}_y-qA_y\right)^2}
\ee
for 2-D motion (gyration) of particles of mass $m$ and electric charge $q$ in a uniform magnetic field, $B$, directed along the $z$-axis of the Cartesian coordinate system. Here, ${\cal P}_{x,y}=p_{x,y}+qA_{x,y}$ are the components of the {\it generalized} momentum \cite[see Ch.16 in][]{landau1975}. Assuming the Landau gauge, $A_x=0,\,A_y=Bx$, for the components of the vector potential, $A_{x,y}$, we can express $p_x\equiv{\cal P}_x$, while conservation of ${\cal P}_y$ for the Hamiltonian function independent on $y$ allows us to consider a group of particles with ${\cal P}_y=0$, so that $x\equiv-\frac{p_y}{qB}$. Their motion is described by the 1-D Hamiltonian function
\be
H(x,p_x)=c\sqrt{m^2c^2+p_x^2+\left(qBx\right)^2}
\ee
with the distribution function depending on $x=-\frac{p_y}{qB}$  and $p_x$. The initial beam distribution function (see the left panel in Fig.~\ref{fig:Tenishev}) equals one  in a narrow cone in momentum directions about the $y$-axis, the momentum range beings $0.01mc\le p\le 10 mc$ and vanishes otherwise.
\begin{figure}[t]
\includegraphics[width=\textwidth]{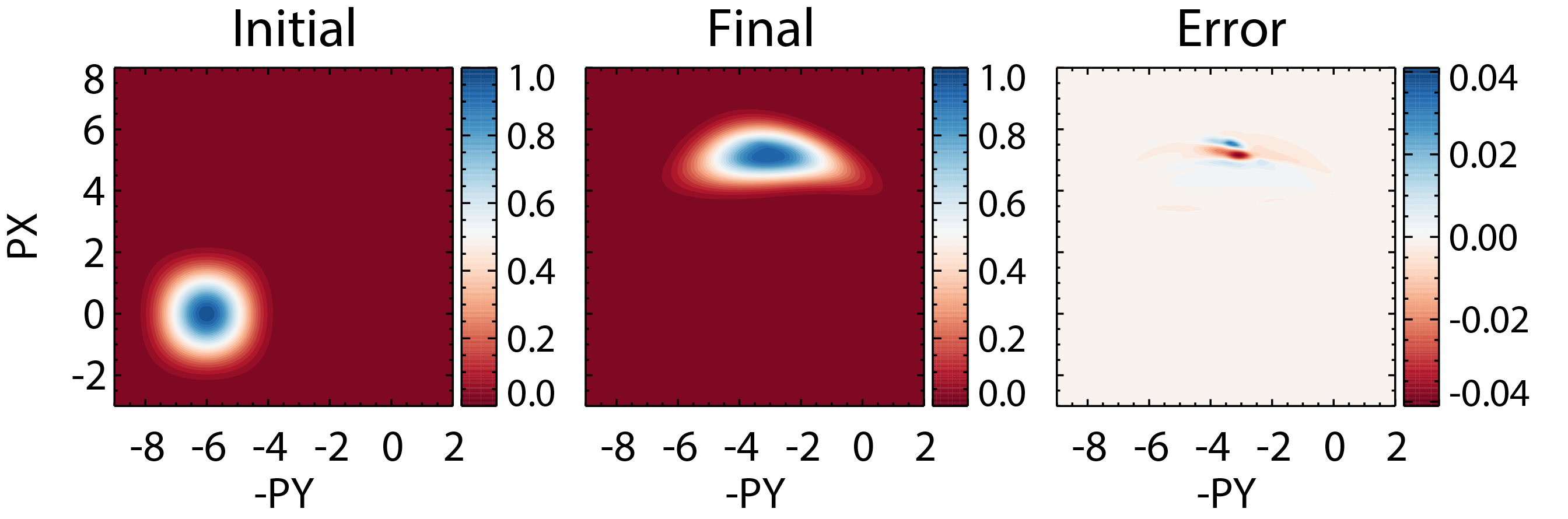}
\caption{Convergence test with the smooth initial distribution. \textbf{Left panel:} Initial distribution. For the time instant, $t=2\pi$, the initial distribution is analytically rotated clockwise by the angle, $t/\sqrt{1+p^2_x+p^2_y}$; \textbf{Middle panel:} Simulation result at this time on a uniform rectangular $300\times300$ grid; \textbf{Right panel:} Error - the difference between numerical and analytical solutions.}
\label{fig:Convergence}
\end{figure}

The simulation results obtained with the scheme given by Eq.~(\ref{eq:TVDScheme}) for $C=0.99$ with the limiter in Eqs.~(\ref{eq:betalimiter},\ref{eq:betalimiter2}) on two different kinds of grid are presented in Fig. \ref{fig:Tenishev} (middle and right panels). The tests are performed with $q=m=B=c=1$. The simulation time, $t=2\pi$, is chosen, at which particles of lower energy, $p_x^2+p_y^2\ll 1$, complete a full Larmor rotation. Due to the relativistic dependence of the gyration frequency on the particle energy, particles with higher energy rotate slower, which results in de-phasing of initially beamed VDF. This effect is important in a gyrotron \citep{Borie1991}, where it is combined with the effect from rotating electric field and results in the electron synchronous gyration and coherent emission. The profile of the distribution function is sharply outlined thanks to the use of suberbee limiter. Thus, the proposed scheme has high accuracy and low diffusion, which makes it suitable for practical use.

We also studied grid convergence in the test with smooth analytical initial distribution:
\be\label{eq:manufactured1}
f_{t=0}=\cos^4(\pi(p_y-6)/6)\cos^4(\pi p_x/6),\quad \text{at}\,\left|p_y-6\right|\le3,\quad \left|p_x\right|\le3,
\ee
and equals 0 otherwise (see Fig.~\ref{fig:Convergence}, left panel). This initial distribution can be analytically rotated by angle $t/\sqrt{1+p^2_x+p^2_y}$ to provide an exact solution, $f_\mathrm{exact}$ at the simulation time, $t=2\pi$, to compare with the numerical solution, $f$ (middle panel). The error was calculated and normalized per the size of area, occupied by the initial distribution: $L_1=\sum_\mathrm{cells}{|f_\mathrm{exact}-f|}V_\mathrm{cell}/(6\times6)$. Thus calculated error for grid of $150\times150$, $300\times300$ and $600\times600$ cells equals $5.500\times10^{-3}$, $1.284\times10^{-3}$, and $2.967\times10^{-4}$ correspondingly. The consequent error ratios, characterizing the convergence rate, are equal to $4.28\approx2^{2.10}$ and $4.32\approx2^{2.11}$, thus demonstrating close to the second order of accuracy. All convergence tests were done at $C_\mathrm{max}=0.5$ with Koren's limiter given by Eqs.(\ref{eq:Koren},\ref{eq:betalimiter2}). The total energy $\int d\Gamma f H$ is conserved to less than 0.01\% relative error even on the coarsest $150\times150$ grid.
\subsection{Energy Non-Conservation}
\label{Sec:Noncons}
In addition to the particle number conservation, the kinetic equation Eq. (\ref{eq:viaPoisson}) also conserves multiple \textit{Casimir invariants}, 
e.g., the total energy, ${\cal E}=\int{Hf\mathrm{d}\mathbf{\Gamma}}$. The pseudo-distribution-function, $H\left(t,\mathbf{x},\mathbf{p}\right)f\left(t,\mathbf{x},\mathbf{p}\right)$, describes the energy density in phase space.  If the Hamiltonian function does not directly depend on time, $\partial_t H\equiv0$, one can multiply Eq. (\ref{eq:viaPoisson}) {by $H$ and account for an evident identity,  $\left\{H,H\right\}\equiv0$, following from Eq. (\ref{eq:Poisson}), which reduces Eq. (\ref{eq:viaPoisson}) to the Liouville equation,
\be\label{eq:EnergyConservation}
\partial_t\left(Hf\right)+\sum_{l=1}^3{\left\{\left(Hf\right);H\right\}_{q_l,p_l}}=0,
\ee
The reason for such reduction is that the value of the Hamiltonian function along the Hamiltonian trajectory is constant and the original Liouville equation advects the unchanged profile of the distribution function along the Hamiltonian trajectory. Therefore, the energy density profile along the Hamiltonian trajectory differs only by a constant factor from the canonical distribution function, so that it is also advected unchanged along the Hamiltonian trajectory. Among consequences from such energy density behavior is the total energy conservation can be proven analogously to the particle number conservation. As long as a numerical method is specifically designed to conserve energy, such as those by \cite{Ortleb2017,Hammett2019a,Hammett2019b}, one can apply scheme Eq. (\ref{eq:4deltaH}) to Eq.  (\ref{eq:EnergyConservation}) to produce an energy conserving scheme:
\bea\label{eq:EnergyConservingScheme}
VH\partial_tf&=&-
\sum_{l=1}^L 
\left\{
\left[
\tilde{H}_l\left(+\frac{\Delta q_l}2,
+ \frac{\Delta p_l}2\right) 
-\tilde{H}_l
\left(+\frac{\Delta q_l}2,-\frac{\Delta p_l}2\right)
\right]
\frac{
f(+\Delta q_l)+f}2
\frac{H(+\Delta q_l)+H}2 
+\right. \nonumber \\
&+&\left[\tilde{H}_l\left(-\frac{\Delta q_l}2,+\frac{\Delta p_l}2\right)-\tilde{H}_l\left(+\frac{\Delta q_l}2,+\frac{\Delta p_l}2\right)\right]\frac{f(+\Delta p_l)+f}2\frac{H(+\Delta p_l)+H}2+
\nonumber\\
&+&\left[\tilde{H}_l\left(-\frac{\Delta q_l}2,-\frac{\Delta p_l}2\right)-\tilde{H}_l\left(-\frac{\Delta q_l}2,+\frac{\Delta p_l}2\right)\right]
\frac{f(-\Delta q_l)+f}2\frac{H(-\Delta q_l)+H}2+
\nonumber\\ 
&+&\left.\left[\tilde{H}_l\left(+\frac{\Delta q_l}2,-\frac{\Delta p_l}2\right)-\tilde{H}_l\left(-\frac{\Delta q_l}2,-\frac{\Delta p_l}2\right)\right]\frac{f(-\Delta p_l)+f}2\frac{H(-\Delta p_l)+H}2\right\}.
\eea
Here the energy conservation follows from the fact that the rate of energy change in each control volume is expressed in terms of the energy fluxes quantifying the energy exchange with the neighboring cells. However, the particle total number with this choice of scheme is not necessarily conserved. 

Specifically, the particle number conserving approach adopted here may potentially result in essential energy non-conservation, so that even the applicability of the term ``conservative'' might be debatable. Particularly, for the general second order scheme in Eq. (\ref{eq:4deltaH}) the energy source in each control volume is given by the RHS of Eq. (\ref{eq:4deltaH}) multiplied by $VH$, while the energy defect per cell may be then evaluated by subtracting the energy-conserving flux divergence in the RHS of Eq. (\ref{eq:EnergyConservingScheme}). 

\bea\label{eq:EnergyDefectSecond}
\left(\frac{\mathrm{d}{\cal E}}{\mathrm{d}t}\right)^\mathrm{(2)}&=&\sum_\mathrm{cells}
\sum_{l=1}^L \frac{V}{\Delta q_l\Delta p_l}\times\nonumber\\
&\times&
\left\{
\left[
H\left(+\frac{\Delta q_l}2,
+ \frac{\Delta p_l}2\right) 
-H
\left(+\frac{\Delta q_l}2,-\frac{\Delta p_l}2\right)
\right]
\frac{
f(+\Delta q_l)+f}2
\frac{H(+\Delta q_l)-H}2 
+\right. \nonumber \\
&+&\left[H\left(-\frac{\Delta q_l}2,+\frac{\Delta p_l}2\right)-H\left(+\frac{\Delta q_l}2,+\frac{\Delta p_l}2\right)\right]\frac{f(+\Delta p_l)+f}2\frac{H(+\Delta p_l)-H}2+
\nonumber\\
&+&\left[H\left(-\frac{\Delta q_l}2,-\frac{\Delta p_l}2\right)-H\left(-\frac{\Delta q_l}2,+\frac{\Delta p_l}2\right)\right]
\frac{f(-\Delta q_l)+f}2\frac{H(-\Delta q_l)-H}2+
\nonumber\\ 
&+&\left.\left[H\left(+\frac{\Delta q_l}2,-\frac{\Delta p_l}2\right)-H\left(-\frac{\Delta q_l}2,-\frac{\Delta p_l}2\right)\right]\frac{f(-\Delta p_l)+f}2\frac{H(-\Delta p_l)-H}2\right\}=\nonumber\\
&=&\frac12\sum_\mathrm{cells}Vf
\sum_{l=1}^L \frac{1}{\Delta q_l\Delta p_l}\times\nonumber\\
&\times&
\left\{
\left[
H\left(+\frac{\Delta q_l}2,
+ \frac{\Delta p_l}2\right) 
-H
\left(+\frac{\Delta q_l}2,-\frac{\Delta p_l}2\right)
\right]
\left[H(+\Delta q_l)-H\right] 
+\right. \nonumber \\
&+&\left[H\left(-\frac{\Delta q_l}2,+\frac{\Delta p_l}2\right)-H\left(+\frac{\Delta q_l}2,+\frac{\Delta p_l}2\right)\right]\left[H(+\Delta p_l)-H\right]+
\nonumber\\
&+&\left[H\left(-\frac{\Delta q_l}2,-\frac{\Delta p_l}2\right)-H\left(-\frac{\Delta q_l}2,+\frac{\Delta p_l}2\right)\right]
\left[H(-\Delta q_l)-H\right]+
\nonumber\\ 
&+&\left.\left[H\left(+\frac{\Delta q_l}2,-\frac{\Delta p_l}2\right)-H\left(-\frac{\Delta q_l}2,-\frac{\Delta p_l}2\right)\right]\left[H(-\Delta p_l)-H\right]\right\}\approx\nonumber\\
&\approx&\frac{1}{8}\sum_\mathrm{cells}
Vf\sum_{l=1}^L \left(\Delta q_l^2\frac{\partial^2}{\partial q_l^2}-
\Delta p_l^2\frac{\partial^2}{\partial p_l^2}\right)\left(\frac{\partial H}{\partial q_l}\frac{\partial H}{\partial p_l}\right).
\eea

The energy defect vanishes if the Hamiltonian function is the total of kinetic energy and potential energy, which are functions of generalized momenta and coordinates, respectively and both are quadratic polynomials, so that $\frac{\partial^3H}{\partial q_l^3}\equiv0$ and $\frac{\partial^3H}{\partial p_l^3}\equiv0$. Otherwise, the non-vanishing energy defect is at least of second order in $\Delta q_l^2$, $\Delta p_l^2$. However, with the first order numerical flux as in Eq. (\ref{eq:firstorder}) the energy defect can be as high as:
\be\label{eq:EnergyDefectFirst}
\left(\frac{\mathrm{d}{\cal E}}{\mathrm{d}t}\right)^\mathrm{(1)}=\sum_\mathrm{cells}f\sum_{j,+}{\delta\tilde{H}^+_j\left(H^{\rm ext}_j-H\right)}.
\ee

Assuming again a rectangular grid and the decomposition of the Hamiltonian function into the sum of potential and kinetic energy Eq. (\ref{eq:EnergyDefectFirst}) can be approximated as follows:
\be\label{eq:EnergyDefect1Approx}
\left(\frac{\mathrm{d}{\cal E}}{\mathrm{d}t}\right)^\mathrm{(1)}\approx\frac12\sum_\mathrm{cells}fV\sum_l{\left(\Delta q_l\frac{\partial^2H}{\partial q_l^2}\left|\frac{\partial H}{\partial p_l}\right|+
\Delta p_l\frac{\partial^2H}{\partial p_l^2}\left|\frac{\partial H}{\partial q_l}\right|\right)}.
\ee
If both the kinetic and potential energies are convex functions, $\frac{\partial^2H}{\partial p_l^2}>0$, $\frac{\partial^2H}{\partial q_l^2}>0$, the energy defect is of first order in grid size and it is positive definite, thus resulting in \textit{numerical heating}.

To test the energy non-conservation and compare it to the overall numerical scheme error for the second order accurate TVD scheme presented in Eq. (\ref{eq:TVDScheme}) we apply it to the kinetic equation of harmonic oscillators with the Hamiltonian $H(q,p)=\frac{q^2}2+\frac{p^2}2$\hl, for which the  energy defect Eq.(\ref{eq:EnergyDefectSecond}) for the second order non-TVD scheme would vanish, while the defect of monotonous flux Eq. (\ref{eq:EnergyDefect1Approx}) is positive. We integrate the kinetic equation over the time interval, $0\le t\le 2\pi$, i. e. through a single period of oscillators, so that the analytically exact distribution function at the time instant, $t=2\pi$, would repeat the initial distribution, $f(q,p,t=0)=1$ at $-10\le p\le 10$, $-1\le q\le 1$. Therefore, the difference between the numerical solution and thus the defined analytical solution, $\sum_\mathrm{cells}{\left|f(q,p,t=2\pi)-f(q,p,t=0)\right|\Delta q\Delta p}/\sum_\mathrm{cells}{f(q,p,t=0)\Delta q\Delta p}$ characterizes $L_1$ norm of deviation of the numerical solution from the analytical one, i.e., the numerical error. On the other hand, the defect ${\cal E}(t=2\pi)-{\cal E}(t=0)$ characterizes the non-conservation of energy, ${\cal E}(t)=\sum_\mathrm{cells}{f(q,p,t)H(q,p)\Delta q\Delta p}$. The discontinuous initial distribution and  comparison of the test numerical solution with the discontinuous analytical solution are designed to artificially increase the error and energy defect to make them noticeable in a comparatively short test simulation.
\begin{figure}[t]
\includegraphics[width=\textwidth]{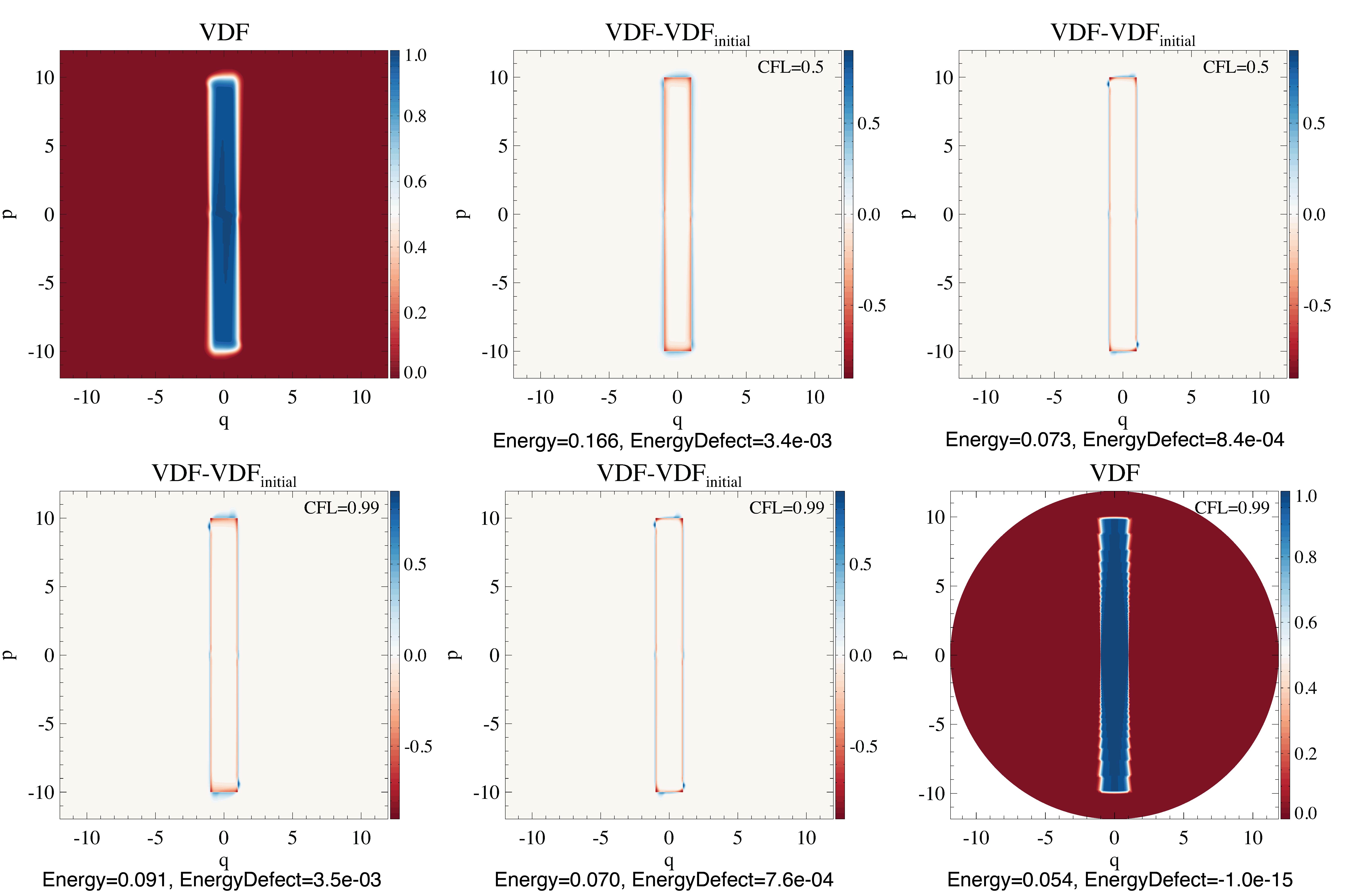}
\caption{Numerical results on a grid of $360\times360$ cells for distribution function of harmonic oscillators, (\textbf{Top left and Bottom right} panels), or for its difference from the exact solution (all other panels).  \textbf{Top left and middle panel} show the result with Koren's limiter Eq.(\ref{eq:Koren}). All other simulations were done with superbee limiter Eq.(\ref{eq:betalimiter},\ref{eq:betalimiter2}), which better maintains sharp boundaries. In high-CFL result in \textbf{Bottom left} panel the larger error is noticeable near points $x=\pm1,y=0$  and $x=0,y=\pm10$, where the sign of $\delta^-f$ changes and the limiter Eq.(\ref{eq:betalimiter2}) adds an extra dissipation. However, if the less restrictive limiter Eq.(\ref{eq:betalimiter}) is improperly used instead, an error reduces as shown in \textbf{Bottom middle panel}, but the TVD property breaks and undershoots occur in the distribution function to negative values $\sim-10^{-3}$.  \textbf{Bottom right panel:} Same problem solved on a grid $300\times360$ in ``action-angle'' coordinates.}
\label{fig:EnergyDefect}
\end{figure}

The numerical simulation result on a grid of $360\times360$ cells in the computational domain, $-12\le q\le12$, $-12\le p\le 12$  is shown in Fig. ~\ref{fig:EnergyDefect}. The top left panel presents the distribution function at the final time instant, $t=2\pi$ and  top middle panel shows its difference from the exact solution, for a test simulation with Koren's limiter Eq.(\ref{eq:Koren},\ref{eq:betalimiter2}). All other simulations were done with superbee limiter Eq.(\ref{eq:betalimiter},\ref{eq:betalimiter2}), which better maintains sharp boundaries. In the top row all results are for CFL number, $C_{\rm max}=0.5$. For CFL number $C_{\max}=0.99$ (bottom left panel) a larger error is noticeable near points $x=\pm1,y=0$  and $x=0,y=\pm10$, where the sign of $\delta^-f$ changes and the limiter Eq.(}\ref{eq:betalimiter2}) adds an extra dissipation. It is easy to outperform the TVD scheme, if the less restrictive limiter Eq.(\ref{eq:betalimiter}) is improperly used instead, as shown in bottom middle panel. However, the TVD property no longer holds in this case, so that undershoots occur in the distribution function to negative values $\sim -10^{-3}$.  By the end of the simulation the relative error of energy $\left[{\cal E}(2\pi) - {\cal E}(0)\right]/{\cal E}(0)=8\times10^{-4}$, which is rather small, compared to the  $7.3\%$ relative error of the solution $f$ in the L1 norm.

To mitigate the energy-non-conservation and to greatly reduce the numerical error, one can choose the generalized coordinate/momentum in such way that the conserved quantity becomes one of the generalized coordinates, for example, using the ``action-angle'' coordinates, $J,\theta$, described by \cite{Landau1976}, which for our test problem can be chosen as follows:
\begin{equation*}
q=\sqrt{2J}\cos{\theta},\quad p=\sqrt{2J}\sin{\theta},\quad H=J=\frac12(q^2+p^2),
\end{equation*}
which also corresponds to the choice of polar coordinates on the phase plane.  The equations for Hamiltonian trajectory in these coordinates becoming: $\frac{\mathrm{d}J}{\mathrm{d}t}\left(=\frac{\partial H}{\partial \theta}\right)=0,\,\frac{\mathrm{d}\theta}{\mathrm{d}t}\left(=-\frac{\partial H}{\partial J}\right)=-1$, so that numerical flux along the coordinate axis of total energy vanishes identically and so does the energy defect.

The test simulation on a somewhat coarser grid of 300 logarithmic-uniform meshes over $J$ ranging from $0.5\cdot10^{-4}$ to $72$ and $360$ meshes over the $2\pi$ range of angle, which corresponds to the same size of computational domain as in the test on rectangular grid. With the same initial condition and simulation time the result is presented in the right panel in Fig.~(\ref{fig:EnergyDefect}) remapped to $q,p$ phase plane. Despite lower resolution, the error in norm $L_1$ noticeably reduces to $5.4\,\%$ and the energy defect vanishes (within roundoff error). That is why in the applications to  the  particle acceleration in the space environment, described below, we always use the distribution function with respect to energy and pitch-angle, rather than the mathematically equivalent formulation with parallel and perpendicular momentum.

\subsection{Simulating Chaos}
\label{Sec:Chaos}
The presented method may be also applied to systems known to demonstrate a chaotic  behaviour, e.g., the Hamiltonian in action-angle coordinates \cite[see][for detailed analysis of this well-known dynamical system just briefly outlined here]{Zaslavski1988,Moon2004}:
\begin{equation}\label{eq:stoc}
H(\theta, J, t)=
\frac{J^2}2+
\omega_0^2
T\cos\theta\sum_{n=-\infty}^\infty{\delta\left(t-nT\right)}.
\end{equation}
The system is $2\pi$-periodic over the $\theta$ coordinate and it is perturbed by a time-periodic impulse, $\mathrm{d} J/\mathrm{d} t = -\partial H/\partial\theta \propto \delta\left(t-nT\right)$,  at $t=nT$, with time period $T$. The impulse has a sinusoidal dependence on $\theta$, so that a particle with the first resonant velocity $J=2\pi/T$ passes through the entire span of the $\theta$ coordinate within a time period $T$.

In the phase plane at velocity $J=0$, there are three stationary points. The ones at $\theta=0$ and $\theta=2\pi$ are unstable, while the one at $\theta=\pi$ is stable. For these angles the impact $\Delta J\propto\sin\theta$ vanishes, so it does not modify the velocity $J=0$ and the particle stays at rest. More detailed pattern of the Hamiltonian trajectories at the phase plane is governed by the parameter
\be
K=\omega_0^2T^2.
\ee
For $K\ll1$, we may average the Hamiltonian in time:
\be
\left\langle H\right\rangle(\theta, J)= 
\frac1T\int\limits_0^T{H(\theta,J, t)\mathrm{d}t}
= \frac{J^2}{2} + \omega_0^2\cos\theta,
\ee
which coincides with the Hamiltonian for 
oscillations. Time-averaging is justified by the observation that the period of oscillations $2\pi/\omega_0=2\pi T/\sqrt{K}$ is much longer than the time interval $T$ between the impacts for $K\ll1$. The \textit{separator} lines $J=\pm2\omega_0\sin\left(\theta/2\right)$ connect two unstable stationary points at $\theta=0$ and $2\pi$ and bound the region of finite motions, i.e. oscillations about the stable stationary point at $\theta=\pi$ from the region of unbounded motion of particles with energies $J^2/2+\omega^2_0\cos\theta=\mathrm{const}>\omega_0^2$ exceeding the peaks of the potential $\omega^2_0\cos\theta$.
\begin{figure}[t]
\includegraphics[width=\textwidth]{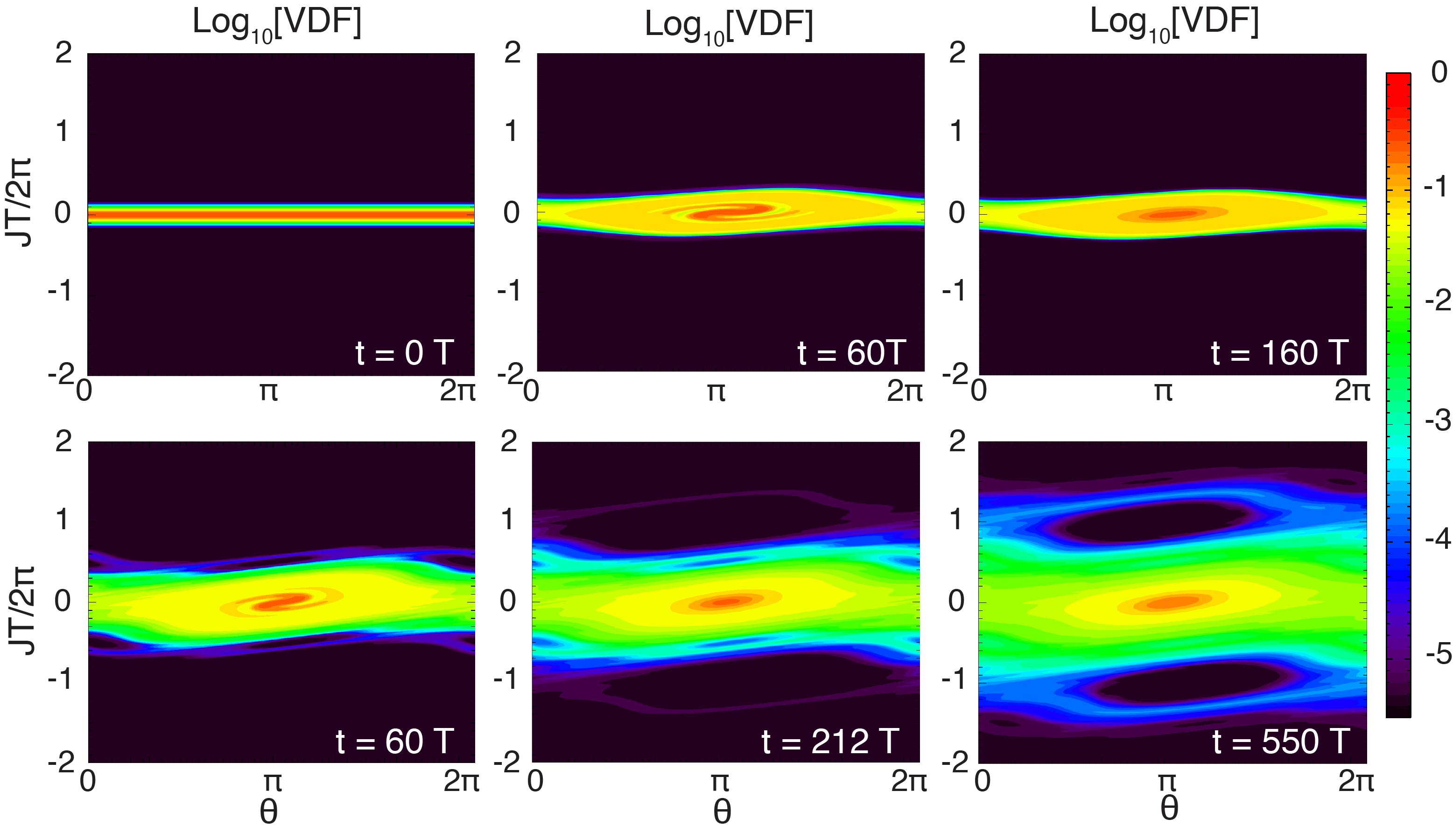}
\caption{Evolution of the distribution function, $\log_{10}f$, of system Eq.~(\ref{eq:stoc}), is simulated on $180\times720$ grid points, with the approximated $\delta-$function, $\delta(t-nT)=1/(0.1T)$ at $nT<t<nT+0.1T$. Horizontal axis: $0\le\theta\le2\pi$, in radians, vertical axis: normalized velocity, $-2\le JT/(2\pi)\le 2$.   \textbf{Top row:} well below stochasticity threshold, $K=0.6<1$: \textbf{Left panel:} $t=0$, initial distribution with small velocity; \textbf{Middle panel:} $t=60T$, domain of finite motion is filled in; \textbf{Right panel:} $t=160T$, steady state is reached. \textbf{Bottom row:} global stochasticity, $K=1.2>1$: \textbf{Left panel:} $t=60T$, stochastic diffusion across the separator fills in the separators of semi-resonances,  $J=\pm\pi/T$, $\theta=0,\pi,2\pi$; \textbf{Middle panel:} $t=212T$ precursors of separators of the first resonance regions start to form; \textbf{Right panel:} $t=550T$, the separators of the first resonances are filled in and separators of new  semi-resonances, $J=\pm3\pi/T$, $\theta=0,\pi,2\pi$ start to form.}
\label{fig:chaos}
\end{figure}

For $K\sim1$ the stronger impacts are essentially discrete, resulting in a qualitative change in the system properties.  Particularly, at the first resonant velocities $J=\pm 2\pi/T$, the $\theta=\pi$ point is also stationary, since the impact amplitude is zero and the particle with such velocity passes the entire coordinate span and arrives back at $\theta=\pi$ (due to the periodicity in $\theta$) by the time of the next impact at $t+T$. 
The new domains of finite motion around these stationary points are bounded by new separators, so that the particle close to one of the stationary points would not be able to approach another one, as long as the system is a regular (not stochastic) dynamical system. However, the system with $K\sim1$ acquires a property of \textit{stochastic diffusion} across the separators. 
Fig.~\ref{fig:chaos} illustrates the behavior of the system. For moderate $K=0.6$ (top row), the distribution function in the middle panel at $t=60T$  fills in the region of zero resonance at the point $(J=0, \theta=\pi)$, then the solution becomes steady state (see the right panel for $t=160T$). The bottom line illustrates the case of $K=1.2$ when the Chirikov criterion $K\ge1$ for global chaos is satisfied. In the left panel ($t=60T$) because of the stochastic diffusion the separators of semi-resonant stationary points $J=\pm\pi/T$, $\theta=0,\pi,2\pi$, are shaped. At the semi-resonant velocity the particle passes from  the point $\theta=0$ to $\theta=\pi$ during the time period $T$ and then passes from $\theta=\pi$ to $\theta=2\pi$ during the next time interval $T$. It passes the entire span within the time $2T$ (rather than $T$ as in the first resonances). The further evolution demonstrates the \textit{resonance overlapping}, resulting in appearance of precursors of first resonances in the middle panel at $t=212T$. 
In the right panel ($t=550T$) in addition to mature separators of the first resonances, the new semi resonances start to form at $J=\pm3\pi/T$, $\theta=0,\pi,2\pi$. We limited the simulation time in these preliminary tests to about $1000T$) ($\sim5\cdot10^5$ iterations) to mitigate effect of numerical heating (see section~\ref{Sec:Noncons}). However, the main idea of \textit{stochastic acceleration} in the course of which some minor fraction of particles with initially low energy may eventually gain unbounded energy, is demonstrated by these tests.
\section{Non-Canonical Kinetic Equations}
\label{Sec:NonCanonical}
The current research is mostly motivated by a need to have an efficient method for modelling the energetic particle acceleration and transport in the solar-terrestrial environment. In the related computational problems the \textit{canonical} Liouville equation is not often used and the Poisson brackets, on the first glance, may look inapplicable. 
The main distinction of the kinetic equations, as applied to space science, is the use of different phase variables $q_1,q_2,\dots,q_M$ for the distribution function, such as the energy or the momentum magnitude, pitch-angle, flux coordinates and so on, which may be not possible to separate for the generalized coordinates and canonically conjugated momenta. 
Therefore, not only the kinetic equation may look quite different from Eq. ~\ref{eq:Liouville}, but even the integral of the particle number $N$ may be different from the expression $N=\int{\mathrm{d}\Gamma f}$, where $\mathrm{d}\Gamma=\prod_l{\mathrm{d}q_l\mathrm{d}p_l}$ is assumed above. 
If such \textit{non-canonical} kinetic equation still can be formulated via the Poisson brackets, the pseudo-Hamiltonian functions $H_l$ in different Poisson brackets may be different and they may or may not have a meaning of the canonical Hamiltonian function. 
The simplest example of such equation with the unmodified integral for particle number is as follows:
\be\label{eq:manyHamiltonians}
\partial_tf+\sum_{l=1}^L{\left\{f;H_l\right\}_{q_{m_{l1}},q_{m_{l2}}}}=0,\qquad N=\int{
\mathrm{d}\mathbf{\Gamma}_Mf},\qquad \mathrm{d}\mathbf{\Gamma}_M=\prod\limits_{m=1}^M{\left(\mathrm{d}q_m\right)},
\ee
in which $L$ Poisson brackets are present with respect to a pair of phase variables, $q_{m_{l1}}$, $q_{m_{l1}}$. In this case the expression in Eq.~(\ref{eq:effective}) for the reduced Hamiltonian function is generalized as follows:
\be\label{eq:effectiveHl}
\widetilde{H}_l(q_{m_{l1}},q_{m_{l2}})=\int\limits_V{\prod\limits_{m\ne m_{l1},m_{l2}}\left(\mathrm{d}q_m\right)H_l(q_1,\dots,q_M)}.
\ee
Another distinction is that, as noticed in section~\ref{Sec:Poisson}, if a phase variable $q_m$ appears in more than one Poisson bracket, i.e. $m=m_{l1}$ or $m=m_{l2}$ for more than one $l$, then more than one difference in the reduced Hamiltonian functions $\delta \tilde{H}_l$ in Eq.~(\ref{eq:4deltaH}) contributes to the numerical flux $\delta\tilde{H}_j$ in Eq.~(\ref{eq:deltaHj}). All other formulae in section~\ref{Sec:Liouville} are applicable to solve Eq.~(\ref{eq:manyHamiltonians}) as well.   

In a more general case, the \textit{Jacobian} $J$ of the transformation from canonical coordinates, $q_l$ and momenta $p_l$ to non-canonical phase coordinates $q_m$ depends on the coordinates, so that:
\be\label{eq:Jacobian}
N=\int{
\prod\limits_{m=1}^M\left(\mathrm{d}q_m\right)J\left(q_1,\dots,q_M\right)f\left(t,q_1,\dots,q_M\right)}.
\ee
The kinetic equation conserves this integral of particle number if it has the following form:
\bea\label{eq:withJacobian}
\partial_t f+\frac1J\sum_{l=1}^L{\left\{f;H_l\right\}_{q_{m_{l1}},q_{m_{l2}}}}=0.
\eea
Similarly to Eq.~(\ref{eq:finitevolume}), to derive the finite volume formulation for Eq.~(\ref{eq:withJacobian}) it should be multiplied by the Jacobian $J$, and integrated over the control phase volume, which is the rectangular box $\prod_m\Delta q_m$:
\be
\frac{\mathrm{d}}{\mathrm{d}t}\int{\mathrm{d}\mathbf{\Gamma}_M J\left(q_1,\dots,q_M\right)f\left(t,q_1,\dots,q_M\right)}=-\sum_l{\oint{\mathrm{d}\tilde{H}_l\left(q_{m_{l1}},q_{m_{l2}}\right)f}},
\ee
with the above expression, Eq.~(\ref{eq:effectiveHl}), for effective pseudo-Hamiltonian, $\tilde{H}_l\left(q_{m_{l1}},q_{m_{l2}}\right)$. 
This equation, again, may be solved using formulae in section 2 if, in addition to the use of modified Eq.~(\ref{eq:effectiveHl}), one also modifies the expression for the size of the control volume:
\be\label{eq:volumeJ}
V=\int\limits_{\prod{\left(\Delta q_m\right)}}{\mathrm{d}\mathbf{\Gamma}_M J\left(q_1,\dots,q_M\right)}.
\ee
\subsection{Transport in an Incompressible Fluid}
\label{Sec:Incompressible}
As an example of a physical system with multiple pseudo-Hamiltonians, 
we discuss the transport of some scalar quantity in an incompressible flow described by the following equation:
\be\label{eq:incomp}
\partial_t\rho+\nabla\cdot\left(\rho\mathbf{u}\right)=0.
\ee
Since the velocity $\mathbf{u}$ is divergence-free for an incompressible flow, $\nabla\cdot\mathbf{u}=0$, Eq.~(\ref{eq:incomp}) can be written in a form similar to the Liouville equation (cf. with Eq.~(\ref{eq:cons6})):
\be\label{eq:incompLiouville}
\partial_t\rho+\left(\mathbf{u}\cdot\nabla\right)\rho=0.
\ee
The divergence-free velocity vector can be written as the curl of a stream function vector $\mathbf{w}$ in 3D (analogous to the scalar stream function in 2D):
\be\label{eq:curlw}
\mathbf{u}=\nabla\times\mathbf{w}.
\ee
Expanding Eqs.~(\ref{eq:incompLiouville} and \ref{eq:curlw}) in Cartesian $x,y,z$ components, one can write Eq.~(\ref{eq:incomp}) as a  non-canonical Liouville equation with Poisson brackets:
\be
\partial_t\rho +\left\{\rho,w_z\right\}_{x,y}+\left\{\rho,w_x\right\}_{y,z}+\left\{\rho,w_y\right\}_{z,x}=0,
\ee
and solve it with the schemes presented here. For a uniformly spaced grid in Cartesian coordinates and for a control volume, $\Delta x\times\Delta y\times\Delta z$ the expression, Eq.~(\ref{eq:effectiveHl}), for effective pseudo-Hamiltonian reads:
\bea
\tilde{H}_z\left(x^{c}\pm\frac{\Delta x}2,y^{c}\pm\frac{\Delta y}2\right)&=&\int\limits_{z^{c}-\frac{\Delta z}2}^{z^{c}+\frac{\Delta z}2}{w_z\left(x^{c}\pm\frac{\Delta x}2,y^{c}\pm\frac{\Delta y}2,z\right)\mathrm{d}z}
,\nonumber\\ 
\tilde{H}_x\left(y^{c}\pm\frac{\Delta y}2,z^{c}\pm\frac{\Delta z}2\right)&=&\int\limits_{x^{c}-\frac{\Delta x}2}^{x^{c}+\frac{\Delta x}2}{w_x\left(x,y^{c}\pm\frac{\Delta y}2,z^{c}\pm\frac{\Delta z}2\right)\mathrm{d}x}
,\nonumber\\
\tilde{H}_y\left(z^{c}\pm\frac{\Delta z}2,x^{c}\pm\frac{\Delta x}2\right)&=&\int\limits_{y^{c}-\frac{\Delta y}2}^{y^{c}+\frac{\Delta y}2}{w_y\left(x^{c}\pm\frac{\Delta x}2,y,z^{c}\pm\frac{\Delta z}2\right)\mathrm{d}y}
,
\eea
so that the formulae from section 2 in this particular example reproduce the main idea of the so-called \textit{staggered grid} by \cite{Yee1968}: the flux of divergence-free velocity vector through the \textit{face} of the control volume is expressed in terms of the contour integral of the stream function $\mathbf{w}$ over the \textit{edges} bounding this face, via Stokes' theorem: $\int{\mathbf{u}\cdot\mathrm{d}\mathbf{S}}=\oint{\mathbf{w}\cdot\mathrm{d}\mathbf{x}}$.

\begin{figure}[t]
\centering
\includegraphics[width=\textwidth]{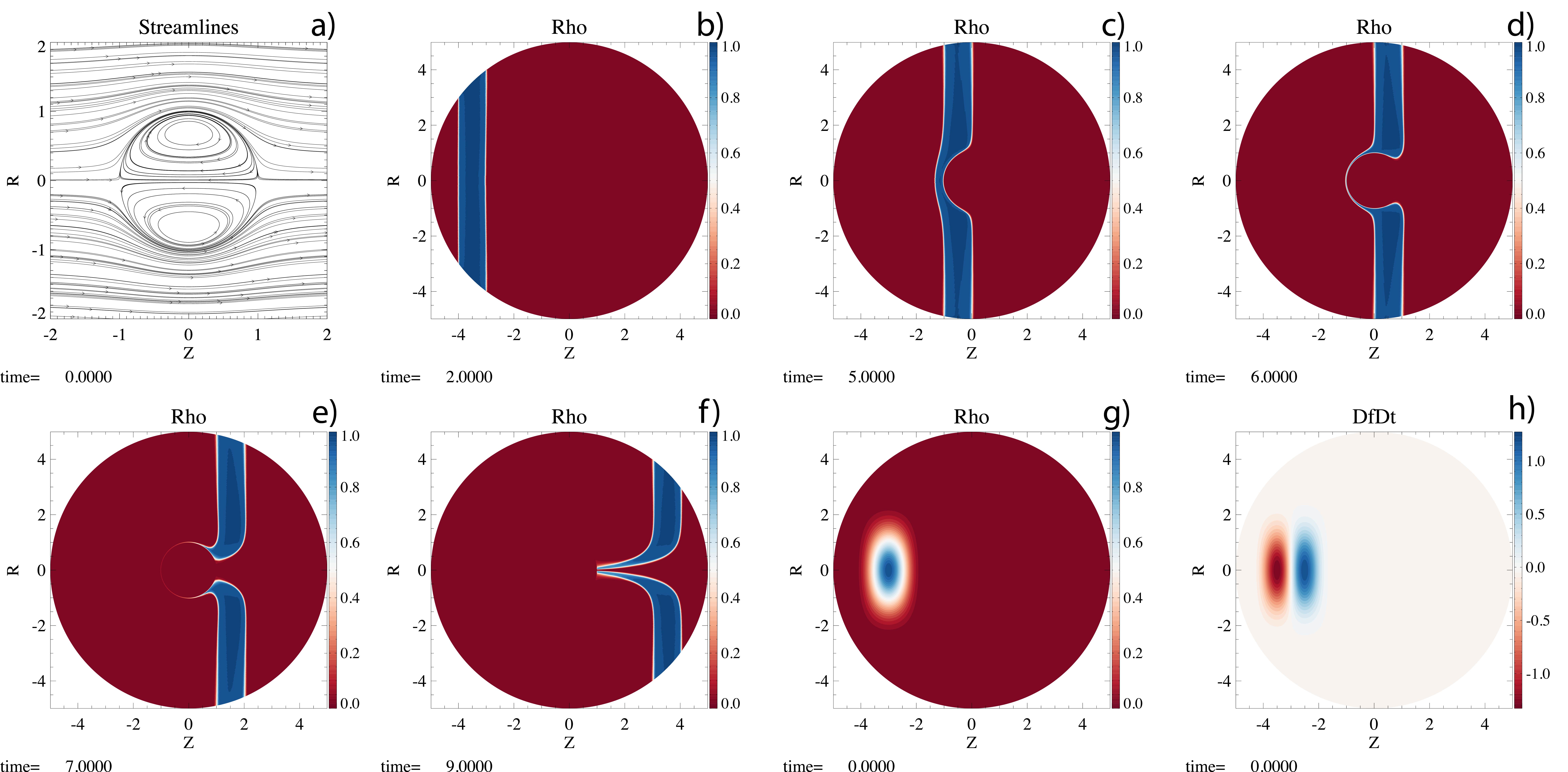}
\caption{Panel a: Streamlines of the spherical Hill vortex of unit radius centered at the origin of the spherical coordinates. Panels b through f: Transport of inked slab at times $t=2, 5, 6, 7$ and 9 is shown in a spherical domain of radius $R=5$ on a uniform grid with 250 points along radius and 360 points in the poloidal angle. The vortex itself inside $R=1$ is impenetrable and stays dark. Panels g and h: The smooth initial distribution (g) and   analytical prediction for its time derivative (h) to be compared with the numerical flux of the semi-discrete TVD scheme.}
\label{fig:HillVortex}
\end{figure} 

In the case of axisymmetric flow the stream function has only one non-zero component $w_\varphi$, where $\varphi$ is the polar angle in cylindrical coordinates or the longitude in spherical coordinates.  
By choosing the control volumes in a form of closed circular rings with rectangular (for cylindrical coordinates) or trapezoid (for spherical coordinates) meridional cross-section, the effective pseudo-Hamiltonian function is obtained by integrating $w_\varphi$ over a circular contour, thus giving the following expressions for the pseudo-Hamiltonian, the Jacobian, and the transport equation, respectively:
\be\label{eq:incompcyl}
\tilde{H}(z,r)=2\pi r w_\varphi(z,r),\quad J=2\pi r, \quad\partial_t\rho+\frac1J\left\{\rho;\tilde{H}\right\}_{z,r}=0,
\ee
in cylindrical coordinates, or
\be\label{eq:incompsph}
\tilde{H}(R,\theta)=2\pi R\sin\theta\, w_\varphi(R,\theta),\quad J=2\pi R^2\sin\theta,\quad \partial_t\rho+\frac1J\left\{\rho;\tilde{H}\right\}_{R,\theta}=0,
\ee
in spherical coordinates. 
The contours of constant $\tilde{H}(z,r)$ in the meridional plane present the \textit{streamlines}, which for a particular case of steady-state flow coincide with the Hamiltonian trajectories for Eq.~(\ref{eq:incompLiouville}). For example, in the top left panel of the Fig.~\ref{fig:HillVortex} the streamlines are shown for a \textit{spherical vortex} by \cite{Hill1894}, in which a free-stream flows around a closed circulating vortex bounded by a spherical surface of a unit radius. For a unit velocity of free-stream, the effective Hamiltonian function of the Hill's vortex is as follows:
\be\label{eq:HillFunction}
\tilde{H}=2\pi R^2\sin^2\theta\times\left\{\begin{array}{ll} \frac34\left(R^2-1\right),\quad \mbox{if $R<1$} \\ \frac12\left(1-\frac1{R^3}\right),\quad\mbox{if $R>1$}\end{array}\right.=2\pi r^2\times\left\{\begin{array}{ll} \frac34\left(z^2+r^2-1\right) \\ \frac12\left(1-(z^2+r^2)^{-3/2}\right).\end{array}\right.
\ee
As a numerical test problem we solve Eq.~(\ref{eq:incompsph}) for the transport of an ``inked'' slab of unit  thickness about the Hill vortex. A
uniform spherical grid of 250 cells over radius, $R$, ranging from 0 to 5, by 360 cells over colatitude, $\theta$, ranging from 0 to $\pi$ is used (the longitude direction is handled by the axisymmetry). When the number of cells over radius (250) is a multiple of 5, the vortex boundary, $R=1$, coincides with the interfaces between cells with the radial index 50 (inside the vortex) and 51 (outside the vortex). The numerical flux vanishes at these interfaces, so that the vortex and surrounding flow are disconnected.  The ink density evolution in the meridional plane is shown at different time instants in Fig.~\ref{fig:HillVortex}. Near the two stagnation points $(z=\pm1,r=0)$, the flow velocity tends to zero, so the parts of the inked slab that contact the vortex region lag behind. Since the inner region of the vortex is impenetrable for the ink, it stays as a dark circle of unit radius in all snapshots. Despite significant gradients both in the direction of the slab propagation and in the transverse direction the TVD scheme provides a solution free of spurious oscillations or excessive diffusion.

The velocity components can be obtained from:
\be
u_z=\frac1J\partial_r\tilde{H},\qquad u_r=-\frac1J\partial_z\tilde{H}.
\ee
Using a smooth initial condition:
\be\label{eq:manufactured}
\rho_{t=0}=\cos^4(\pi(z+3)/3)\cos^4(\pi r/6),
\ee
provides a convenient framework to test semi-discrete scheme Eq.(\ref{eq:incompLiouville}) in the way which is insensitive to the choice of time-integration scheme, in comparing the numerical flux by itself, rather than the result of its time integration. Indeed, from Eq.~(\ref{eq:incompLiouville}) the time derivative at the initial time instant, $\left(\partial_t\rho\right)_{t=0}=-\left(\mathbf{u}\cdot\nabla\right)\rho_{t=0}$, may be calculated analytically. The initial distribution and thus calculated time derivative are shown in Fig.~\ref{fig:HillVortex}. The latter can be compared with the numerical flux, $[\partial_tf]^\mathrm{(TVD)}$, of the semi-discrete scheme, calculated according to Eq.~(\ref{eq_22}). To make comparison more representative, the test combines the easy to treat analytical formulae in \textit{cylindrical} coordinates with the numerical flux for the flow Eq.~(\ref{eq:HillFunction}) calculated in \textit{spherical} coordinates, so that the features in the  initial distribution are not aligned with the coordinate axes. We evaluate the error in the L1 norm defined as $L_1=\sum_i{
\left|\partial_t
\rho_{t=0}-\left[\partial_tf\right]^\mathrm{(TVD)}\right|V_i}/\sum_i{V_i}$.

The results are the following. On the same grid as described above the second order flux approximation, $\left[\partial_tf\right]^{(2)}$ (see Eq.~(\ref{eq:deltaHj})), in which all face values are arithmetic averages of the neighboring cell values, results in the error $L_1=1.1\times10^{-5}$. The combination of the third-order (at smooth monotone solutions) Koren's $\beta$-limiter Eq.~(\ref{eq:Koren}) with $\beta=2$, (see Eq.~(\ref{eq:betalimiter}) for semi-discrete scheme),  with $\gamma$-limiter provides a better accuracy, $L_1=7.803\times10^{-6}$. At twice higher resolution, the error reduces to $L_1=2.004\times10^{-6}$. The ratio of the errors at the two grid resolutions characterizing the convergence rate is $3.894$, corresponding to the convergence rate $\log_23.894=1.96$. Therefore, the accuracy of the TVD semi-discrete scheme with Koren's limiter is not worse than that of the second order flux approximation.

\subsection{Time-Dependent Jacobian and Poisson Bracket}
\label{Sec:TimeJacobian}
In a more general and challenging case the Jacobian of transformation to the non-canonical phase coordinate may explicitly depend on time, $J=J\left(t,q_1,\dots,q_M\right)$, so that
\be\label{eq:Jacobiant}
N=\int{\mathrm{d}\mathbf{\Gamma}_MJ\left(t,q_1,\dots,q_M\right)f\left(t,q_1,\dots,q_M\right)}.
\ee
For simplicity, we consider a particular case, when the Jacobian is independent of one of the generalized coordinate $q_{m^\prime}$ and the kinetic equation conserving the integral~(\ref{eq:Jacobiant}) is as follows:
\be\label{eq:withJacobiant}
\frac1J\left(\left\{f;H_0\right\}_{t,q_{m^\prime}}+\sum_{l=1}^L{\left\{f;H_l\right\}_{q_{m_{l1}},q_{m_{l2}}}}\right)=0,\qquad H_0=J\,q_{m^\prime}.
\ee
To obtain the finite volume scheme for Eq.~(\ref{eq:withJacobiant}), we multiply it by $J$, integrate over the control phase volume $\prod_m\Delta q_m$ and over time from $t$ to $t+\Delta t$, and then divide it by $\Delta t$. Similarly to Eqs.~(\ref{eq:effectiveHl}), (\ref{eq:4deltaH}) and (\ref{eq:deltaHj}), on introducing the effective Hamiltonian,
\be
\tilde{H}_0=\frac{Vq_{m^\prime}}{\Delta t\Delta q_{m^\prime}},
\ee
and by reducing the two-dimensional integration over $\mathrm{d}t\mathrm{d}q_{m^\prime}$ to the contour integral, we find:
\bea\label{eq:2delta}
\frac{
V(+\Delta t) f(+\Delta t)
}{\Delta t}&-&\frac{
V f
}{\Delta t} +\sum_j\delta\tilde{H}_j\frac{f+f^{\rm ext}_j}2+\nonumber\\
&+&\left[\tilde{H}_0\left(+\Delta q_{m^\prime}/2\right)-\tilde{H}_0\left(+\Delta t,+\Delta q_{m^\prime}/2\right)\right]\frac{f+f(+\Delta q_{m^\prime})}2+\nonumber\\
&+&\left[\tilde{H}_0\left(+\Delta t,-\Delta q_{m^\prime}/2\right)-\tilde{H}_0\left(-\Delta q_{m^\prime}/2\right)\right]\frac{f+f(-\Delta q_{m^\prime})}2=0.
\eea
The differences $\delta\tilde{H}_0$ of the Hamiltonian function values in Eq.~(\ref{eq:2delta}) may be expressed as follows, 
\begin{eqnarray}
\delta\tilde{H}_{0,1}=\tilde{H}_0\left(+\Delta q_{m^\prime}/2\right)-\tilde{H}_0\left(+\Delta t,+\Delta q_{m^\prime}/2\right)&=&-\frac{\Delta V}{\Delta t}\frac{q_{m^\prime}+\Delta q_{m^\prime}/2}{\Delta q_{m^\prime}},\\
\delta\tilde{H}_{0,2}=\tilde{H}_0\left(+\Delta t,-\Delta q_{m^\prime}/2\right)-\tilde{H}_0\left(-\Delta q_{m^\prime}/2\right)&=&\frac{\Delta V}{\Delta t}\frac{q_{m^\prime}-\Delta q_{m^\prime}/2}{\Delta q_{m^\prime}}\\
\mathrm{with}\qquad\frac{\Delta V}{\Delta t}=\frac{V(+\Delta t)-V}{\Delta t}
\end{eqnarray}
and used to calculate the contribution to the numerical flux through the faces along the direction of $q_{m^\prime}$ coordinate, in the same way we did before. Now, we assume that in the finite volume formulation the quantities $\delta\tilde{H}_j$ already include these contributions:
\be\label{eq:finitevolumeJt}
\frac{
V(+\Delta t) f(+\Delta t)
}{\Delta t}-\frac{Vf}{\Delta t} +\sum_j\delta\tilde{H}_j\frac{f+f^{\rm ext}_j}2=0.
\ee
However, 
the sum of the two $\delta\tilde{H}_0$ terms 
defined above is non-zero: 
\be
\sum^2_{j=1}\delta\tilde{H}_{0,j}=-\frac{\Delta V}{\Delta t},
\ee 
This results in non-zero total of $\delta\tilde{H}_j$ because of the contributions from $\delta\tilde{H}_0$:
\be\label{eq:Hplusminust}
\sum_j\delta\tilde{H}_j=\sum^2_{j=1}\delta\tilde{H}_{0,j}=-\frac{\Delta V}{\Delta t}, \qquad 
\sum_{j,+}{\delta \tilde{H}^+_j}+\frac{\Delta V}{\Delta t}=-\sum_{j,-}{\delta \tilde{H}^-_j}.
\ee
Thus, the presence of the Poisson bracket with respect to time not only contributes to the
numerical flux, which is easy to account for, but also breaks Eqs.~(\ref{eq:deltaHj})--(\ref{eq:Hplusminus}). However, all the consequences from these equations derived in section 2 hold valid for Eq.~(\ref{eq:withJacobiant}) as well, although the appearance of the formulas may be different. Particularly, we can rewrite Eq.~(\ref{eq:finitevolumeJt}) via $\delta^\pm f$ as defined in Eq.~(\ref{eq:deltaf}):
\be
\frac{f(+\Delta t)-
f
}{\Delta t}=\frac{\left(\sum_{j,-}{\delta \tilde{H}^-_j}\right)\delta^- f- \left(\sum_{j,+}{\delta \tilde{H}^+_j}\right)\delta^+ f}{2V(+\Delta t)},
\ee
which is a straightforward generalization of Eq.~(\ref{eq:secondorder}). By adding the numerical diffusion
\be\label{eq:diffusiont}
D=
\frac{\sum_j{\left\vert \delta\tilde{H}_j\right\vert\left(f^{\rm ext}_j -f\right)}}{2V(+\Delta t)}=\frac{\left(\sum_{j,-}{\delta \tilde{H}^-_j}\right)\delta^- f+ \left(\sum_{j,+}{\delta \tilde{H}^+_j}\right)\delta^+ f}{2V(+\Delta t)}
\ee
that is similar to  (\ref{eq:diffusion}), we obtain the first order monotone numerical flux
\be\label{eq:firstordert}
\left(\frac{f(+\Delta t)-f}{\Delta t}\right)^{(1)}=
\frac{\left(\sum_{j,-}{\delta\tilde{ H}^-_j}\right)\delta^- f}{V\left(+\Delta t\right)}=
\frac{\left(\sum_{j,-}{\delta\tilde{ H}_j^-}\right)f-\sum_{j,-}{\delta\tilde{H}^-_jf^{\rm ext}_j}}{V\left(+\Delta t\right)}
\ee
that is a generalization of (\ref{eq:firstorder}). The flux in Eq.~(\ref{eq:firstordert}) is monotonous if 
\be\label{eq:CFLt}
C^\prime=-\frac{\Delta t\left(\sum_{j,-}{\delta\tilde{ H}^-_j}\right)}{V\left(+\Delta t\right)}=\frac{\Delta t}{V\left(+\Delta t\right)}\left\vert\sum_{j,-}\delta\tilde{H}^-_j\right\vert \le 1.
\ee
The relationship between $C'$ and $C$ defined in ({\ref{eq:CFL})  (using Eq.~\ref{eq:Hplusminust}) is 
\be
\left(C^\prime-1\right)=\frac{C-1}{1+\Delta V/V}.
\ee
Hence, if the requirement, $C^\prime\le1$ is satisfied, the criterion $C\le1$ for equation with time-dependent Jacobian is satisfied too. Therefore, the maximum time step $(\Delta t)_{\max}$, still satisfying the stability criterion $C^\prime\le1$, may be solved for \textit{explicitly}:
\be\label{eq:dtmax}
\left(\Delta t\right)_{\max}= \min\limits_\mathrm{cells}\frac{V}{\sum\limits_{j,+}\delta^+\tilde{H}},
\ee
rather than from an implicit equation, $\max\left(C^\prime\right)=1$. 
\subsection{TVD Scheme for Kinetic Equation with Time-Dependent Jacobian}
\label{Sec:TimeTVD}
Now, we define TV for Eq.~(\ref{eq:withJacobiant}), which can be also written in a quasi-linear form as follows:
\be\label{eq:qltimedependent}
\partial_tf+\frac1J\left(-(\partial_tJ)q_{m^\prime}\frac{\partial f}{\partial q_{m^\prime}}+\sum_{l=1}^L{\left\{f;H_l\right\}_{q_{m_{l1}},q_{m_{l2}}}}\right)=0.
\ee
To use the results presented above, we extend the set of $M$ phase coordinates by another fictive coordinate, $0\le\xi\le1$. At the initial time instant $\partial_\xi f\equiv0$, we also assume floating boundary condition, $\partial_\xi f=0$ at $\xi=1$. Now, Eq.~(\ref{eq:qltimedependent}) may be written in the form similar to  Eq.~(\ref{eq:withJacobian}):     
\be\label{eq:withxi}
\partial_tf+\frac1J\left(\left\{f;(\partial_tJ)\xi q_{m^\prime}\right\}_{\xi,q_{m^\prime}}+\sum_{l=1}^L{\left\{f;H_l\right\}_{q_{m_{l1}},q_{m_{l2}}}}\right)=0,
\ee
which can be also rewritten as $J\partial_tf+\mathbf{u}_{M+1}\cdot\nabla_{M+1}f=0$ or $J\partial_tf+\nabla_{M+1}\cdot\left(f\mathbf{u}_{M+1}\right)=0$. The TV function given by Eq.~(\ref{eq:TVPoisson}) can be generalized for Eq.~(\ref{eq:withxi}) as follows: 
\be\label{eq:TVwithxi}
T=\int\limits_0^1{\mathrm{d}\xi\int\limits_{V}{
\mathrm{d}\mathbf{\Gamma}_M\left\vert\mathbf{u}_{M+1}\cdot\nabla_{M+1} f\right\vert}}.
\ee
Similarly to Eq.~(\ref{eq:TVpmPoisson}) one can express TV in Eq.~(\ref{eq:TVwithxi}) in terms of integrals, $T=T^++T^-$, over monotonicity regions, $V^\pm$ depending on the sign of gradient of distribution function
$\mathbf{u}_M\cdot\nabla_M$:
\be\label{eq:TVprimepm}
T^{\pm}=\pm\left[\,\,\int\limits_{V^\pm}{\mathrm{d}\mathbf{\Gamma}_M\nabla_M\cdot\left(\mathbf{u}_M f\right)}+\int\limits_{V^\pm}{\mathrm{d}\mathbf{\Gamma}_M
(\partial_t J)f}\right].
\ee
Here, the first term results from integrating over $0\le\xi\le 1$ the $M$ terms in divergence that are independent of $\xi$. 
The second term results from integrating the $\xi$ partial derivative with respect to $\xi$, which results in the difference of the primitive function at the external boundary surfaces at  $\xi=1$ and $\xi=0$. At the $\xi=0$ boundary $(\partial_t J)\xi$ vanishes so only the $\xi=1$ boundary contributes to the second term.
Note, that we assume the floating boundary condition at $\xi=1$, $\left(\partial_\xi f\right)_{\xi=1}=0$. We can no longer prove TVD property as we did in Section~\ref{Sec:TVD} for the fixed boundary condition, since an extra integral over $\xi=1$ boundary contributes to the time derivative of TV, according to Eq.~(\ref{eq:dTdtfixed}):
\bea\label{eq:dTdtwithxi}
\frac{\mathrm{d}T}{\mathrm{d}t}&=&\int\limits_{\partial V}{\mathrm{d}\mathbf{S}_M\cdot\mathbf{u}_M\mathrm{sign}\left(\mathbf{u}_M\cdot\nabla_Mf\right)\partial_tf}
+\int\limits_V{\mathrm{d}\mathbf{\Gamma}_M\left(\partial_tJ\right)\mathrm{sign}\left(\mathbf{u}_M\cdot\nabla_Mf\right)\partial_tf}\nonumber\\
&\le& \int\limits_V{\mathrm{d}\mathbf{\Gamma}_M\left(\partial_tJ\right)\mathrm{sign}\left(\mathbf{u}_M\cdot\nabla_Mf\right)\partial_tf}.
\eea
The sign of the RHS in Eq.~(\ref{eq:dTdtwithxi}) is not definite, thus making TV defined by Eq.~(\ref{eq:TVwithxi}) is inapplicable to formulate the TVD principle. Therefore, we re-define the TV in the following way:
\bea\label{eq:TVpmwithxi}
\tilde{T}&=&T-\int\limits_V{\mathrm{d}\mathbf{\Gamma}_M\left(\partial_tJ\right)\mathrm{sign}\left(\mathbf{u}_M\cdot\nabla_Mf\right)f}=\tilde{T}^++\tilde{T}^-, 
\nonumber\\
\tilde{T}^{\pm}&=&\int\limits_{V^\pm}{
\mathrm{d}\mathbf{\Gamma}_M\left[\left\vert\mathbf{u}_{M}\cdot\nabla_{M} f\right\vert\mp(\partial_t J)f\right]}=\pm\int\limits_{V^\pm}{\mathrm{d}\mathbf{\Gamma}_M\nabla_M\cdot\left(\mathbf{u}_M f\right)}.
\eea
However, this modification does not fully cancel out the time derivative of TV in Eq.~(\ref{eq:dTdtwithxi}):
\bea\label{eq:dtildeTdt}
\frac{\mathrm{d}\tilde{T}}{\mathrm{d}t}&=&\frac{\mathrm{d}T}{\mathrm{d}t}
-\frac{\mathrm{d}}{\mathrm{d}t}\int\limits_V{\mathrm{d}\mathbf{\Gamma}_M\left(\partial_tJ\right)\mathrm{sign}\left(\mathbf{u}_M\cdot\nabla_Mf\right)f}
\le\int\limits_V{\mathrm{d}\mathbf{\Gamma}_M\left(\partial_tJ\right)\mathrm{sign}\left(\mathbf{u}_M\cdot\nabla_Mf\right)\partial_tf}-\nonumber\\
&-&\frac{\mathrm{d}}{\mathrm{d}t}\int\limits_V{\mathrm{d}\mathbf{\Gamma}_M\left(\partial_tJ\right)\mathrm{sign}\left(\mathbf{u}_M\cdot\nabla_Mf\right)f}
=
\sum_{+,-}\mp\int\limits_{\partial V^\pm}{\mathrm{d}\mathbf{S}_M\cdot\mathbf{u}_\mathrm{b}\left(\partial_tJ\right)f}.
\eea
The re-defined TV does not increase  with time only if in the time derivative of the integral in (\ref{eq:dtildeTdt}) one neglects  the motion of the interface separating $\partial V^\pm$ regions, with the velocity $\mathbf{u}_\mathrm{b}=\mathbf{u}_M/J$, or, equivalently, if in calculating this time derivative only the distribution function in the integrand is differentiated over time, but the sign function is not. Otherwise, the RHS in (\ref{eq:dtildeTdt}) calculated similarly to Eq.~(\ref{eq:Reynolds}) does not vanish, since the finite magnitude integrand, $\pm\left(\partial_tJ\right)f$, in the definition of modified TV in Eq.~(\ref{eq:TVpmwithxi}), changes its sign in a discontinuously when the interface passes. By neglecting the interface variation, the TVD property preserves monotonicity by prohibiting the appearance of opposite monotonicity region, $\delta V^\pm$, inside the existing monotonicity region, since this would require the positive increase in TV,
\be
\delta \tilde{T}^{\pm}=2\int\limits_{\delta V^\pm}{
\mathrm{d}\mathbf{\Gamma}_M\left\vert\mathbf{u}_M\cdot\nabla_Mf\right\vert}>0.
\ee
Although the assumption of a non-moving interface between $V^\pm$ regions is a severe limitation of the proved TVD principle for kinetic equation with time-dependent Jacobian, we still can apply it to the semi-discrete finite volume numerical scheme. Indeed, the assumption that the interface between cells with opposite sign of the distribution function gradient is not displaced by a finite cell size within an infinitesimal time interval, is legitimate. Therefore, for the set of control volumes we introduce $\tilde{T}^{\pm}$ by combining Eqs.~(\ref{eq:TVPlusminus}) and (\ref{eq:TVprimepm}). For the regions of monotonicity, $\mathrm{cells}^\pm$, in which $\mathrm{sign}\left(\delta^-f\right)=\pm1$, we define $\tilde{T}^{\pm}$ for $\frac{\Delta V}{\Delta t}\ne0$ as follows: 
\bea\label{eq:TVPlusminust}
\tilde{T}^{\pm}&=&-\sum_{\mathrm{cells}^\pm}{
\left(\sum_{j,-}{\delta\tilde{H}^-_j}\right)\vert \delta^- f\vert}\mp \sum\limits_{\mathrm{cells}^\pm}{\frac{\Delta V}{\Delta t}f}=\pm\sum_{\mathrm{cells}^\pm}{\left[\left({\sum_{j,+}}{\delta\tilde{H}^+_j}\right)f+{\sum_{j,-}}{\delta\tilde{H}^-_jf^{\rm ext}_j}\right]}=\nonumber\\
&=&\pm\sum_{\mathrm{boundary\, cells}^\pm}{\left[\left({\sum_{j,+}}^\prime{\delta\tilde{H}^+_j}\right)f+{\sum_{j,-}}^\prime{\delta\tilde{H}^-_jf^{\rm ext}_j}\right]}.
\eea
After this derivation, it appears that Eqs.~(\ref{eq:TV}) and (\ref{eq:TVPlusminus}) hold for the TV defined in Eq.~(\ref{eq:TVPlusminust}) and all statements made in section~\ref{Sec:TVD} are also valid for this re-defined TV. Particularly, the TV reduces to the sum over boundaries of the monotonicity regions. Its time derivative is non-positive as long as the numerical scheme satisfies Eq.~(\ref{eq:TVDCrit}), thus ensuring the TVD property. The first order scheme (\ref{eq:firstordert}) with monotone numerical flux is TVD. 

The TVD property eliminates monotonicity breaking. To create an isolated cell inside the monotonicity region with the sign of $\delta^-f$ being opposite to that in the monotonicity region, the TV defined in Eq.~(\ref{eq:TVPlusminust}) would have to gain a positive increment, $-2\left(\sum_{j,-}{\delta\tilde{H}^-_j}\right)\vert \delta^- f\vert$, i. e. should increase, thus contradicting the TVD property. Note, that the choice of $\mp$ sign of the term $\mp\frac{\Delta V}{\Delta t}f$ in the definition is governed by the sign of $\delta^-f$ in the monotonicity region, rather than that in the presumably appearing isolated cell with broken monotonicity.  
The second order TVD scheme applicable to the non-canonical kinetic equations with the time-dependent Jacobian may be chosen in the following way:
\bea\label{eq:TVDSchemeWithJacobian}
f(+\Delta t)&-&f=-C^\prime\,\delta^-f-\frac{\Delta t}{V(+\Delta t)}\left((1-C^\prime\right)\frac{\left(\delta^+f\right)^\mathrm{lim}}2\sum_{j,+}{\delta\tilde{H}^+_j}-\nonumber\\
&-&\frac{\Delta t}{V(+\Delta  t)}\sum_{j,-}{\delta\tilde{ H}^-_j\left\{\gamma^{\rm ext}_j\Psi^{\rm ext}_\beta\left(\frac{f-f_j}2\right)-C^{\prime\mathrm ext}_j\left[\frac{\left(\delta^+f\right)^\mathrm{lim}}2\right]^{\mathrm ext}_j\right\}}.
\eea
It satisfies the TVD criterion (\ref{eq:TVDCrit}) once the CFL number does not exceed one. The scheme in (\ref{eq:TVDSchemeWithJacobian}) is analogous to that 
in Eq.~(\ref{eq:TVDScheme}), however, the CFL number $C'$ should be defined according to Eq.~(\ref{eq:CFLt}) and the time-dependent volume should be calculated at the end of the time step. 

\section{Non-Canonical Kinetic Equation for SEPs with Poisson Bracket}
\label{Sec:Parker}
Here, we show how the Poisson bracket based scheme may be used for modelling the Solar Energetic Particle (SEP) acceleration by the interplanetary shock waves as well as their transport toward the Earth orbit \cite[see][and papers cited therein]{boro18}. To simulate the fluxes of shock-accelerated SEPs, the two competing approaches are employed, which differently treat the shock region. Particularly, the shock wave may be thought of as a prescribed source of accelerated particles, derived from semi-analytical or semi-empirical models. In this case the kinetic model is designed to just solve an upstream transport of already accelerated particles thorough the heliosphere. An alternative approach is to solve the kinetic equation throughout the whole computational domain including the shock wave region too, so that the diffusive shock acceleration mechanism by \cite{axford77,krymsky77,Bell1978a,Bell1978b,blandford78,axford81} is the part of the SEPs model. For the latter application, it is important to use a particle conserving scheme, otherwise the prediction for the SEP flux may be compromised by the fake particle production due to approximation errors at high spatial gradients at the shock.

Usually, in application to the SEPs modelling, one introduces their {\it gyrotropic} distribution function, $\tilde{f}\left(t, {\bf x}, p^3/3,\mu\right)$, in a magnetized moving turbulent plasma, which is defined in a frame of reference, co-moving with the local velocity of interplanetary plasma, 
${\bf u}({\bf x},t)$, at the given point, ${\bf x}$. On introducing spherical coordinates, 
$(p=|{\bf p}|,\,\mu={\bf b}\cdot{\bf p}/p,\,\varphi)$,
in the momentum space, such that the polar axis is aligned with the direction, ${\bf b}=\mathbf{B}/B$,
of the Interplanetary Magnetic Field (IMF), $\mathbf{B}({\bf x},t)$, the gyrotropic distribution function is averaged over $\varphi$, i.e. over the phase of particle Larmor gyration: 
\be
\tilde{f}\left({\bf x}, \frac{p^3}3,\mu,t\right)=\frac1{2\pi}\int\limits_0^{2\pi}{d\varphi \,f({\bf x}, p,\mu,\varphi,t)}, \quad dN=2\pi d^3{\bf x}\int\limits_0^\infty{d\frac{p^3}3\int\limits_{-1}^1{d\mu \,\tilde{f}\left({\bf x}, \frac{p^3}3,\mu,t\right)}},
\ee
where $dN$ is the particle number in the phase volume element.
In section~\ref{Sec:More}
we provide the governing equation and numerical results for the SEP model, which keeps dependence on the cosine of pitch-angle, $\mu$ within the framework of the so-called {\it focused transport equation}. However, here we make a further simplification and deal with the {\it isotropic (omnidirectional)} distribution function, $f_0\left({\bf x}, p^3/3,t\right)$, which is averaged over the whole solid angle:
\be
f_0\left({\bf x}, \frac{p^3}3,t\right)=
\frac1{2}\int_{-1}^1{d\mu\, \tilde{f}\left({\bf x}, \frac{p^3}3,\mu,t\right)},\quad dN=4\pi d^3{\bf x}\int_0^\infty{d\frac{p^3}3 \,f_0\left({\bf x}, p,t\right))}.
\ee
The kinetic equation for the isotropic distribution function was introduced by \cite{Parker1965}:
\begin{equation}
\label{eq:parker}
\partial_t f_0 + 
\left(\mathbf{u}\cdot\nabla\right)f_0 -
\left(\nabla\cdot\mathbf{u}\right)\frac{p^3}{3}
\frac{\partial f_0}{\partial\left( p^3/3\right)} = 
\nabla\cdot\left(\varkappa\cdot\nabla f_0\right),
\end{equation}
where $\varkappa=D_{xx}\mathbf{b}\mathbf{b}$ 
is the tensor of parallel spatial diffusion along the IMF, $D_{xx}$ being the parallel diffusion coefficient.
In this approximation,  the cross-field diffusion of particles is neglected.
The Parker equation \ref{eq:parker} accounts for effects from the IMF and other background parameters of the solar wind on the SEP acceleration and transport in the solar atmosphere.
The term proportional to the divergence of $\mathbf{u}$ describes the adiabatic cooling, for $\left(\nabla\cdot\mathbf{u}\right)>0$, or (the first order Fermi) acceleration in shock compression waves, for $\left(\nabla\cdot\mathbf{u}\right)<0$.
\subsection{Flux/Lagrangian Coordinates }
\label{sec:Lagrangian}
\cite{sokolov04} and \cite{Kota2005} showed how the Lagrangian coordinates may be applied to the kinetic equation for SEP acceleration and transport along the IMF. This approach is based on the assumption that particles don't decouple from their field lines. The particle motion consists of: (a) displacement of particle's guiding center along some IMF line; and (b) joint advection of both the guiding center and the IMF line together with plasma into which the field is frozen. 

Mathematically, the method employs Lagrangian coordinates \cite[see, e.g.,][]{Landau1959}, ${\bf x}_L$, 
which stay with advecting fluid elements rather 
than with fixed positions in space. 
As each fluid element moves, its Lagrangian coordinates, ${\bf x}_L$, 
remain unchanged,
while its spatial location, ${\bf x}\left({\bf x}_L,t\right)$,
changes in time in accordance with the local velocity of plasma,~$\mathbf{u}(\mathbf{x},t)$:
\begin{equation}
\label{eq:DxDt}
\frac{D{\bf x}({\bf x}_L,t)}{Dt}={\bf u}({\bf x},t)
\end{equation}
The coefficients in equation \ref{eq:parker} can be expressed in term of the Lagrangian derivatives and spatial derivative along the IMF line ($\partial / \partial s = \textbf{b} \cdot \nabla$). Herewith, we denote the full time derivative operator and the derivative along the magnetic field line as:
\be
\frac{D}{Dt}=\frac{\partial}{\partial t}+ \mathbf{u}\cdot\nabla,\quad 
\mathbf{b}\cdot\nabla = \frac{\partial}{\partial s}=\frac1{\delta s}\frac{\partial}{\partial s_L},\quad\delta s=\frac{\partial s}{\partial s_L},
\ee
where $s_L$ is any Lagrangian coordinate marking the fluid particles along the given IMF line. The partial time derivative at constant 
${\bf x}_L$ is denoted as $\frac{D}{Dt}$ or $\partial_\tau$, while the notation $\frac\partial{\partial t}$ or $\partial_t$ are used to denote the partial time derivative at constant Eulerian coordinates $\mathbf{x}$. 

A way to discretize this kinetic equation on the grid of multiple moving IMF lines (herewith enumerated with two indexes, $j,k$) had been described by \cite{boro18}.  With this approach, we arrive at a multitude of independent equations describing a time evolution of the isotropic distribution function, $f_{j,k}(s_L,p^3/3,\mu)$, for a particle group assigned to a given moving magnetic field line. Using the plasma MHD equations, particularly, the continuity equation,
\begin{equation}\label{eq:cont}
\partial_\tau\rho+\rho(\nabla\cdot\mathbf{u})=0,
\end{equation}
for the plasma density, $\rho(\mathbf{x},t)$, and the scalar product of the induction equation,
\begin{equation}\label{eq:induction}
\partial_\tau\mathbf{B}+\mathbf{B}(\nabla\cdot\mathbf{u})=(\mathbf{B}\cdot\nabla)\mathbf{u},
\end{equation}
by the direction vector, $\mathbf{b}$, we obtain:
\be
\label{eq:Lagr:induction}
\left(\mathbb{I}-\mathbf{bb}\right):\nabla\mathbf{u} = - \left(\ln{B}\right)_\tau,
\ee
where $\mathbb{I}$ is the identity matrix. 
The equation for time-dependent
distance, $\delta s$, between two neighboring Lagrangian meshes on the moving IMF line
may be found in \cite{Landau1984}:
\bea
\label{eq:Lagr:distance}
\partial_\tau\left(\ln{\delta s}\right) =\mathbf{bb}:\nabla\mathbf{u}
\eea
Using Eqs.~(\ref{eq:cont}~\ref{eq:Lagr:induction}), \ref{eq:Lagr:distance}, one can express 3-D flow divergence term, $(\nabla\cdot\mathbf{u})$, present in Eq.~(\ref{eq:parker}):
\begin{equation}
\label{eq:Lagr:distance:Brho}
\partial_\tau\left[\ln\left(\frac{\delta s}B\right)\right] =-\partial_\tau\left(\ln\rho\right)=(\nabla\cdot\mathbf{u}).
\end{equation}
On the other hand, using the solenoidal constraint, $\nabla \cdot \textbf{B} = 0$, one can reduce the RHS of Eq.~(\ref{eq:parker}) to 1-D diffusion along the IMF line. Finally, in terms of the Poisson bracket, Eq.~(\ref{eq:parker}) reads:
\begin{equation}
    \label{eq:Parker:bracket}
   \frac{B}{\delta s}\left\{f_{j,k};\frac{\delta s}{B}\frac{p^3}3\right\}_{\tau,p^3/3}=\frac{B}{\delta s}\frac\partial{\partial s_L}\left(\frac{D_{xx}}{B\delta s}\frac{\partial f_{j,k}}{\partial s_L}\right).
\end{equation}

To confirm that the LHS of Eq.~(\ref{eq:Parker:bracket}) belongs to the class of non-canonical equations (\ref{eq:withJacobiant}), with time-dependent Jacobian, $J\propto \delta s/B$, and may be solved with the scheme described in Section~\ref{Sec:NonCanonical}, we need to analyze, how the integration over physical volume (i.e. over $\mathrm{d}^3\mathbf{x}$) is transformed in the Lagrangian coordinates. In application to SEPs one can introduce the heliocentric spherical {\it source surface} at the heliocentric distance of $R_{\rm SS}=2.5-3.5\,R_{\odot}$ beyond which we assume that all IMF lines are {\it open}. The grid points, $\mathbf{x}_{j,k}$ on the spherical surface, or on some part of it,  may be introduced, enumerated with two indexes, $j,k$. The IMF at each grid point, $\mathbf{B}_{j,k}$, at the initial time instant together with the surface area element, $\mathrm{d}\mathbf{S}_{j,k}$, associated with each grid point, may be used to introduce the unsigned flux, $\psi_{j,k}=\vert\mathbf{B}_{j,k}\cdot\mathrm{d}\mathbf{S}_{j,k}\vert$, through each grid element.

At the initial time instant, the IMF line can be traced through each grid point at the source surface, by solving the ODE, $\mathrm{d}\mathbf{x}/\mathrm{d}s=\pm\mathrm{b}(\mathbf{x})$. With both choices of sign we obtain both the outward directed part of the IMF line and that connected to the Sun. In the course of line tracing, the sequence of points belonging to the IMF line is obtained, which may be sequentially enumerated with index, $i$, for each line, enumerated with the indexes, $j,k$. A set of three integers $s_L\equiv i,j,k$ is the most convenient vector of Lagrangian mark for the Lagrangian grid point. The time evolution of thus initially constructed grid points is governed by equations \ref{eq:DxDt}, to be solved for realistic 3-D solar wind motion. The easy-to-find distance between neighboring Lagrangian grid points, $s_{i+1}-s_i\equiv\left\|\mathbf{x}_{i+1}-\mathbf{x}_i\right\|$ is nothing but an approximation for $\delta s_{i+1/2}=\frac{\mathrm{d}s}{\mathrm{d}i}=\frac{s_{i+1}-s_i}{(i+1)-i}$. In this way all Hamiltonian functions and control volumes in equations similar to~(\ref{eq:Parker:bracket}) can be efficiently computed with an account of instantaneous values of the IMF intensity and plasma density in the grid point locations or their Lagrangian time derivatives. Note a useful consequence from Eq.~(\ref{eq:Lagr:distance:Brho}):
\be\label{eq:Lagrangian:invariant}
\partial_\tau\left(\frac{\rho\delta s}{B}\right)=0,\qquad \frac{\rho\delta s}{B}=\left(\frac{\rho\delta s}{B}\right)_{i,j,k}, 
\ee
that is the combination of quantities in Eq.~(\ref{eq:Lagrangian:invariant}) is a \textit{Lagrangian invariant}, which can be once calculated for any Lagrangian grid point and then reused, if desired. On multiplying by $\psi_{j,k}\mathrm{d}s_L$ this combination becomes equal to the conserved element of mass, $\rho\frac{\partial s}{\partial s_L} \mathrm{d}s_LS$, enclosed between two close cross-sections of the flux tube, located at the Lagrangian coordinates, $s_L$ and $s_L+\mathrm{d}s_L$, and moving with the plasma.   

To derive the Jacobian, the IMF line may be thought of as the central line of some flux tube of small cross section, ${\mathrm d}S$. Since the magnetic flux, $\psi_{j,k}=B{\mathrm d}S$, is constant along the flux tube, the phase volume element, ${\mathrm d}\Gamma_{j,k}$, for a given IMF line may be expressed as follows:
\be
{\mathrm d}\Gamma_{j,k}=4\pi{\mathrm d}S\,{\mathrm d}s\,{\mathrm d}\frac{p^3}3=\left(4\pi\psi_{j,k}\right)\frac{\delta s}B{\mathrm d}s_L\,{\mathrm d}\frac{p^3}3,
\ee
the factor, $(4\pi\psi_{j,k})$, being constant along the flux tube. The particle number integral becomes:
\be\label{eq:Parker:N}
N_{j,k}=\int{d\Gamma_{j,k}\,f_{j,k}}=\left(4\pi\psi_{j,k}\right)\int{{\mathrm d}s_L{\mathrm d}\frac{p^3}3\left[\frac{\delta s}Bf_{j,k}\left(t,s_L,\frac{p^3}3\right)\right]},
\ee
where the integration over $\mathrm{d}s_L$ in effect reduces to summation over $i\equiv s_L$.
For a grid with multiple magnetic field lines (tubes) the particle number should be summed over all tubes. It is easy to see that Eq.~(\ref{eq:Parker:bracket}) conserves the particle number integral, given by Eq.~(\ref{eq:Parker:N}) and that the expression for Jacobian, $J=4\pi\psi_{j,k}\frac{\delta s}B\propto\frac{\delta s}B$, justifies that Eq.~(\ref{eq:Parker:bracket}) with zero RHS is a particular case of Eq.~(\ref{eq:withJacobiant}).

\subsection{Numerical Result}
\label{Sec:ResultParker}
Eq.~(\ref{eq:Parker:bracket}) combining Poisson bracket and 1-D diffusion operator can be solved using the Strang splitting method. This means that to advance the numerical solution of Eq.~(\ref{eq:Parker:bracket}) through the time step, from $t$ to $t+\Delta t$, we alternate the stage at which we solve equation,
\begin{equation}\label{eq:ParkerNoDiffusion}
   \frac{B}{\delta s}\left\{f_{j,k};\frac{\delta s}{B}\frac{p^3}3\right\}_{\tau,p^3/3}=0,
\end{equation}
with the Poisson bracket and no diffusion, using the scheme in (\ref{eq:TVDSchemeWithJacobian}) as discussed above, followed by the stage at which we solve the spatial diffusion equation:
\begin{equation}\label{eq:ParkerDiffusionOnly}
\frac{\delta s}B
(f_{j,k})_\tau=
\frac{\partial}{\partial s_L}
\left(\frac{D_{xx}}{B\delta s}\frac{\partial f_{j,k}}{\partial s_L}\right),
\end{equation}
using a fully implicit scheme. In the latter scheme the diffusion operator in the RHS is discretized via cell-centered values of the distribution functions at the \textit{upper} time level, $t+\Delta t$. 
As long as the CFL condition for solving Eq.~(\ref{eq:ParkerNoDiffusion}) is satisfied, the overall numerical scheme for Eq.~(\ref{eq:Parker:bracket}) appears to be stable, since the implicit scheme for Eq.~(\ref{eq:ParkerDiffusionOnly}) is unconditionally stable, no matter how high the diffusion coefficient could be. On the other hand, the implicit scheme for 1-D diffusion operator in the RHS of Eq.~(\ref{eq:ParkerDiffusionOnly}) reduces to the system of linear equations with a tri-diagonal matrix, because the numerical solution in a given cell depends only on the solution in two neighboring cells. Such system can be explicitly solved with a single iteration of the Gauss-Seidel method.

As a numerical test we consider the particle acceleration in 1-D shock wave, propagating with the unit speed along a uniform magnetic field (see Fig.~\ref{fig:Parker}). The Lagrangian points at the initial time instant, $t=0$, when the shock wave front is at $s=0$, form equidistant grid, the initial grid point coordinates being $s_i=i-0.5$. The $i$th grid point is at rest till the shock wave reaches it at the time instant, $t=t_i=i-0.5$. Then, during the unit time, the grid point accelerates to the speed, $0.75$, so that $s_i=i-0.5+0.375(t-t_i)^2$, for $0\le t-t_i\le 1$. After this, the point keeps moving with the constant speed, so that $s_i=i-0.5+0.375+0.75(t-t_i-1)$, for $t-t_i\ge 1$. 
In Fig.~\ref{fig:Parker}, left panel presents coordinates, $s_i$, (horizontal axis) and the inverse of distance, $1/V_i=2/(s_{i+1}-s_{i-1})$, (vertical axis) for the time instant, $t=60000$. The particles starting from the index, $i=6001$, are not yet reached by the shock wave and are still equidistant with the unit mesh, $(\delta s)_{i+1/2}=s_{i+1}-s_i=1$. The particle with the index $i=6000$ and the coordinate, $s_{6000}=5999\frac{19}{32}$, still close to its initial value, $s^{(0)}_{6000}=5999.5$, has been just passed by the shock wave and starts to accelerate. All particles with the lower value if index are behind the shock wave front and move at the speed equal to $0.75$, the distance between the neighboring particles being $\delta s=0.25$. Accordingly, the inverse distance proportional to the bulk plasma density increases by a factor of 4 behind the shock, corresponding to the compression ratio $4=(\gamma+1)/(\gamma-1)$, across the strong shock in plasma with the adiabatic index, $\gamma=5/3$.

\begin{figure}[htb]
\centering
\includegraphics[height={2.1in},width={3in}]{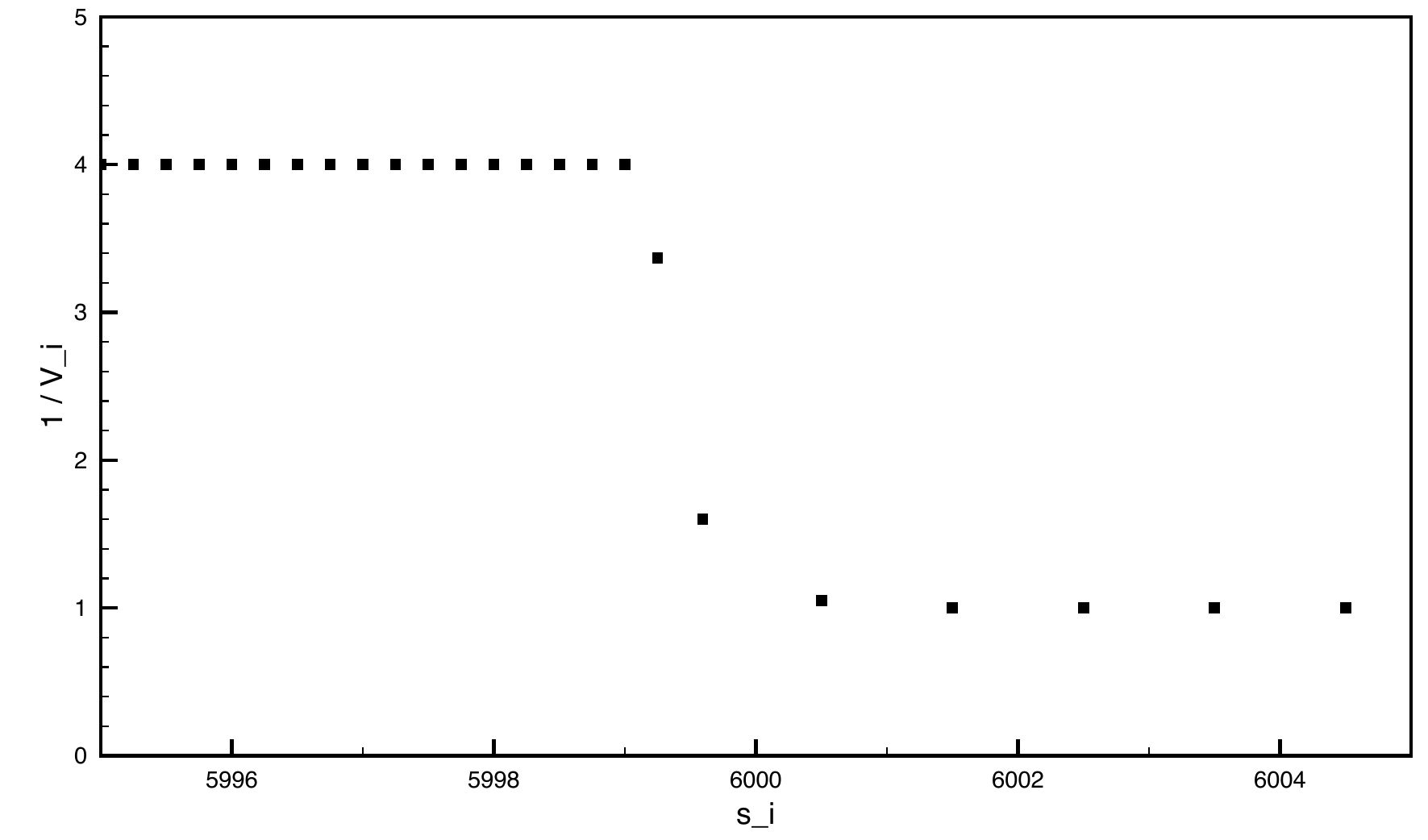}
\includegraphics[width={3in}]{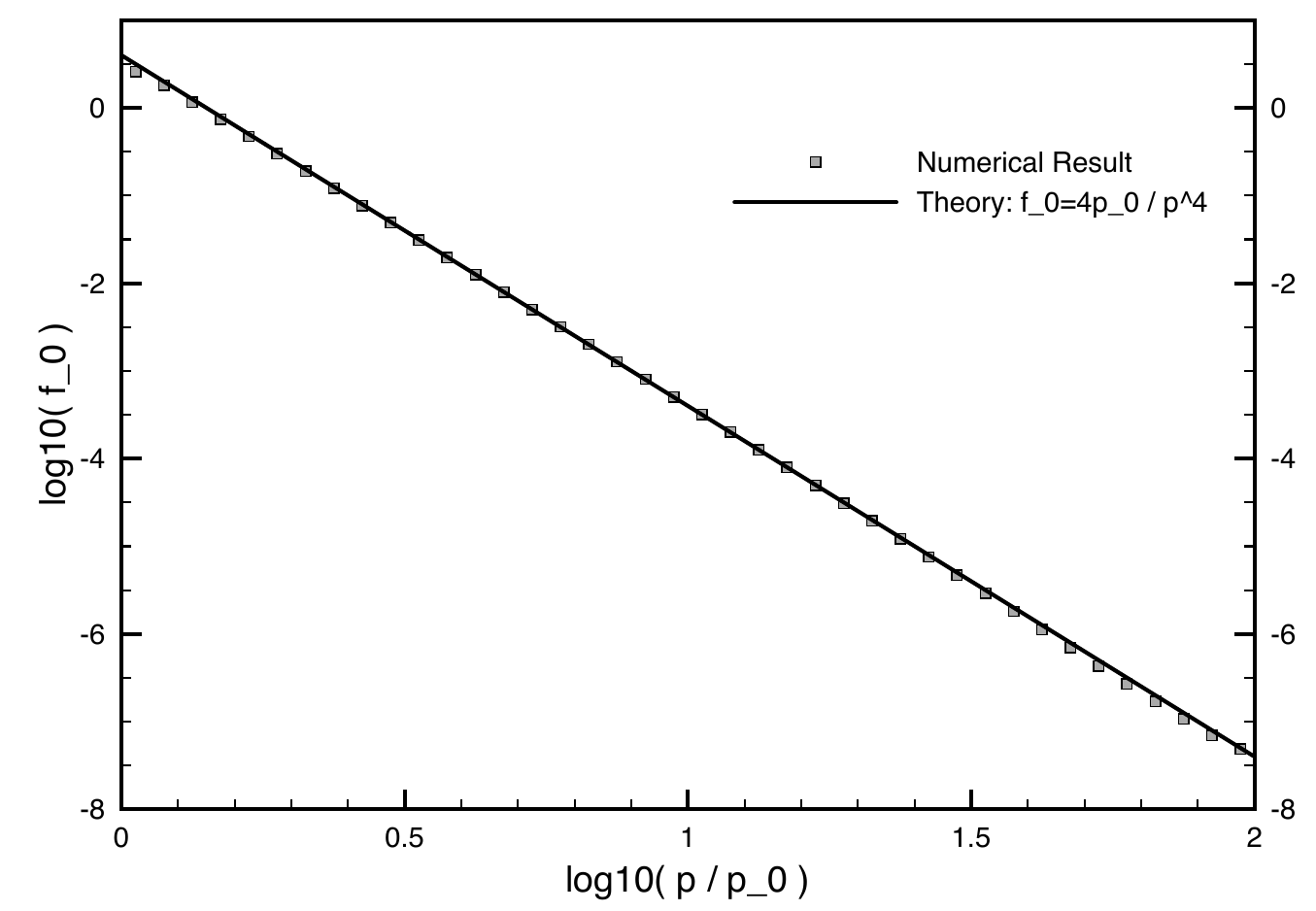}
\caption{
{\bf Left panel:} Coordinates, $s_i$, (horizontal axis) and the inverse of distance, $1/\delta s_i=2/(s_{i+1}-s_{i-1})$, (vertical axis) of the Lagrangian grid points near the model shock front. {\bf Right panel:}  Comparison between a numerical result for the particle distribution function behind the shock wave front (symbols) with the theoretical prediction (line).}
\label{fig:Parker}
\end{figure}
The numerical result for the particle distribution function behind the shock wave front (at i=5000) is shown (with symbols) in the right panel of Fig.~\ref{fig:Parker}. 10000 Lagrangian points are used with log-uniform grid with 40 intervals for the momentum ranging by two orders of magnitude: $p_0\le p\le100p_0$. The diffusion coefficient is $D_{xx}=100$. We assume zero diffusion flux through the spatial boundaries at the first and last Lagrangian points as well as zero distribution function at the momentum below $p_0$, therefore, in neglecting minor particle flow across the maximum momentum boundary, the boundaries of the computational phase space do not disturb the total particle number conservation. The seed particles were initially put into the first momentum bin, their number per the Lagrangian point being normalized per one. Under these assumptions the theoretical prediction for the distribution function behind the front of the 1-D shock wave with the compression ratio of 4 is \cite[see][]{axford77,krymsky77,Bell1978a,Bell1978b,blandford78,axford81}:
\be
f_0=\frac{4p_0}{p^4}.
\ee
This theoretically predicted spectrum is shown with the line in the right panel in Fig.~\ref{fig:Parker}. A perfect agreement with the numerical result is achieved thanks  to the use of numerical scheme based on exact \textit{integral} relations. That is why the method works even in the case when the shock wave profile is in fact discontinuous and no \textit{differential} relation can be properly approximated.
\begin{figure}[htb]
\centering
\includegraphics[height={2.4in}]{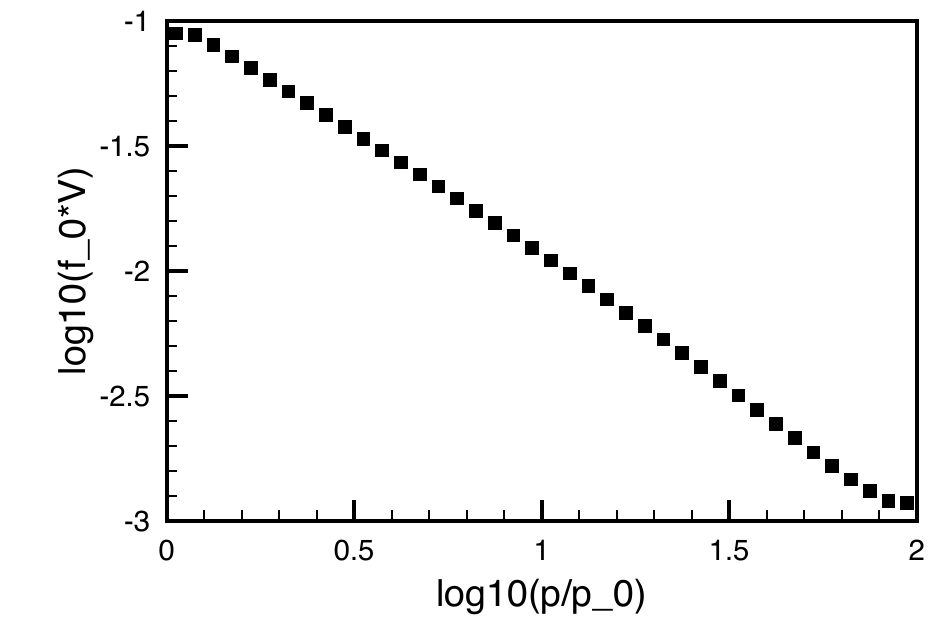}
\caption{For the test presented in Fig.~(\ref{fig:Parker}): a fraction of particles, $f_0(p)V$, falling into a given control volume, as a function of cell-centered value of momentum, $p$. Only about $10^{-3}$ of particles reach the maximal energy.}
\label{fig:StatWeight}
\end{figure}

The test results provide some opportunity to demonstrate the advantage of the proposed method versus particle simulations. Using data shown in the right panel of Fig.~(\ref{fig:Parker}) one can calculate the values of the product, $f(p)V$, for a chosen set of control volumes. Since the initial distribution function is normalized per one particle per cell of the lowest energy, these values characterize the fraction of particles accelerated to the energy range represented by the given control volume. From Fig. (\ref{fig:StatWeight}) one can see that the fraction of particles reaching the maximal energy is as low as $10^{-3}$. To reproduce this result within the usual framework of particle methods more than a thousand of macroparticles per cell should be initially distributed to get at least one particle at maximum energy, which is still insufficient for statistical reliability of the result. Alternatively, in Monte Carlo particle methods the problem of low statistical weight can be overcome by assigning individual statistical weights to each model particle \cite[see, e.g.][who applied Monte-Carlo method to the shock wave acceleration problem similar to what is discussed here]{tenishev05}. Specifically, the individual weights of high-energy macro-particles are artificially reduced in increasing accordingly their number, thus achieving statistical representation. At the same time the number of the low-energy particles is artificially reduced in increasing their individual weights. This approach allows a construction of numerical algorithm in which the entire particle spectrum is statistically well represented with no significant increase in the particle number to be used in simulation.     Particle splitting and merging techniques are dynamically used to ensure a good statistical quality of the simulation results of the stochastic acceleration process over the entire energy range of interest. The individual particle weights are accounted for while calculating moments of the distribution functions, which are the measurable quantities. These techniques, however, are both laborious and computationally expensive. 

The scheme under consideration is free of these complications and better suited to simulate particle acceleration to high energy in space environment. It allows simulating small fluxes of hazardous high-energy particles with no need to care about their small statistical weight.

\section{More Particle Transport Equation with Poisson Brackets}
\label{Sec:More}
Here, we provide more examples of physically relevant transport equations, which can be formulated via the Poisson brackets and solved with the proposed method. 
\subsection{Focused Transport Equation}
\label{Sec:FTE}
Beyond the framework of Parker equation the SEPs are described by the pitch-angle dependent gyrotropic distribution function.  A {\it focused transport equation} describing an evolution of such VDF in a turbulent interplanetary magnetic field had been published by \citet{Skilling1971}. A detailed view on different aspects of  particle propagation along magnetic-field lines, cooling/heating, and focusing
can be found in \citet{Kota:1997,Kota:2004}. A novelty of our current approach to this equation is that we formulate it in terms of the Poisson brackets:
\bea
\label{eq:skilling1}
&&
\partial_t \tilde{f} 
+ \left\{
\tilde{f};\frac{p^3\mathbf{u}}
{3}\right\}_{\mathbf{x},p^3/3}
+\left\{
\tilde{f};\frac{\left(\mu^2-1\right)p\mathbf{b}}
{2m_i}\right\}_{\mathbf{x},\mu}
+ \nonumber\\
&+& \left\{\tilde{f};\frac{1-\mu^2}{2}\left[\frac{\mu p^3}3
\left(3(\mathbf{bb}:\nabla\mathbf{u})
- (\nabla\cdot\mathbf{u})\right)+
p^2m_i\left(\mathbf{b}\cdot\frac{D\mathbf{u}}{Dt}\right)
\right]\right\}_{p^3/3,\mu}=I^{({\rm s})}.
\eea
The scattering integral, $I^{({\rm s})}$ in the RHS describes the particle interaction with the MHD turbulence. 
Similarly to the transition from Eq.~\ref{eq:parker} to Eq.~\ref{eq:Parker:bracket} in Section~\ref{Sec:Parker}, one can proceed to Lagrangian coordinates in equation \ref{eq:skilling1}. In this way we arrive at the equation describing a time evolution of the gyrotropic distribution function, $f_{j,k}(t,s_L,p^3/3,\mu)$, for a particle group relating to a given moving magnetic field line, which, again, can be formulated in terms of the Poisson brackets:
\bea
\label{eq:kota:bracket}
&&\frac{B}{\delta s}\left\{f_{j,k};\frac{p^3\delta s\, }{3B}\right\}_{\tau,p^3/3}+\frac{B}{\delta s}\left\{f_{j,k};\frac{\left(\mu^2-1\right)p}{2m_iB}\right\}_{s_L,\mu}+\\
&&+\frac{B}{\delta s}\left\{f_{j,k};\frac{\left(1-\mu^2\right)p^2}{2}\left[
 \mu p\left(\frac{2}3\partial_\tau\left(\frac{\delta s}B\right)-\delta s\partial_\tau\left(\frac1B\right)\right)+\frac{m_i\delta s}B\partial_\tau\left(\mathbf{b}\cdot\mathbf{u}\right)
\right]\right\}_{p^3/3,\mu}=
I^{({\rm s})}.\nonumber
\eea
The particle number integral can be formulated similarly to that in Eq.~\ref{eq:Parker:N}:
\be\label{eq:kota:N}
N_{j,k}=
\int{\mathrm{d}\Gamma_{j,k} f_{j,k}\left(t,s_L, \frac{p^3}3,\mu\right)}=
\left(2\pi\psi_{j,k}\right)\int{
{\mathrm d}s_L\,{\mathrm d}\frac{p^3}3\,{\mathrm d}\mu\left[\frac{\delta s}Bf_{j,k}\left(t,s_L,\frac{p^3}3,\mu\right)\right]}.
\ee
Eq.~(\ref{eq:kota:bracket}) conserves the particle number integral, given by Eq.~(\ref{eq:kota:N}). 
\subsection{Particle Transport in Steady-State Stream-Aligned Interplanetary Magnetic Field}
\label{Sec:StreamAligned}
The models discussed so far are based on numerical solutions of the transport equation using moving Lagrangian grids. They are well suited to solve the SEP acceleration in non-steady-state plasmas, such as \textit{coronal mass ejections}, which might result in SEP acceleration at interplanetary shock waves. However, for many space science applications, particle transport through a ``quiet'' heliosphere, or even through a steady-state heliosphere might be of interest. Indeed, the quiet solar wind is close to steady-state in the coordinate frame co-rotating with the Sun. In this frame solar wind sources that rotate with the Sun (i.e. active regions and coronal holes), do not move and evolve slowly, thus resulting in a steady-state solar atmosphere. The transport of Galactic Cosmic Rays (GCRs) through such steady-state background can be described to relate the evolution from their interstellar distribution to the observed flux deep inside the heliosphere. 

Recent Fermi-LAT gamma-ray observations of the solar disk \citep{Ng2016,Linden2018,Tang2018} have shown some unexplained features, such as temporal and spacial variability on the solar disk over the solar cycle, the high energy ($>200~\text{GeV}$) spectrum, and a gamma-ray flux higher than expected \citep[e.g.,][]{Seckel1991}. These solar gamma-ray photons are likely the result of the interactions of hadronic GCRs with the solar atmosphere. Additionally, the inverse-Compton halo about the Sun is induced by interactions between GCR electrons and solar photons, tracing the distribution of GCR electrons in the inner heliosphere \citep{Abdo2011}. Studying these processes require describing the propagation of GCRs through the heliosphere to the photosphere. Steady-state solar wind solutions can be applied to these problems, as these observations occur on time-scales of months or even years.

Another possible field of application is the particle acceleration in Corotating Interaction Regions \cite[CIRs, see, e.g.,][]{Schwadron2021a}, which result in the time variation of the energetic particles observed in-situ, even though they are steady-state in the co-rotating frame. Finally, the \textit{seed population} of energetic particles, that can be further accelerated by some dynamical processes, are generated and transported in the quiet (steady-state) solar wind. This variety of steady-state particle transport problems are challenging to simulate on moving Lagrangian grids.    

In the \cite{parker58} model, the steady-state magnetized solar wind is thought of as plasma motion along the magnetic field, or, alternatively, as tstream-aligned MHD motion, where the magnetic field is always parallel or anti-parallel to the velocity:
\be\label{eq:align}
\mathbf{B}=\alpha\mathbf{u},\qquad \mathbf{B}\times\mathbf{u}\equiv0.
\ee
\cite{Sokolov2021a} demonstrated how a realistic 3-D steady-state solar atmosphere can be numerically solved, constrained with observed photospheric magnetograms. Here, we show how the focused transport equation can be formulated in terms of Poisson brackets for such background. 

As described in sections~\ref{sec:Lagrangian} and \ref{Sec:FTE}, we introduce a grid of IMF lines enumerated with two indexes, $j,k$ as well as the distribution functions, $f_{j,k}$ for the groups of particles within the corresponding flux tube. In this case the coordinate along the line is just $s$, not $s_L$, so that in Eq.~(\ref{eq:kota:bracket}) 
the no longer used derivative over $s_L$ should be expressed as follows:
$\partial_{s_L}=\delta s\partial_s$. The time derivative at constant $\mathbf{x}_L$ for the distribution function should be expressed as $D_t f=f_t+uf_s$, while for the background time-independent quantities this expression simplifies: $D_t(\dots) = u \mathrm{d}_s(\dots)$. In the particle number integral (\ref{eq:kota:N}) the differential, $\mathrm{d}s_L$, needs to be eliminated:
\be\label{eq:kotaSS:N}
N_{j,k}=
\left(2\pi\psi_{j,k}\right)\int{
{\mathrm d}s\,{\mathrm d}\frac{p^3}3\,{\mathrm d}\mu\left[\frac1Bf_{j,k}\left(t,s_L,\frac{p^3}3,\mu\right)\right]}.
\ee
One can use Eq.~(\ref{eq:Lagrangian:invariant}) to eliminate $\frac{B}{\delta s}$, and $\left(\frac{\delta s}B\right)_\tau$ in  Eq.~(\ref{eq:kota:bracket}):
\bea
\label{eq:kota:bracketSS}
&&\partial_t f_{j,k} + B\left\{f_{j,k}; \left(\frac{\rho u}B\right)_{j,k}\frac{p^3 } {3\rho}\right\}_{s,p^3/3} + B\left\{f_{j,k}; \frac{\left(\mu^2-1\right)p} {2m_iB}\right\}_{s,\mu} + \\ 
&& + B\left\{f_{j,k}; \frac{\left(1-\mu^2\right)p^2}{2}
\left[\mu p\left(\frac23 \left(\frac{\rho u}B\right)_{j,k} \partial_s\left(\frac1\rho\right) -u \partial_s\left( \frac1B\right)\right) +\frac{m_iu}{B}\partial_su\right]
\right\}_{p^3/3,\mu}= I^{({\rm s})}.\nonumber
\eea
Herewith, the combination of quantities, $\left(\frac{\rho u}B \right)_{j,k}$, is constant throughout the steady-state IMF line, therefore, it is denoted with indices, $j,k$. This parameter represents a constant value of the mass flux per unit magnetic flux tube area, since both the mass flux, $\rho u S$, and the magnetic flux, $\psi_{j,k}=BS$, are constant along the IMF line with their ratio being independent of $S$ \cite[see][for more detail]{Sokolov2021a}. The version of the Parker equation (\ref{eq:parker}) for stream-aligned MHD background is:
\be\label{eq:ParkerSAMHD}
    \partial_tf_{j,k}+B\left\{f_{j,k};\left(\frac{\rho u}B\right)_{j,k}\frac{p^3 }{3\rho}\right\}_{s,p^3/3}=B\frac\partial{\partial s}\left(\frac{D_{xx}}B\frac{\partial f_{j,k}}{\partial s}\right).
\ee
\subsection{Numerical result}
\label{Sec:SAMHDResult}
To test particle transport in a steady-state stream-aligned MHD flow we performed a two-stage simulation run. For the first stage we repeated the numerical solution of Eq.~(\ref{eq:Parker:bracket}) describing particle acceleration by a travelling shock wave (see section~\ref{Sec:ResultParker}) for $t=6000$. The only difference from the earlier solution was the increase of the diffusion coefficient (from $D_{xx}=100$ to $D_{xx}=200$), reducing the acceleration rate $\propto D_{xx}^{-1}$ by a factor of $0.5$. Starting with the distribution function calculated at the end of first stage we simulated the evolution of distribution function described by Eq.~(\ref{eq:ParkerSAMHD}) for another $t=6000$. The particle acceleration was due to a steady-state shock wave in the \textit{boosted frame} of reference, co-moving with the shock wave, i.e. moving to the right with unit speed. In this frame of reference the uniform stream in front of the shock wave has the transformed velocity, $u=-1$, so that $\left(\frac{\rho u}B\right)=-1$. According to Eq.~(\ref{eq:Lagrangian:invariant}), the values of inverse density on spatial faces of control volumes in this test can be calculated as $\rho^{-1}_{i+1/2}=s_{i+1}-s_{i}$. The dependence of face density on the face coordinate, $s_{i+1/2}=\left(s_i+s_{i+1}\right)/2$ shown in the left panel of Fig.~\ref{fig:ParkerSA} is sufficient to compute the Hamiltonian values.
\begin{figure}[t]
\centering
\includegraphics[height={2.1in},width={3in}]{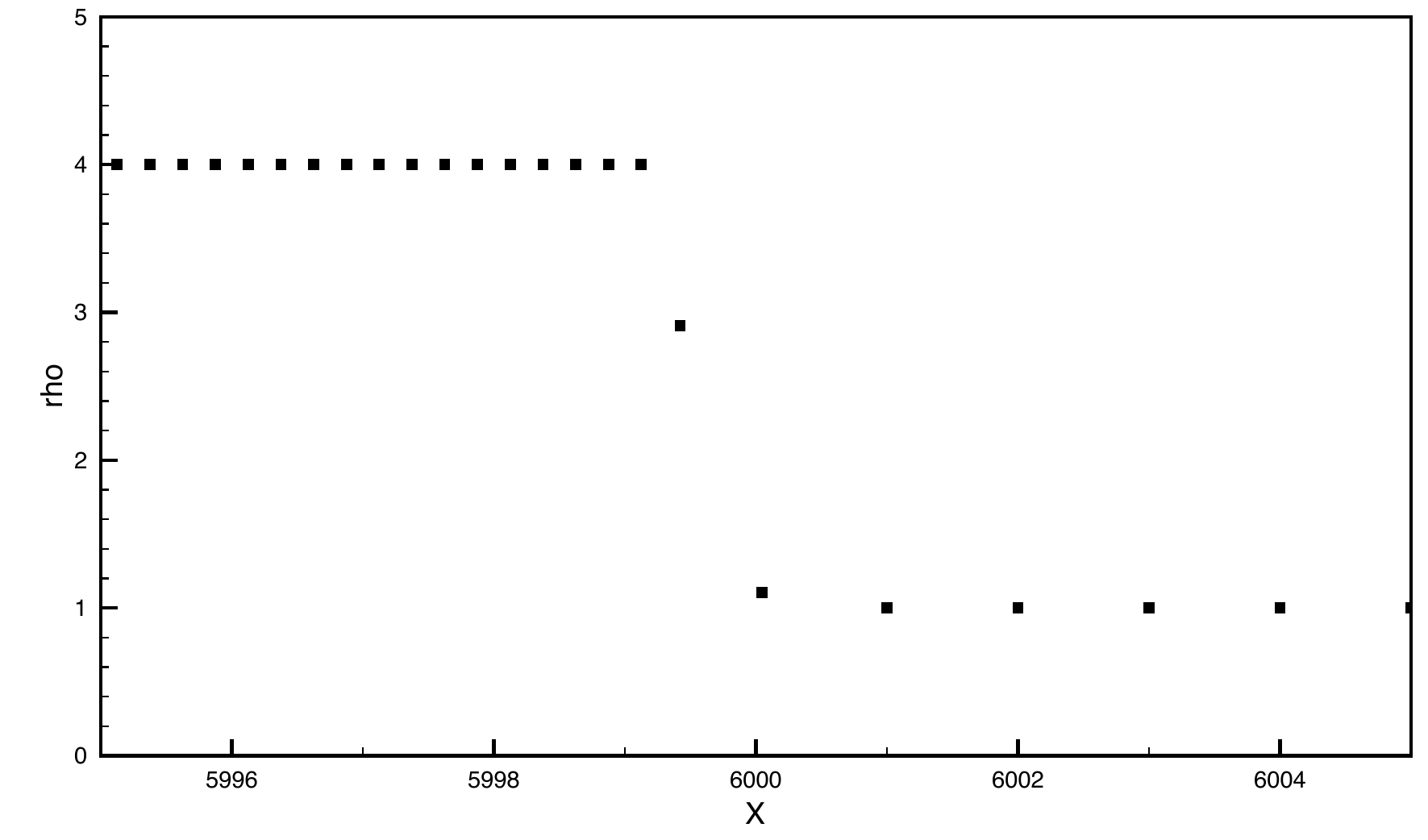}
\includegraphics[height={2.1
in},width={3in}]{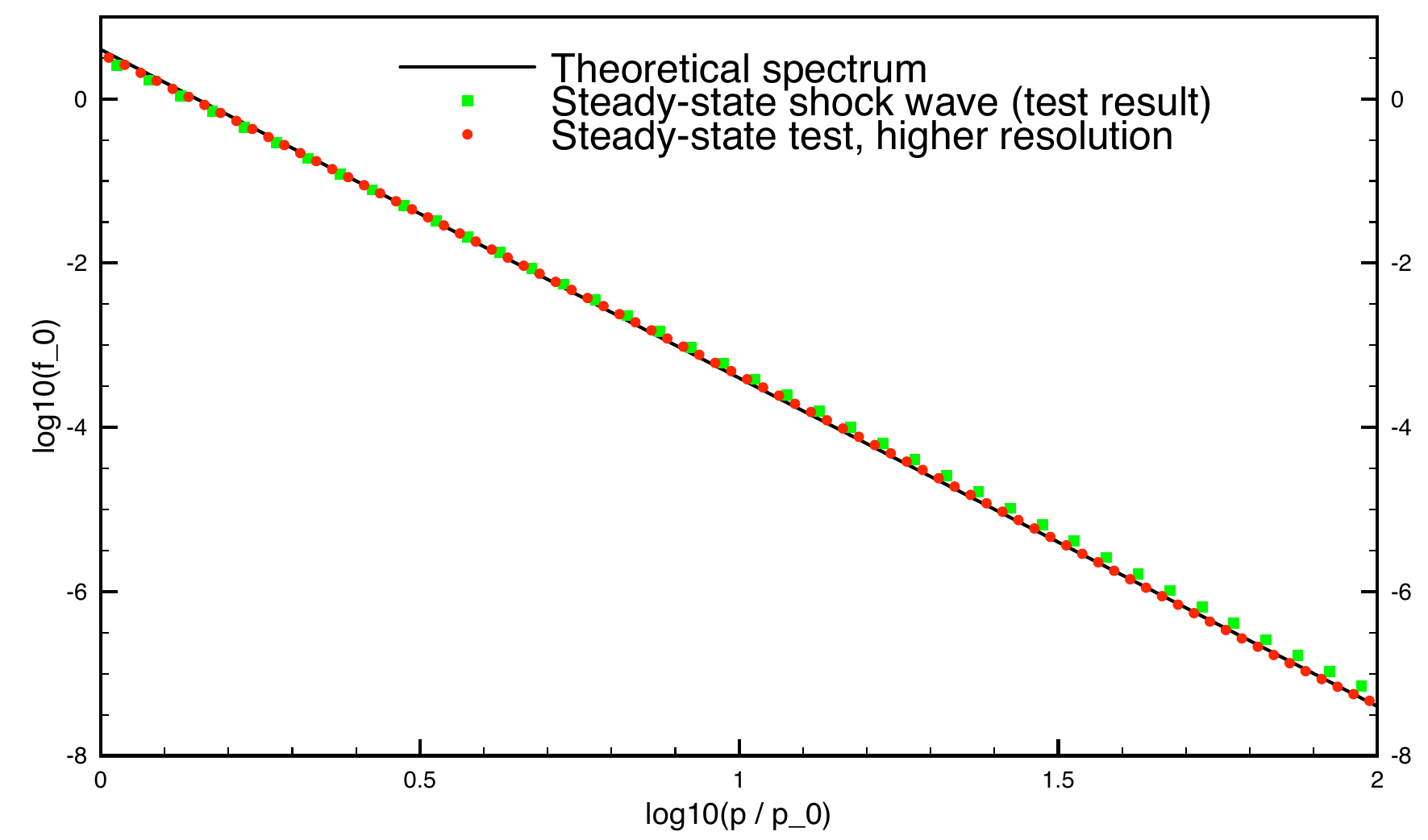}
\caption{
{\bf Left panel:} Face coordinates, $\left(s_i+s_{i+1}\right)/2$, (horizontal axis) and the density at faces, $\rho_{i+1/2}=1/\left(s_{i+1}-s_{i}\right)$, (vertical axis) near the steady state shock front. {\bf Right panel:}  Comparison between a numerical result for the particle distribution function behind the shock wave front (green symbols and red symbols for lower and higher resolution respectively) with the theoretical prediction (line).}
\label{fig:ParkerSA}
\end{figure} 

The simulation results obtained using the same grid as in section~\ref{Sec:ResultParker} are shown in the right panel in Fig.~\ref{fig:ParkerSA} with green squares. At high energies the simulated spectrum is harder than the theoretical one shown with a black curve, thus overestimating the acceleration efficiency due to finite mesh effects. However, with twice higher resolution (80 meshes over the momentum range) the simulated spectrum shown with red circles perfectly agrees with theoretical predictions.
\section{Conclusions and Discussion}
\label{Sec:Conclusion}
\subsection{Summary}

We have shown that several systems of equations of practical importance can be cast in a form using Poisson brackets. These equations can be discretized in a manner that satisfies conservation of phase space density
to round-off errors. Second or higher order accurate schemes with ad-hoc (including direction-by-direction TVD) limiters can be easily employed to solve these equations.

As an alternative, we have introduced a total variation (TV) functional appropriate for the multi-dimensional Liouville equation. We proved that the TV does not increase in time and eliminates the possibility of spurious numerical oscillations. Based on this concept, we have constructed a new high resolution finite volume scheme that satisfies the Total Variation Diminishing (TVD) property. 


Among the practical applications is the  kinetic transport equation, which is directly applicable to solar energetic particle (SEP) simulations even in the simplest version of the \cite{Parker1965} equation. A more refined SEP model based on the focused transport equation with multiple Poisson brackets for the gyrotropic distribution function is briefly discussed here with more detailed explanation and test results being delegated to 
the  Appendix.

\subsection{Prospective for Future Work}

We plan to employ the proposed scheme for simulating SEP acceleration, propagation, and transport within the SWMF framework designed for modeling space weather \cite[see][]{Gombosi:2021rev}. Apart from solving the SEP transport using   Eqs.~(\ref{eq:kota:bracket} and
\ref{eq:kota:bracketSS}) on separate magnetic field lines, we are going to solve the Poisson bracket version of Parker's equation Eq.~(\ref{eq:parker}):
\be
\partial_t f_0 + \left\{
f_0;\frac{p^3\mathbf{u}}
{3}\right\}_{\mathbf{x},p^3/3} 
 = 
\nabla\cdot\left(\varkappa\cdot\nabla f_0\right),
\ee
on top of realistic 3-D model for solar wind velocity, $\mathbf{u}$, and the magnetic field, $\mathbf{B}$.
\section{Acknowledgement}
This research is supported by the National Aeronautics and Space
Administration under Grant No. 80NSSC20K1354 issued through the Heliophysics Supporting Research Program. It was also supported by NSF INSPIRE grant PHY-1513379, NSF ANSWERS grant GEO-2149771, NASA R2O2R grant 80NSSC22K0269, NASA LWS grants 80NSSC20K1778 and 80NSSC22K0892, and by NASA Heliophysics DSC grant 80NSSC20K0600. We would also like to acknowledge high-performance computing support from: (1) Yellowstone
provided by NCAR's Computational and Information Systems Laboratory, sponsored by the  NSF, and (2) Pleiades operated by NASA's Advanced Supercomputing Division.
\begin{appendix}
\section{Test Results for Focused Transport Equation}
\begin{figure}[htb]
\label{shm6}
\includegraphics[width={3.2in}]{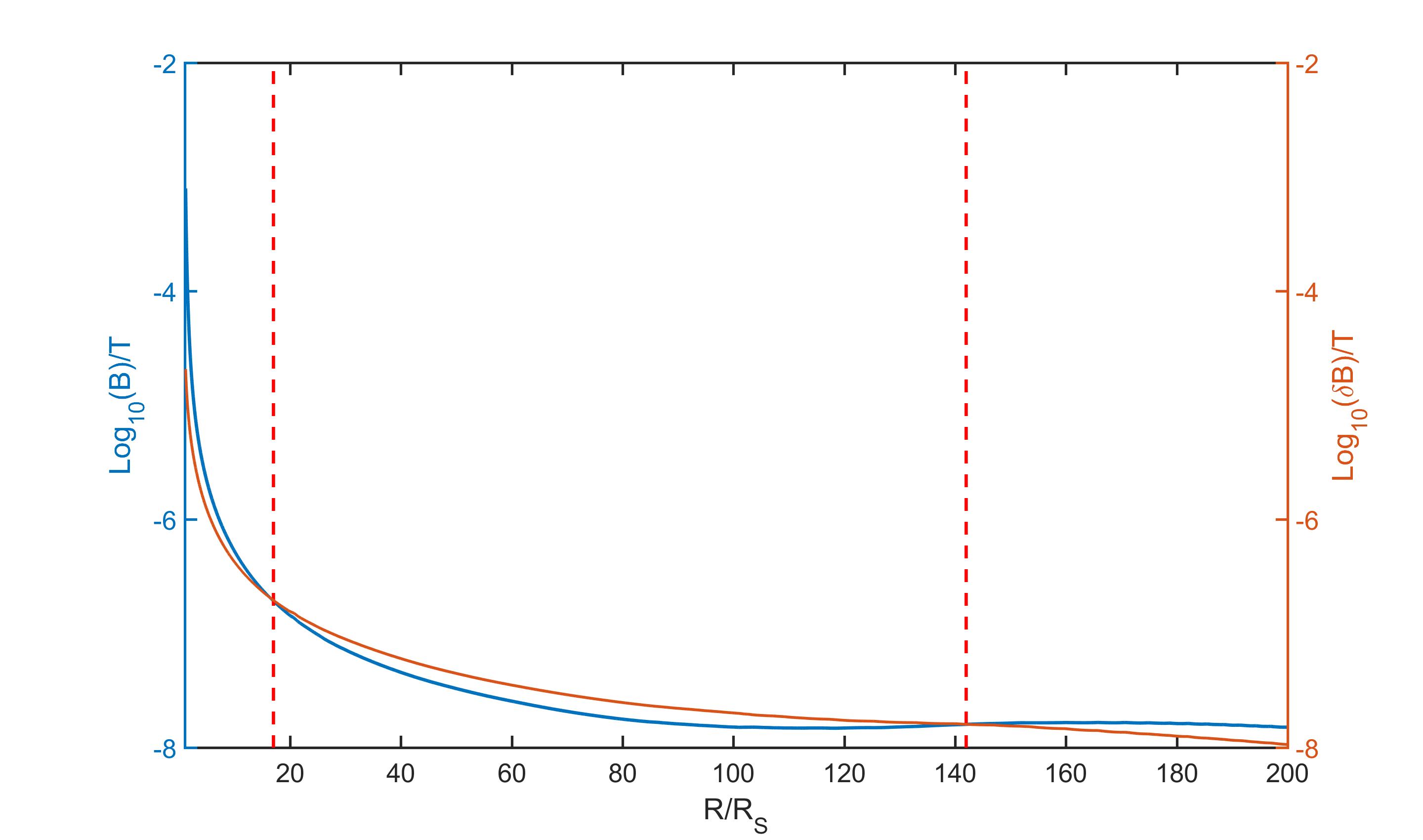}
\centering
\includegraphics[width={3.2in}]{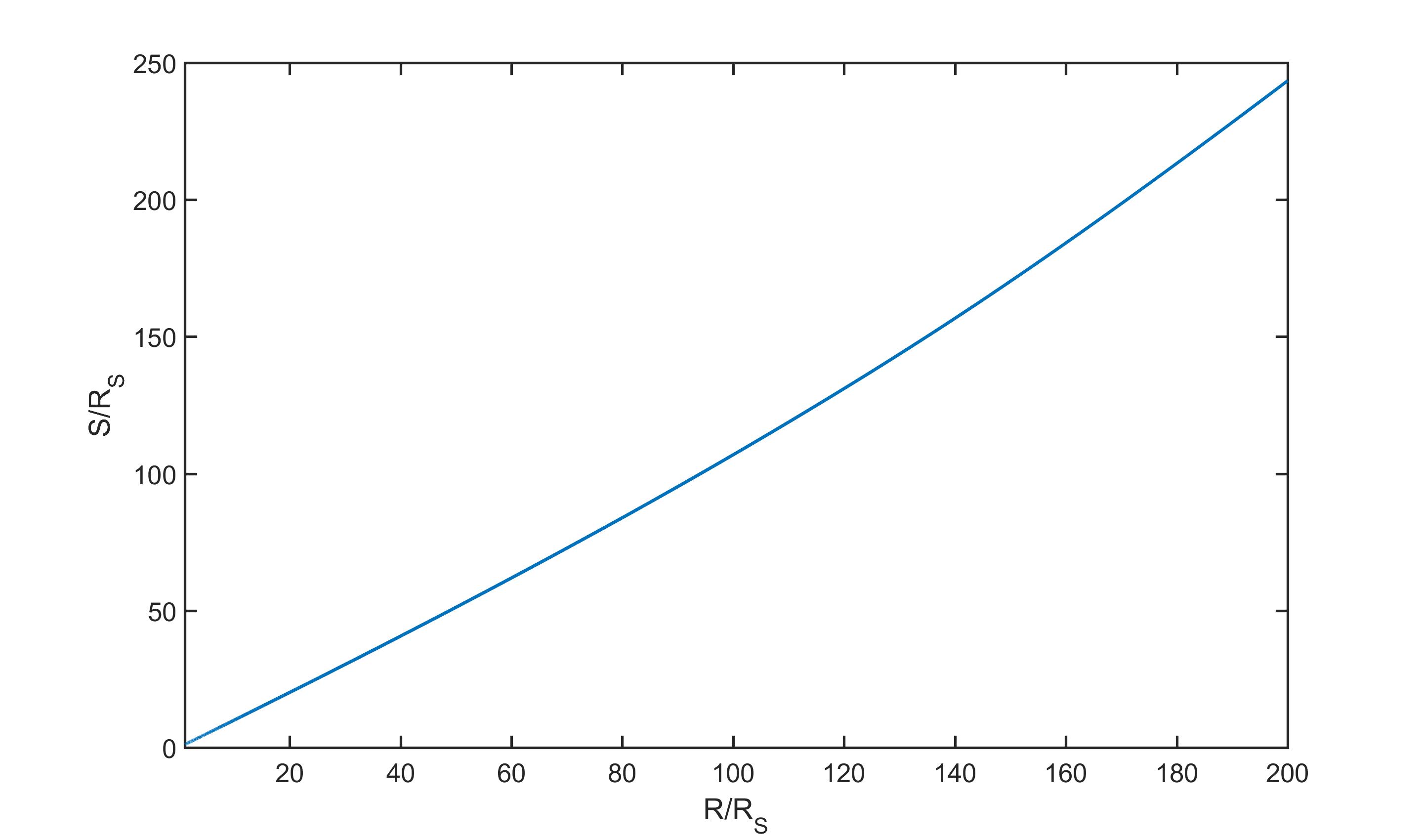}
\centering
\caption{\textbf{Left panel:} Regular IMF field $B[{\rm T}]$ (blue curve) controlling the particle focusing and \alf wave turbulence field,$\delta B[{\rm T}]$ (orange curve), controlling the particle scattering are shown as functions of heliocentric distance $R/R_\sSun$. The value of $\delta B/B$ can be larger than "1" in the region $R=17-142R_\sSun$, as is denoted by the red dashed lines in the figure. \textbf{Right panel:} $S/R_\sSun$ as a function of Heliocentric distance $R/R_\sSun$.}
\end{figure}

The test simulations of the SEP VDF are performed within the SWMF framework by \cite{Toth:2005swmf,Toth2012} on top of the realistic state of the coronal plasma, solar wind, and IMF simulated using the Alfven Wave turbulence driven Solar atmosphere Model (AWSoM)  \citep{Sokolov2013,oran13,Holst2014,Gombosi2018}. 
The Multiple Field Line Advection Model for Particle Acceleration \cite[MFLAMPA, see][]{boro18} is employed allowing tracing multiple field lines dynamically and extracting time-dependent profiles of the MHD data along the Lagrangian grid on these lines. With these dynamical sets of data, the pseudo-Hamiltonian functions in kinetic equation are calculated and evolution of the VDF is obtained by solving numerically Eq.~\ref{eq:kota:bracket}. 

Here, we provide results only for the simplest case, of the  steady-state IMF field and the curve length, $s/R_S$, being shown in Fig. A.5 as functions of a heliocentric distance $R/R_\sSun$.
\subsection{Numerical Result for Single Poisson Bracket}
The first Poisson bracket in Eq.~\ref{eq:kota:bracket} is identical to that in Eq.~\ref{eq:Parker:bracket} and has been studied above. The second bracket describes the particle propagation accompanied with the adiabatic focusing. For test simulation, we assume a steady-state background, neglect the scattering term in the RSH of Eq.~\ref{eq:kota:bracket}, and solve the following equation: 
\be\label{eq_extra}
\frac{\partial f}{\partial t}+\frac{B}{\delta s}\left\{f,\frac{(-1+\mu^2)p}{2m_iB}\right\}_{s_L,\mu}=0.
\ee
The initial population of SEPs is assumed to be independent of $\mu$, concentrated at small heliocentric distances $R\approx2\div 4R_\sSun$, and having a power law energy spectrum, $f(p)\propto p^{-5}$. The evolution of VDF is solved from Eq.~\ref{eq_extra} using the scheme ~\ref{eq:TVDScheme}.
\begin{figure}[htb]
\label{shm1}
\centering
\includegraphics[height={2.2in},width={2.1in}]{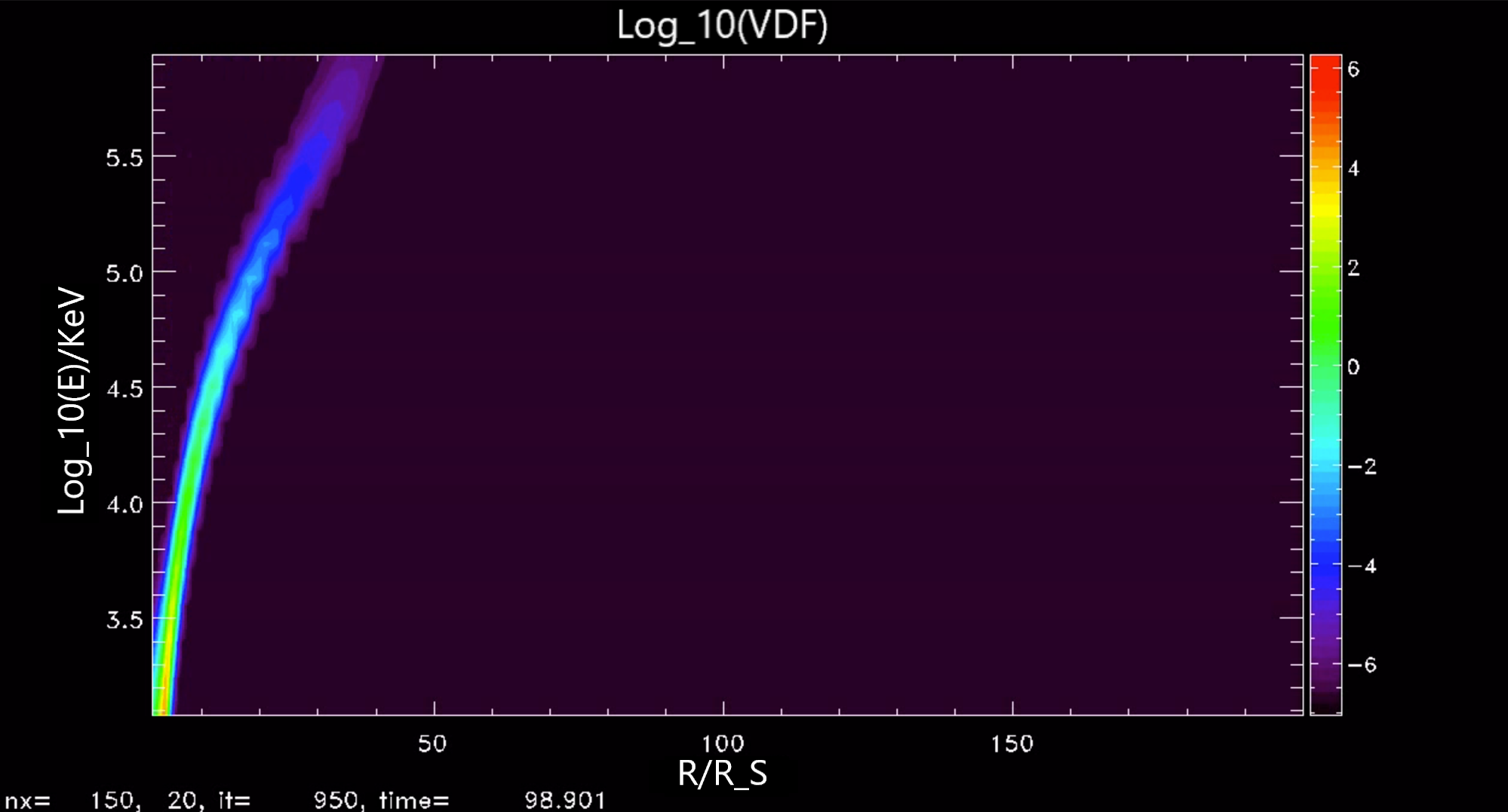}
\centering
\includegraphics[height={2.2in},width={2.1in}]{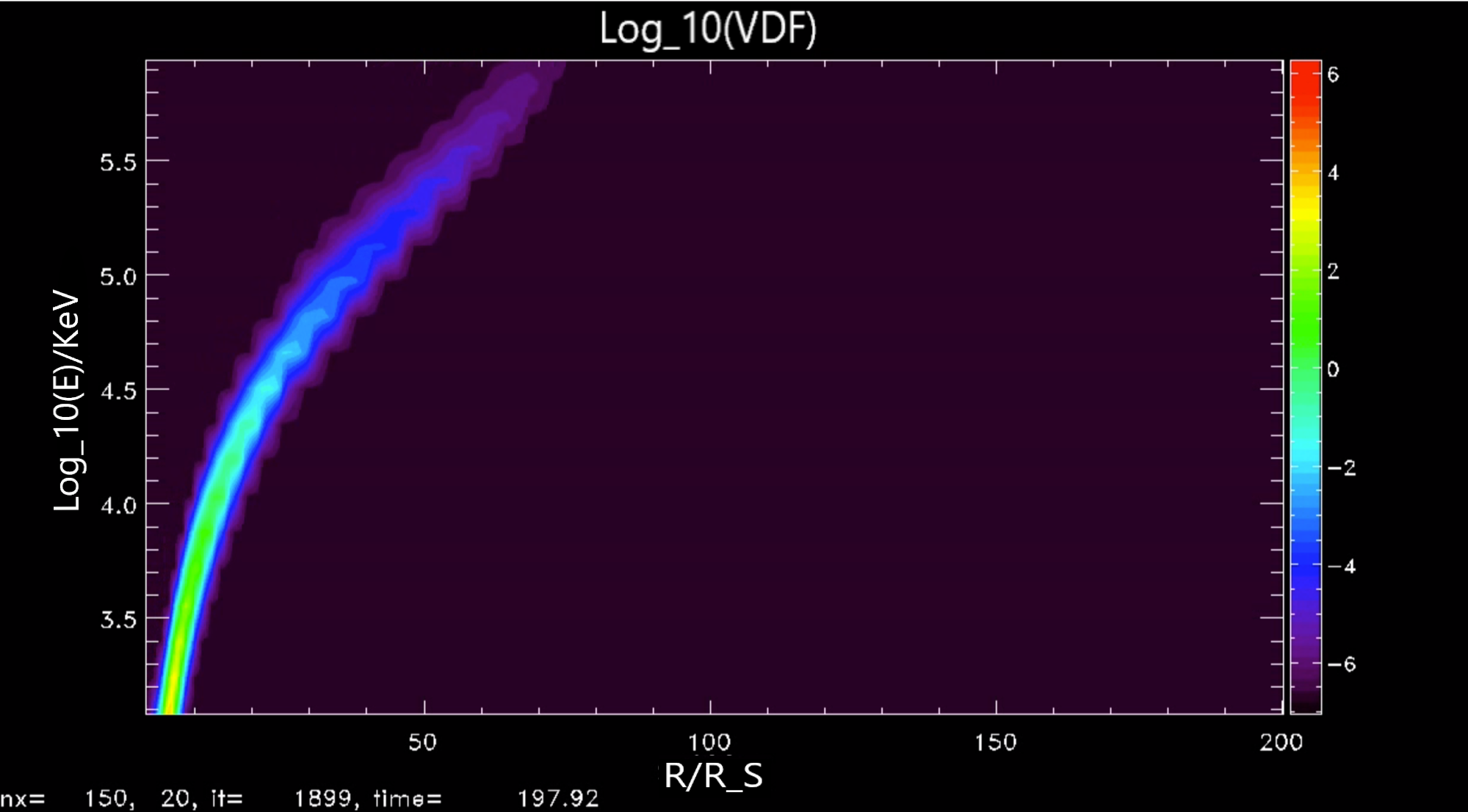}
\centering
\includegraphics[height={2.2in},width={2.1in}]{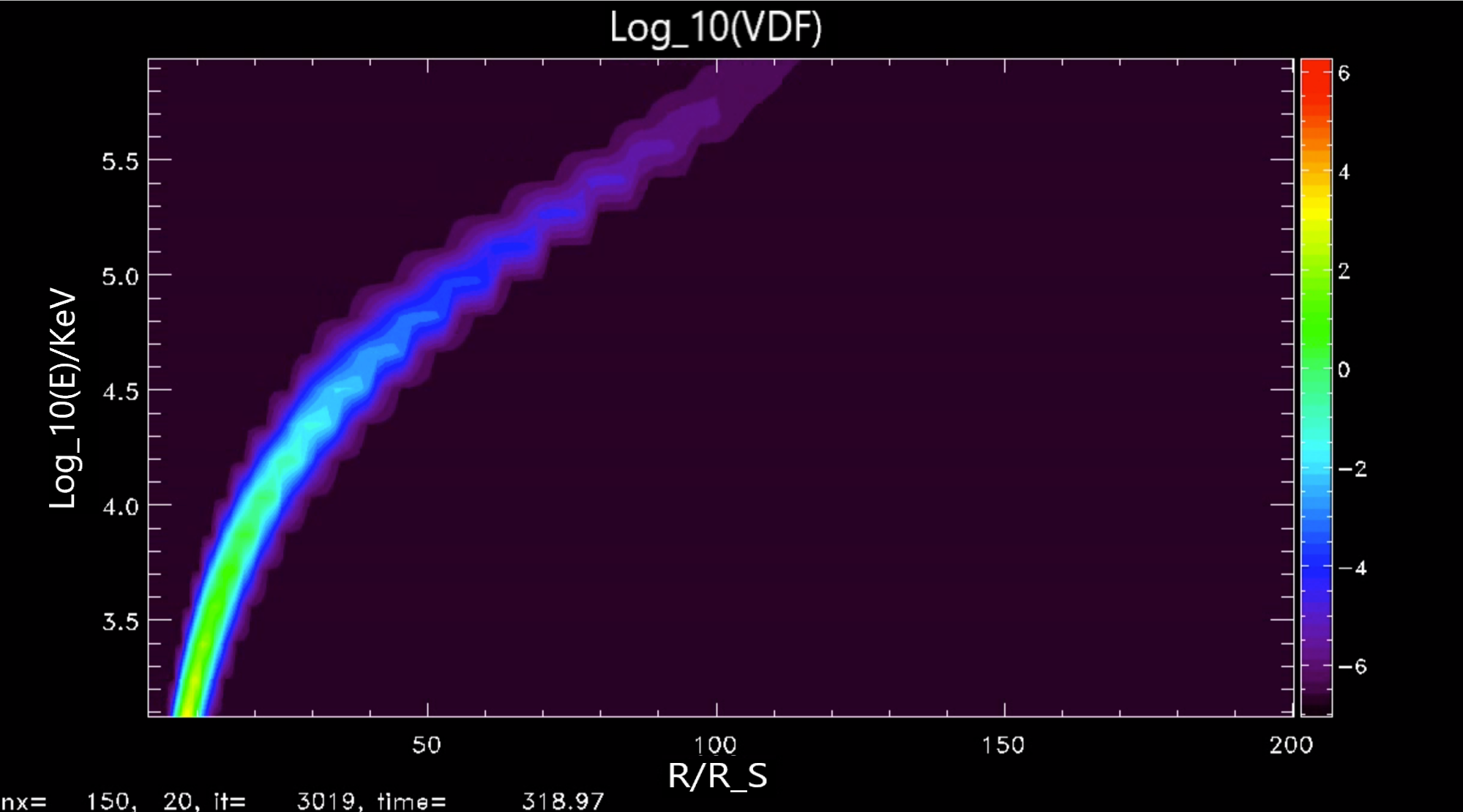}
\caption{Time evolution of the VDF as a function of the heliocentric distance normalized per $R_{Sun}$ and energy in KeV. \textbf{Left panel:} Distribution function at t=100s \textbf{Middle panel:} Distribution function at t=200s \textbf{Right panel:} Distribution function at t=320s.}
\end{figure}

Fig. A.6 shows the VDF averaged over $\mu$ as a function of $R$ normalized per the solar radius, $R_S$, and the energy in KeV. The particles of higher energies propagate faster. At the front of pulsed SEP flux (where the field is weaker) the particle parallel velocity is greater than that for bulk particles (for which the field is stronger). 
 Although initially the VDF is uniform in $\mu$, at larger $R$ the particles tend to concentrate at $\mu\rightarrow1$. These effects are due to conservation of the adiabatic invariant, $\frac{p^2_{\perp}}{2B}$, in the absence of scattering. The magnetic field decays with $R$ (see Fig.~A.5), therefore, while the particles propagate run away, their the perpendicular momentum reduces, $p_{\perp}\propto \sqrt{B}$, and the parallel one increases, $p_{\parallel}=\sqrt{p^2-p_\perp^2}$. As the result, the particles at larger $R$ move faster ($p_\|\rightarrow p$) and their momentum vector closer aligns with the magnetic field. 
\subsection{Focused Transport Combined with Particle Scattering}
\begin{figure}[htb]
\label{shm3}
\centering
\includegraphics[height={2.5in},width={3.2in}]{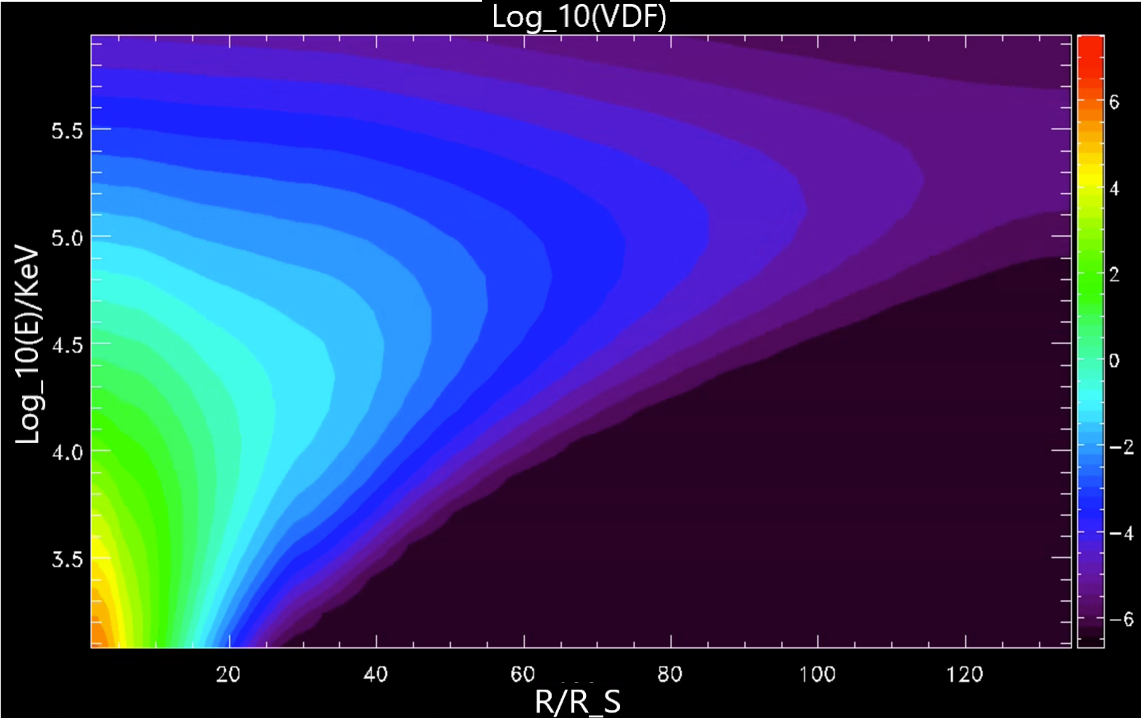}
\includegraphics[height={2.5in},width={3.2in}]{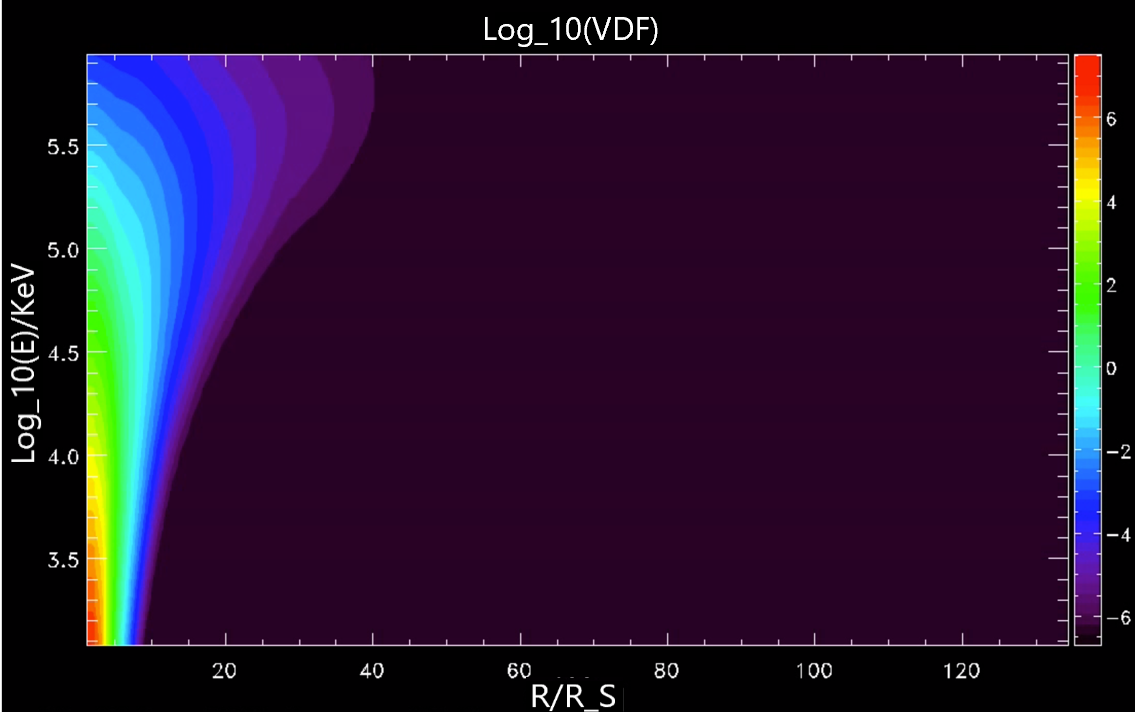}
\centering
\includegraphics[height={2.5in},width={3.2in}]{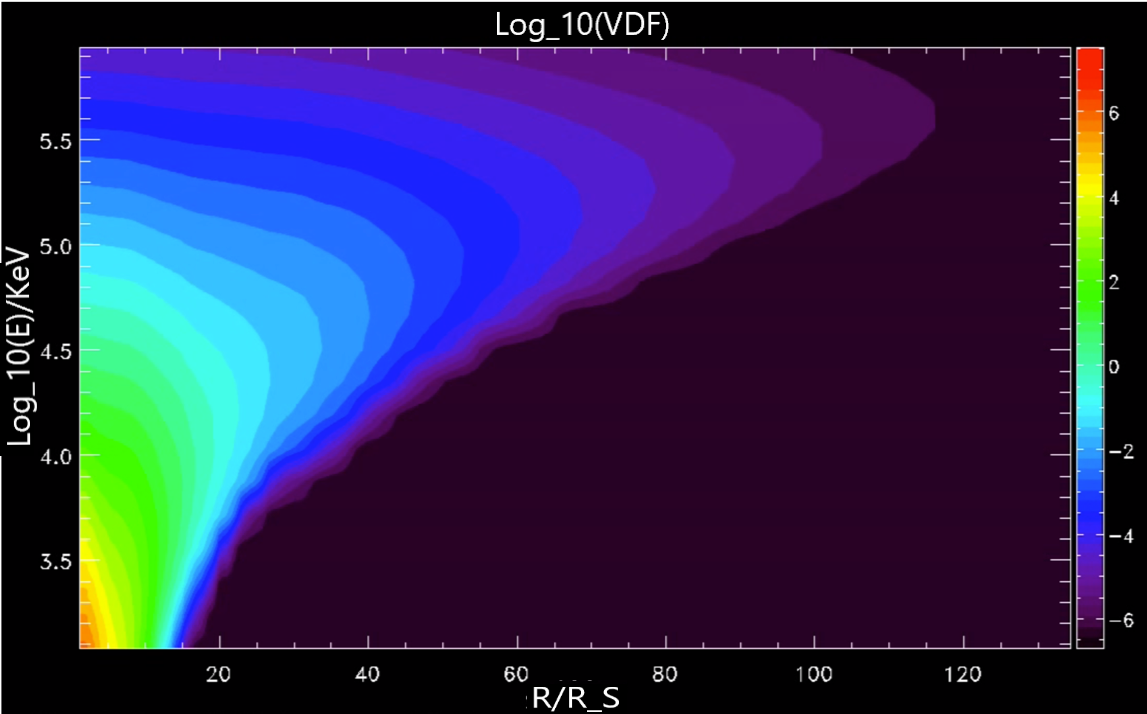}
\includegraphics[height={2.5in},width={3.2in}]{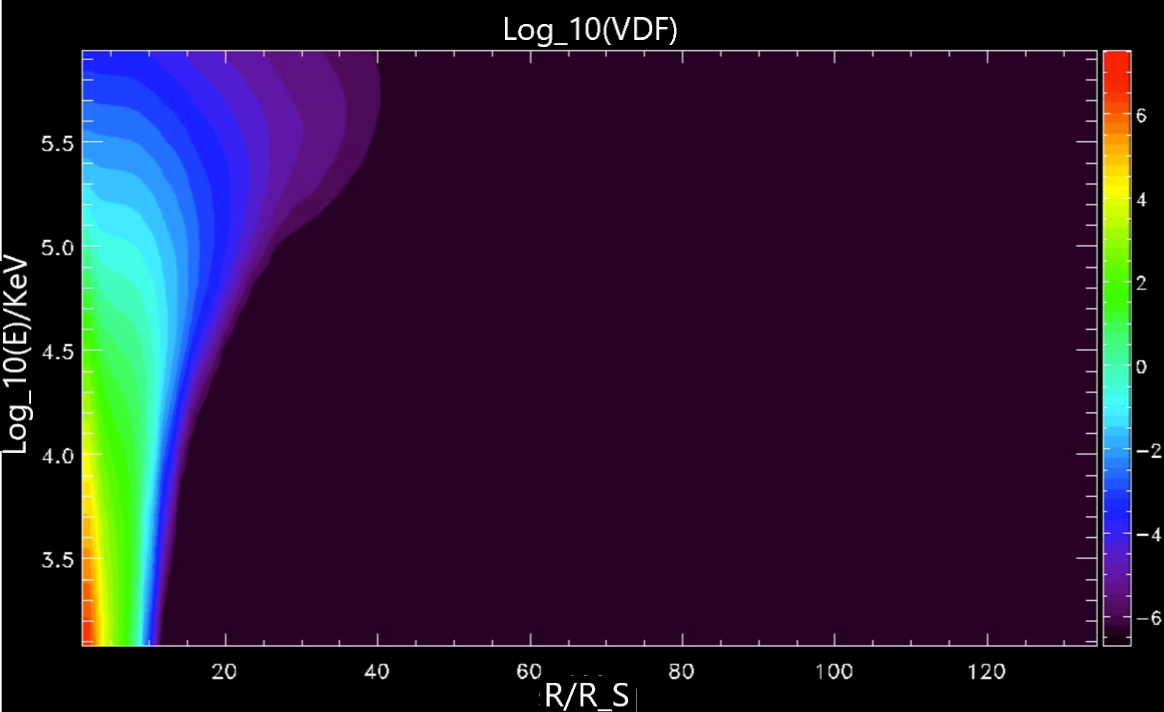}
\caption{Comparison between a spatial diffusion approximation given by Eq.~\ref{eq_47} (\textbf{Upper panels}) and more accurate pitch-angle diffusion equation  Eq.~\ref{eq_scatter} (\textbf{Lower panels}), at time $t=850s$. \textbf{Left panel set:} for larger mean free path ($L_{max}=0.8R$). \textbf{Right panel set:} for smaller mean free path ($L_{max}=0.03R$).}
\end{figure}
The combined effect of the focused transport and the particle pitch-angle scattering may be studied by including the scattering integral in the Fokker-Planck approximation,
\be\label{eq_43}
\left(\frac{\delta f}{\delta t}\right)_{scat}=\frac{\partial}{\partial\mu}\left(D_{\mu\mu}\frac{\partial f}{\partial\mu}\right), 
\ee
where $D_{\mu\mu}$ is the pitch-angle diffusion coefficient. Here, we consider the particles scattering due to the \alf wave turbulence in a quasi-linear approach and assume the Kolmogorov's spectrum of turbulence. Under these assumptions, we get an expression for pitch-angle scattering rate:
\be\label{eq_44}
D_{\mu\mu}=\frac{v}{\lambda_{\mu\mu}}\left(1-\mu^2\right)|\mu|^{2/3},\quad \lambda_{\mu\mu}=\frac{6}{\pi}\frac{B^2}{(\delta B)^2}r_{L}^{\frac{1}{3}}\left(\frac{L_{max}}{2\pi}\right)^{2/3}
\ee
\cite[see detail in][]{boro18}, in which $r_{L}=\frac{p}{eB}$ is the Larmor radius of proton, $L_{max}$ is the maximal spatial scale of  turbulence (assumed to be proportional to the heliocentric distance, $R$, and ranging from $\sim$0.03R to $\sim$0.8R), $(\delta B)^2=\mu_0w$ relates to the wave energy density, $w$, with $\mu_0$ being a vacuum magnetic permeability. Within the quasi-linear approach, in Eq.~\ref{eq_44} a condition, $\delta B\ll B$, is assumed meaning that the perturbation of magnetic field is much smaller than its averaged magnitude. However, from the upper panel of Fig. 4, we see that in the range of heliocentric distance, $R=(17\div142)R_S$, marked the red dashed vertical lines, the said inequality is reversed, $\delta B\ge B$, thus breaking applicability of the quasi-linear approach. This phenomenon has been recently observed by the Parker Solar Probe. So in order to keep using quasi-linear theory in this case, instead of the background magnetic field squared in a numerator in Eq.~\ref{eq_44}, we use the total magnetic field: $(\delta B)^2+B^2$.

The equation combining focused transport with scattering may be used to validate the {\it diffusive approximation}. This approximation is used to solve equations similar to \ref{eq_scatter} for high scattering rates (large $D_{\mu\mu}$) by means of representing the total VDF as $f(s_L,p,\mu)=f_0(s_L,p)+\delta f(s_L,p,\mu)$, where the omni-directional part,  $f_0(s_L,p)=\frac12\int_{-1}^1{f(s_L,p,\mu)}$ dominates over the $\mu$-dependent one, $\delta f(s_L,p,\mu)$, at high scattering rate.  In this limit,
Eq.~\ref{eq_scatter} reduces to the Parker equation Eq.~\ref{eq:Parker:bracket} with no adiabatic losses:
\bea
\label{eq_47}
\frac{\partial f_0}{\partial t}=B\frac{\partial}{\partial s}\left(\frac{D_{xx}}{B}\frac{\partial f_0}{\partial s}\right),
\eea
where the spatial diffusion coefficient is expressed in terms of the scattering rate:
\bea
\label{eq_48}
D_{xx}=\frac{v^2}{8}\int_{-1}^1\frac{(1-\mu^2)^2}{D_{\mu\mu}}d\mu.
\eea:
\be
\label{eq_scatter}
\frac{\partial f}{\partial t}+\frac{B}{\delta s}\left\{f,\frac{(-1+\mu^2)v}{2B}\right\}_{s_L,\mu}=\frac{\partial}{\partial\mu}\left(D_{\mu\mu}\frac{\partial f}{\partial\mu}\right), 
\ee
Eqs.~\ref{eq_47},\ref{eq_scatter} are solved numerically, using the Strang splitting method as we discussed in Section~\ref{Sec:ResultParker} and the data set as in the simulation above (see Fig.~A.5) and for the same initial condition. Fig.~A.7 shows the comparison between the spatial diffusion approximation Eq.~\ref{eq_47} (top panel) and full $\mu$-dependent solution of Eq.~\ref{eq_scatter}, at the time, $t=850$ s, assuming larger  $L_{max}=0.8R$ (hence, lower scattering rate). We see that the diffusive approximation tends to overestimate the transport of higher energy particles. Fig.~8 shows the analogous comparison, with taking much lower $L_{max}=0.03R$ (higher scattering rate). In this the results are almost identical, which demonstrates a validity of the diffusion approximation. 
\subsection{Discussion}
With numerical simulations for a simplified case, we found that the focusing effect plays an important role in the kinetic transport of SEPs. We also found that the diffusive approximation is a good one when $L_{max}$ is small, while for larger $L_{max}$ it tends to overestimate the transport of higher energy particles. Here, we do not present a complete solution to the SEP transport equation \ref{eq:kota:bracket} with three Poisson brackets. These findings will be published elsewhere.
\end{appendix}






\end{document}